\documentclass[opre,nonblindrev]{informs3} %

\DoubleSpacedXI %
\usepackage{endnotes}
\usepackage{natbib}
\bibpunct[, ]{(}{)}{,}{a}{}{,}%
\def\bibfont{\small}%
\TheoremsNumberedThrough     %
\ECRepeatTheorems

\EquationsNumberedThrough    %

\usepackage{comment}
\usepackage{wasysym}
\usepackage{graphicx}
\usepackage{algorithm}
\usepackage{algorithmic}
\usepackage{enumitem}

\usepackage{tikz}
\usepackage{subfigure}
\usepackage{pgfplots}

\usepackage{pifont}%

\newcommand{\rowdot}{\cdot}
\usepackage{utopia}
\usepackage[normalem]{ulem}
\usepackage{soul}

\def\En{{\mathbb{E}_n}}

\def\hWk{\hat{W}_{k,\cdot}}

\def\Pk{\bar{\Psi}_{k,\cdot} }
\def\Pii{\bar{\Psi}_{i,\cdot} }
\def\Phk{\hat{\Psi}_{k,\cdot} }

\def\tWkdot{\tilde W_{k,\cdot}}
\def\hWkdot{\hat W_{k,\cdot}}

\def\Pkdot{\Psi_{k,\cdot}}
\def\hPkdot{\hat \Psi_{k,\cdot}}
\def\bPkdot{\bar{\Psi}_{k,\cdot} }

\def\hPkdot{\hat{\Psi}_{k,\cdot} }

\def\Hinvkdot{H^{-1}_{k,\cdot}}
\def\bWkdot{\bar W_{k,\cdot}}

\definecolor{PennBlue}{RGB}{001,031,091}
\definecolor{PennRed}{RGB}{153,0,0}
\usepackage{hyperref}

\usepackage[blocks]{authblk}

\begin{document}
\TITLE{Latent Agents in Networks:\\ Estimation and Targeting}

\ARTICLEAUTHORS{%
	\AUTHOR{Baris Ata}
	\AFF{University of Chicago, Booth School of Business, \EMAIL{baris.ata@chicagobooth.edu}} %
	\AUTHOR{Alexandre Belloni}
	\AFF{Duke University, The Fuqua School of Business, \EMAIL{abn5@duke.edu}} %
	\AUTHOR{Ozan Candogan}
	\AFF{University of Chicago, Booth School of Business, \EMAIL{ozan.candogan@chicagobooth.edu}} %
} %

\ABSTRACT{\begin{center}
	\end{center}
 
We consider a network of agents. Associated with each agent are her covariate and  outcome. Agents influence each other’s outcomes according to a certain connection/influence structure. A subset of the agents participate on a platform, and hence, are observable to it. The rest are not observable to the platform and are called the latent agents. The platform does not know the influence structure of the observable or the latent parts of the network. It only observes the data on past covariates and decisions of the observable agents. Observable agents influence each other both directly and indirectly through the influence they exert on the latent agents.

We investigate how the platform can estimate the dependence of the observable agents’ outcomes on their covariates, taking the latent agents into account. First, we show that this relationship can be succinctly captured by a matrix and provide an algorithm for estimating it under a suitable approximate sparsity condition using historical data of covariates and outcomes for the observable agents. We also obtain convergence rates for the proposed estimator despite the high dimensionality that allows more agents than observations. Second, we show that the approximate sparsity condition holds under the standard conditions used in the literature. Hence, our results apply to a large class of networks. Finally, we apply our results to two practical settings: targeted advertising and promotional pricing. We show that by using the available historical data with our estimator, it is possible to obtain asymptotically optimal advertising/pricing decisions, despite the presence of latent agents.	

	}

\maketitle

\section{Introduction} \label{se:Intro}
{ 
Network effects are relevant in many social and economic settings.
Recent literature has empirically characterized the strength of these effects in a wide variety of domains ranging
from consumption to risk sharing and
 from education to crime
(see, e.g.,
\cite{calvo2009peer,bramoulle2009identification,patacchini2009juvenile,fletcher2012peer,blume2015linear,de2017econometrics,patacchini2017heterogeneous,angelucci2018consumption,de2020consumption}).
A different strand of the literature
complemented this line of work by shedding light on how information on the network structure can be used to improve decision making. For instance,
motivated by the
prevalence of online social networks,
\citet{candogan2012optimal,belloni2015mechanism,bimpikis2016competitive} and \citet{zhou2016targeted}
have  focused on understanding how a seller can use
the available social network information to
target agents in the social network with
 improved
pricing/seeding/marketing
decisions.

The first line of research often assumes that
historical data on all of the agents in the network are readily available.
The second line of research makes a stronger assumption and assumes that the decision maker fully knows the underlying network/influence structure.
These  informational assumptions can be too strong  in practice.
This paper asks
the following fundamental    questions:
(i) How can network effects be estimated in the  absence of data on some agents?
(ii) How can a decision maker  leverage such estimates  to improve her targeting decisions?

\subsection{Estimating Network Effects in the Presence of Latent Agents}
First, we focus on the estimation question.
Specifically, Section   \ref{se:model} presents a social network model, where the agents' outcomes depend linearly on their neighbors' outcomes, as well as their (agent-specific) covariates.
The underlying network is weighted, and the entries of the associated (weighted) adjacency matrix capture how much the outcomes of agents influence each other.
Historical data on a subset of the agents, hereafter referred to as \emph{observable agents},
are available.
The remaining agents in the network are called the \emph{latent agents}.
A priori there is  no information on the latent agents; e.g., neither their influence on the observable agents, nor the number of latent agents is known.
Crucially, the latent agents still influence (and are influenced by) the observable ones.
However,
influence structure among observable agents is also unknown.

Our first contribution is
to
provide an algorithm to
estimate
the relationship between the
observable agents' outcomes and   covariates (Section \ref{se:estimation}).
In our model, this relationship is linear and is captured by a matrix which is closely related to the underlying network structure.
Specifically, this matrix is a
sub-block (corresponding to the observable agents) of the
inverse of the matrix given by the difference of a diagonal matrix and the (weighted) adjacency matrix.
We investigate how to estimate this matrix, whose $i$th row captures the change in the outcome of agent $i$ due to changes in the covariates of this agent and the remaining agents.

Social networks often involve a large number of agents that are sparsely connected.\footnote{See, e.g., \citet{ugander2011anatomy}, who report that in May 2011, the Facebook graph consisted of 721 million active nodes, and  users had on average  190 Facebook friends.}
Consequently, our  problem is a high-dimensional estimation problem.
A natural way to deal with this high dimensionality
and obtain efficient estimators
is
to exploit the underlying sparsity.
That being said,
even when the underlying network is sparse,
the aforementioned matrix need not be sparse (since it is given by the \emph{inverse} of a matrix related to the structure of the network; see Section \ref{se:estimation}).

We get around this difficulty by using  a notion of approximate sparsity for this matrix
(see Definition~\ref{def:appxSparse}).
Intuitively, this sparsity notion posits that even though the matrix is not sparse, it admits a sparse approximation. Under the approximate sparsity assumption, we provide an estimator for this matrix, and characterize its convergence rates in matrix $1$ and $\infty$-norms.
Controlling the estimation errors in both matrix norms is methodologically challenging, but needed in our setting. This is because, for many decision problems where the payoff of the decision maker depends on the underlying network structure, using estimators with small errors in  both norms enables constructing near-optimal  policies (see the  relevant discussion in Section \ref{subse:applications}). Such guarantees are not possible  for estimators that control errors only in one or the other norm.  From the statistical perspective this amounts to controlling errors  not only within each model  (that captures the dependence of the outcome of a \emph{single} agent on all covariates -- which is standard) but also \emph{across} linear models.  Achieving this necessitates   leveraging  a different approximate sparsity condition that encodes sparsity across linear models, and using  thresholded estimators after de-biasing. 
We also provide  distributional limits for our estimator that allow for constructing valid confidence intervals.

It is not a priori clear  when the aforementioned approximate sparsity condition holds.
Our second contribution is to provide conditions on the network structure and edge weights that ensure that this condition is satisfied (Section \ref{se:estimationExamples}).
Using ideas from the theory of banded matrices, approximation theory,   spectral theory, and
Banach algebras of matrices and their off-diagonal decay properties,
we establish that our approximate sparsity assumption holds
for large classes of networks
under  (diagonal dominance) assumptions that are standard in the social networks literature.
For such classes of networks, we prove that
our estimators yield small approximation errors
even when the number of observations scales logarithmically with the number of observable agents, thereby making these estimators suitable for large networks.

\subsection{Applications: Improving Targeting Decisions} \label{subse:applications}

Our third contribution (Section \ref{se:applications}) is to
illustrate how the estimation framework can improve targeting decisions of a decision maker.
We focus on two applications:  targeted
advertising  and
pricing.
In both applications,
agents consume a divisible product that exhibits positive network externalities.
Due to the presence of network externalities, each agent's consumption decision depends on those of her neighbors in the underlying social network.
Hence, an ad shown or a discount offered to  an agent impacts the consumption decision of not only  this agent, but also those of her neighbors (as well as their neighbors and so on).

A subset of these agents (observable ones)
purchase the product
from an online seller who offers it
through a social networking platform.
While the online seller does not know how much agents influence each other,
she has historical data on  the
targeting/consumption decisions for the observable agents.
We investigate how this seller should target these agents with ads or discounts to maximize her payoff, which depends on the sales to observable agents.

The targeting question here is partly motivated by  firms that
offer targeted prices/ads through online social networks.
	For instance,
prominent social networking platforms such as Facebook allow retailers to target individuals who previously shopped with them
via ads
through their
``Custom Audiences''
(see, \cite{facebookCustomAudience}).
Moreover, these ads can take the form of ``Offer Ads'', where
the targeted individuals
  receive unique  promo codes
(see \cite{facebookOfferAds} and \cite{facebookUniqueCodes}).
This type of ``custom audience'' targeting is not unique to Facebook, and it is common to all major social networking platforms (see, e.g., \cite{instagram, twitter}).

Consider the
customers who shopped with an online seller in the past (i.e., the ``custom audience'').
The seller can target them with
informative ads (which help increase sales, as in our advertising application) or
promo codes (which correspond  to price discounts in our pricing application). These individuals correspond to the observable agents in our model.
Any agent who is not in the custom audience is latent.
These agents can correspond to other participants of the social networking platform, or agents who
do not participate in such platforms but still
influence others through offline channels.\footnote{These cases are mathematically equivalent. For sake of exposition, when we present our results in Section \ref{se:applications}, we frame the problem through the latter case.}
For products that exhibit network externalities (e.g., fashion items), the optimal targeting decisions (advertising intensities and/or discounts involving promo codes) depend on the underlying network structure. This information is not readily available to the online  seller.
That said, information on whether/how the targeted individuals engage with ads (e.g., whether the individual uses the ``unique code'' and what they purchase with it -- see \cite{facebookUniqueCodes})
is available. In the context of our model, this means that data on past  targeting decisions as well as  outcomes are available.
Note that  advertising  a product
may increase the consumption of observable agents,
but it may also increase the  latent agents'
incentives to obtain similar products
(possibly
through  alternative channels/sellers) due to the presence of network externalities.
The latter effect could  in turn incentivize other observable agents to consume more.
How should the seller optimize her advertising/pricing decisions for the observable agents?

We show that the seller can use the available historical data to estimate network effects, and construct targeting decisions that
would be optimal if the network effects were precisely captured by the estimated quantities.
Leveraging the convergence rates of our estimator in matrix
$1$ and $\infty$-norms, we show that the constructed targeting decisions are asymptotically optimal
in the sense that the ratio of
 the payoff under these
 to  the optimal payoff that would be achieved if the seller  knew the underlying influence structure, converges to one as the number of samples increases.
Once again, we
show that if the number of observations is at least logarithmic in the number of observable agents, then
a  small payoff gap can be guaranteed  and we provide a precise characterization of this gap in terms of a measure of approximate sparsity.
These results indicate that even with the
limited amount of available data, the seller
can exploit approximate sparsity to construct targeting decisions that are near-optimal.

To the best of our knowledge, the proposed network setting with latent agents is new. Collectively, our results shed light on how in such settings
the mapping between the
covariates and
 outcomes of the observable agents  can be estimated, and how such  estimates can be used to construct approximately optimal targeting decisions.
}

\subsection{Related Literature}

Our paper is closely related to the literature on
identification and estimation of network effects, and targeting.
The application we cover in Section \ref{subse:estimationPrices} also relates to the pricing literature.

\paragraph{Identification and estimation of peer effects:}
There is a growing body of literature on identification and estimation of peer effects, where
it is often assumed that observations on outcomes and covariates of \emph{all} agents are available \citep[e.g.,][]{sacerdote2001peer,calvo2009peer,epple2011peer}.
The canonical model is the linear specification presented in \citet{manski1993identification}, which is
similar to the one considered in the  present paper (although \citet{manski1993identification} considers additional covariates).  \citet{manski1993identification}
points out
an identification issue for the estimation of peer effects (the reflection problem):
it is not possible to disentangle the endogenous effects (which are given by the average decisions of individuals in a group) and the exogenous (contextual) effects (which are given by the average of the covariates of individuals in the group).
\citet{bramoulle2009identification} and \citet{blume2015linear} show that  when the underlying network lacks a regular structure this identification problem disappears.
In the present paper, we assume that
an agent's   outcome is impacted  by the outcomes of her neighbors rather than their covariates. Hence, the exogenous effects do not play a role, and the aforementioned identification issue is not relevant.

\citet{manresa2013estimating} and \citet{rose2017identification} consider the identification and estimation of peer effects when the network is sparse. These papers rely on panel data, and leverage  lasso or instrumental variable approaches
\citep[e.g., due to][]{gautier2014high} to the estimation problem.
A relevant recent survey of \citet{de2015econometrics} discusses the estimation of the network structure via penalization methods like lasso, SCAD, and others.
In the presence of latent agents the underlying aggregate influence structure among observable agents is no longer sparse
(see Section~\ref{se:estimation}).
Moreover, as opposed to estimating the influence structure, our focus is on estimating the inverse of a matrix  related to it (which also is not sparse).
Hence, the results from this line of literature are not readily applicable.
It is also worth mentioning that
the methodology in these papers can be used to obtain
estimators that yield small estimation errors for each row of the matrices of interest.
However, in this paper our estimator yields small estimation errors simultaneously for both rows and columns, which is essential for
ensuring asymptotic optimality in the subsequent decision problems.

\paragraph{Decision making with incomplete network information:}
A relevant stream of the social networks literature assumes that
as opposed to knowing the precise network structure, a decision maker has access to some summary statistics about the network; e.g., she knows  the degree distribution in the network, which
partially reflects the extent of externalities.
In such a setting,
each agent's degree might be modeled as her private information and
the decision maker's challenge is to
elicit the relevant information and
decide on how to optimally target
agents
using this information.
The literature provides various mechanisms
for eliciting agents' private information and
optimizing
targeting decisions in the context of
pricing, advertising, or product referrals \citep[e.g.,][]{hartline2008optimal,campbell2013word,galeotti2010,lobel2015customer,yingJu2016}.
By contrast,  we assume that past data on individual outcomes are available for some (observable) agents,
and shed light on how the available data can be employed to better understand the underlying influence structure and improve targeting decisions.

\paragraph{Learning and pricing problems:}
There is also a related  literature on the question of learning optimal prices through dynamic price experimentation, which is of interest even in the absence of social interactions
\citep[e.g.,][]{harrison2012bayesian,keskin2014dynamic,besbes2015surprising}.
This literature assumes that a parameter of the underlying demand system is unknown,
and develops price experimentation policies
that do not have a large  performance gap relative to
the optimal policy derived in a setting where the  underlying demand system is fully known.
Similarly, in our application in Section \ref{subse:estimationPrices}
the  demand system is not fully known,
since the underlying network is not known.
On the other hand,  due to network externalities, the seller now finds it optimal to price differentiate agents. Thus,  the seller needs to learn an optimal price vector as opposed to a single price, which leads to a more complex learning problem. Although it is beyond the scope of the present work, it would be interesting to study how this dynamic price experimentation can be done in a way that minimizes the long-run performance gap in the seller's profits measured over the entire experimentation horizon.

\section{Model and Preliminaries}\label{se:model}

We consider a social network  with a set of agents  $V$ and a set of connections among agents  $E\subset V\times V$.
We represent the social network with a directed graph
${\cal G}=(V, E)$, where $V$ and $E$ respectively correspond to nodes and edges.
In the social network, the set of agents who are connected to agent $i$ are referred to as the neighbors of $i$, denoted by $N(i):=\{j|(i,j)\in E\} \cup \{j | (j,i)\in E \}$.

Each agent $i$ is associated with an outcome ($y_i$), which linearly depends on an agent-specific covariate ($p_i$)  as well as the outcomes of her neighbors.
Specifically, we denote the outcome of agent $i$ at time $t$ by
	\begin{equation}\label{eq:dependence}
	y_i^{(t)} = \frac{1}{\lambda_i}  \left(a_i +  \sum_j G_{ij} y_j^{(t)} -  p_i^{(t)}  + 	\xi_i^{(t)}\right).
\end{equation}
Here $a_i$ is an agent-specific intercept term, and
 $\xi_i^{(t)}$ denotes an {(idiosyncratic)}  shock for agent $i$ at time $t$, which
impacts her outcome  but is not observable to the researcher.
The term $\lambda_i$ denotes an agent-specific scaling factor (which will be set equal to $1$ in one of our applications, and its effect will be estimated in the other one; see Section \ref{se:applications}).
We assume that $\{\xi_i^{(t)}\}_{i,t}$ have zero mean, and are independent over time,  but are possibly correlated across agents.
Each edge $(i,j)\in E$ is associated with a weight
$G_{ij}\in \mathbb{R}$
that captures
how much  the outcome of $j$ influences that of agent $i$ ($G_{ij}=0$ if agents $i$ and $j$ are not connected, i.e., $(i,j)\notin E$).
Unless  otherwise noted, we do not require the weights to be symmetric; i.e., in general $G_{ij}\neq G_{ji}$.
Intuitively this allows an agent to influence her neighbors more than she is influenced by them.
Until Section \ref{se:applications}, we also allow the weights to be positive or negative.
We refer to the set of weights $\{G_{ij}\}_{i,j \in S}$ as the influence structure in $S\subset V$.

We assume that  a subset $V_O\subset V$ of the agents
participate in an online platform.
These correspond to the  {observable agents} in our model, and we assume that  data on past
covariates and outcomes of these agents are available.
By contrast, no such  information  about the remaining (latent) agents is available.
We denote the set of latent agents by $V_L$, and assume that $V_O\cup V_L=V$ and $V_O \cap V_L = \emptyset$.

For a given set of parameters $\{v_i\}_{i\in V}$, we denote the associated column vector by ${v}$, e.g., ${a,p,y}$ respectively stand for $\{a_i\}_{i\in V}$,   $\{p_i\}_{i\in V}$, and~$\{y_i\}_{i\in V}$.
For any vector ${v}\in \mathbb{R}^{|V|}$, we represent its entries corresponding to observable and latent nodes by ${v}_O$ and ${v}_L$ respectively, i.e., ${v}=[{v}_O; {v}_L]$.\footnote{For any column vectors ${v}_1,{v}_2$, we denote by $[{v}_1; {v}_2]$ the column concatenation,
	and by $[{v}_1, {v}_2]$ the row concatenation if they are the same size. We similarly denote row/column concatenation for matrices.}
Similarly, for any matrix $A\in \mathbb{R}^{|V|\times|V|}$, we express the blocks corresponding to observable and latent components as follows:
\begin{equation*}
	A=
	\begin{bmatrix}
		A_{OO}, & A_{OL}\\ A_{LO}, & A_{LL}
	\end{bmatrix}.
\end{equation*}
We denote by $A_{k,\rowdot}$ the $k$th row of $A$,
by
$A_{\rowdot,k}$
its $k$th column.
We use ${e} \in \mathbb{R}^{|V|}$ to denote a vector of ones.

In our applications in Section \ref{se:applications}, the covariates associated with the latent agents are identical and constant over time (e.g., as they represent the advertising intensity/price discounts through the online platform; which are equal to zero for the agents that are not on the platform).
{Motivated by this, and to simplify the exposition, we let
$p_i^{(t)} = {p}_0$ for all $i\in V_L$ and $t$, where $p_0\in \mathbb{R}$  is a constant.
We emphasize that our estimation results hold under weaker conditions (e.g. it suffices to have, $E[ p_L - E[p_L] | p_O ]=0$). We revisit this point in
Remark \ref{rem:latent}. }

Let $\Lambda\in \mathbb{R}^{|V| \times |V|}$ denote a diagonal matrix whose $i$th diagonal entry is given by $\lambda_i$, and let $G\in \mathbb{R}^{|V| \times |V|}$ denote a matrix whose $i,j$th entry is $G_{ij}$. Let $M \in \mathbb{R}^{|V| \times |V|}$ be such that
$$M:=\Lambda-G.$$
Throughout the paper we index the  entries of matrices $G$ and $M$  (as well as other network-related matrices) by the nodes of the underlying network.

Using  matrix notation, and rearranging terms in \eqref{eq:dependence}, we obtain the following relationship among the outcomes and covariates:
\begin{equation} \label{eq:firstOrder}
	{y}^{(t)}= M^{-1} ({a} + {\xi}^{(t)} -{p}^{(t)}).
\end{equation}

We next focus
on observable agents
and restate this relationship
 more explicitly for these agents.
To this end, we introduce the following matrices:
\begin{equation}\label{eq:sMatDef}
	\begin{aligned}
		S_{OL}:=  M_{OL}M_{LL}^{-1} \mbox{\quad and \quad}
		H:= M_{OO}-M_{OL}M_{LL}^{-1}M_{LO},
	\end{aligned}
\end{equation}
and also the following vectors:
\begin{equation}\label{eq:defVo}
	{v}_O := H^{-1}  {a}_O-H^{-1} S_{OL} ( {a}_L - {p}_L) \mbox{\quad and \quad}
	\varepsilon_O^{(t)} := H^{-1} \xi_O^{(t)}-H^{-1} S_{OL}  {\xi}^{(t)}_L.
\end{equation}

\begin{lemma} \label{lem:firstOrderO}
	In period $t$, the observable
	agents' outcomes are given by
	\begin{equation}\label{eq:def:noise1}
		{y}_O^{(t)}
		=  {v}_O-H^{-1} {p}_O^{(t)} +\varepsilon_O^{(t)}.
	\end{equation}
\end{lemma}
{The preceding discussion implicitly assumes the invertability of $M$, its sub-block $M_{LL}$, %
	as well as $H$.
	This is a mild condition that will be imposed in the remainder of our analysis.
	We also assume that  the maximum absolute row and column sums of $M^{-1}$
	(equivalently the $1$ and $\infty$ norms of this matrix)
	are bounded.
	It is worth noting that
	in our applications, we will focus on settings where $M$ satisfies some diagonal dominance condition (see Assumption \ref{ass:DiagDominance}), which readily implies all of these assumptions (see Section \ref{se:applications} and Lemmas \ref{lem:MInv},  \ref{lem:matBounds}).
	Finally,
	we
	conduct our analysis under the assumption that  the
	parameters
	$\{a_i, \lambda_i\}$ as well as the covariates $\{p_i^{(t)}\}$ are bounded, i.e.,
	$|a_i| \leq \bar{a}$,
	$|\lambda_i| \leq \bar{\lambda}$, and $|p_i^{(t)}| \leq \bar p$ for
	all 		$i\in V$ and $t$, and
	some constants $\bar a, \bar \lambda, \bar p \in \mathbb{R}_+$.
}

Lemma \ref{lem:firstOrderO} suggests that the entries of $H^{-1}$ capture how changes in the covariates of the observable  agents impact their outcomes.
Intuitively, the  components
$M_{OO}$ and $M_{OL}M_{LL}^{-1}M_{LO}$
of $H$, respectively represent the direct influence  of observable agents' outcomes on each other
and their indirect influence through the latent agents.
We refer to $H=M_{OO}-M_{OL}M_{LL}^{-1}M_{LO}$ as the \emph{aggregate externality structure}  among observable agents.

{In our applications in Section \ref{se:applications}, the covariates will correspond to
prices offered for a product that exhibits network externatilites or advertising intensity for this product; and the outcome will capture the agents' purchase quantities.
We will investigate how the decision maker should set these covariates to maximize an objective  of interest, e.g., expected revenues or sales.
Lemma~\ref{lem:firstOrderO} suggests that knowing  $H^{-1}$ is critical
for deciding how to target observable agents so as to maximize the aforementioned objectives.
We will focus on a setting where the decision maker
does not know $H^{-1}$, but
has historical information on the past covariates and outcomes.
The crucial questions are whether using the aforementioned data the matrix $H^{-1}$
can be estimated and whether any such  estimates can be used for improving pricing/advertising decisions. }

{Linear models have a long history of facilitating empirical research in various fields.
	As such, linear models similar to  \eqref{eq:dependence} are prevalent in the
	network estimation literature.
	\cite{topa2015neighborhood} provides an
overview of research on  social networks and their
role in shaping behavior and economic outcomes.
In particular,	the authors discuss local-aggregate and local-average network models,
	which
	have a similar structure to ours and assume that the outcome of each agent depends linearly on the outcomes of other agents as well as some covariates.
	The difference between these model is in the specification of edge weights.
The first class of models
assume that  all edges have identical weights, whereas the  second one scales down the edge weights adjacent to each node by the degree of that node.
Our model allows for more general influence weights among agents than identical weights or degree-scaled weights.	
\cite{topa2015neighborhood} summarizes various papers that use   these models and
highlight applications in education
\cite{calvo2009peer,de2010identification,lin2010identifying,bifulco2011effect,boucher2014peers,patacchini2017heterogeneous},
crime
\cite{patacchini2009juvenile,liu2012criminal,lindquist2014key},  labor
\cite{patacchini2012ethnic},
consumption
\cite{de2020consumption},
smoking
\cite{fletcher2010social,bisin2011empirical}, alcohol consumption
\cite{fletcher2012peer},
and risk sharing
\cite{angelucci2018consumption}.
More recent applications of these models in other domains, e.g., R\&D networks, have also appeared in the literature (e.g., see \cite{konig2019r}).
 \cite{de2017econometrics} also highlights the prevalence of linear models and
argues that
``The canonical representation for the joint determination of outcomes mediated by social
interactions builds on the linear specification''.

This literature has largely assumed  away the presence of latent agents, despite their prevalence in  network data.
The next section provides a framework for estimating network effects
(summarized through the $H^{-1}$ matrix)
in the presence of latent agents, by focusing on
the linear model
in \eqref{eq:dependence}.
In doing so,
we assume that
aside from past covariates/outcomes of observable agents,
no
additional information is available. In particular, we do not assume  the knowledge of the underlying network structure (or the $\{G_{ij}\}$ parameters), the set $V_L$ of latent agents (or its cardinality),  parameters $\{a_i,\lambda_i\}_{i\in V}$, or taste shocks $\{\xi_i^{(t)}\}_{i,t}$.
}

\begin{remark}[Identification issues]
		Observe that in the absence of latent agents $H^{-1}=M^{-1}$, and hence its estimate readily reveals  the dependence of outcomes on covariates (by~\eqref{eq:firstOrder}).
		However, when there are latent agents
		it is not possible to identify   $M^{-1}$, and hence we restrict attention to estimation of $H^{-1}$.
		To see this,
		consider a network with latent agents, where
		the mapping between outcomes and covariates is as in Lemma \ref{lem:firstOrderO}.
		Note that
		another network, which has no latent agents and admits an
		influence structure $\bar{G}_{ij}=-H_{ij}$ for $i\neq j$ for all observable agents,
		exhibits the same relationship between covariates and outcomes.
		Thus, either network could explain covariate/outcome observations, and it is not possible to identify the true $M^{-1}$ matrix.~$\hfill\square$
	\end{remark}

\section{An Estimator for Large Networks} \label{se:estimation}
This section focuses on the estimation of
the $H^{-1}$ matrix
 from
    panel data on the observable agents,
   $\{(y_O^{(t)},p_O^{(t)} ) \}_{t\in[n]}$, where
   $[n]:=\{1,\dots,n\}$.
We are particularly interested in large networks, where the number of observable agents
$|V_O|$ can exceed the number of periods $n$ observed in the data.
The linear specification in Lemma \ref{lem:firstOrderO}
can be exploited for the estimation of $H^{-1}$
provided that covariates $p_O^{(t)}$ and taste shocks $\xi^{(t)}$ (similarly $\varepsilon_O^{(t)}=H^{-1} \xi_O^{(t)}-H^{-1} S_{OL} \mathbf{\xi}^{(t)}_L$)
are orthogonal. That is,
\begin{equation}\label{eq:moment}
{E}[(\varepsilon_O^{(t)})(1;p_O^{(t)})^T]=
{E}[\{y_O^{(t)} - (v_O - H^{-1}p_O^{(t)})\}(1;p_O^{(t)})^T]=0.
\end{equation}

To estimate the coefficients of interest (and in particular, the entries of  $H^{-1}$ and $v_O$)
 we use the moment condition (\ref{eq:moment}) with an
$\ell_1$-regularization procedure (which results in a variant of the Dantzig selector; see \citet{CandesTao2007,belloni2017pivotal}).
The use of the $\ell_1$-penalty is often motivated by  big data applications where
underlying models involve
many different variables,
yet the available sample size is substantially smaller. Estimation in such settings becomes possible only if a relatively small number of variables matter, i.e., if the underlying coefficient vector is sparse.
Employing $\ell_1$-penalty guarantees the sparsity of the estimator, and  allows for estimating  the relevant coefficients.

The number of entries of the matrix $H^{-1}$ is large ($|V_O|^2$ -- and hence scales quadratically with the number of observable agents),
which naturally results in a high-dimensional estimation setting.
Yet, this matrix is not necessarily sparse  even if the underlying influence structure is sparse (i.e., $G$ has a small number of nonzero entries).
In fact, as we establish in
Lemma \ref{lem:MInv} of Appendix \ref{subse:MatrixLemma},	
if the network is (strongly)
connected (and satisfies an additional assumption imposed in  our applications),
then all entries of $M^{-1}$ as well as $H^{-1}$ are nonzero.
There are two effects that contribute to this nonsparsity.
First,  even if the underlying influence matrix $G$ is sparse, the related matrices obtained after a matrix inversion operation
(such as $M^{-1} = (\Lambda-G)^{-1}$ as well as $H^{-1}$)
are not sparse.
Second, in the presence of latent agents, the aggregate externality structure
		$H= M_{OO}-M_{OL}M_{LL}^{-1}M_{LO}$ is not sparse (due to the term $M_{OL}M_{LL}^{-1}M_{LO}$).

To tackle this issue,
in Section \ref{subse:approximateSparsity},
we introduce an ``approximate sparsity'' condition on $H^{-1}$.
In Section \ref{subse:estimationAlgorithm},
 we provide our estimation algorithm for $H^{-1}$, and
in Section  \ref{subse:estimationBounds},
we   obtain rates of convergence
for our algorithm
under the approximate sparsity condition.
We revisit this condition in Section \ref{se:estimationExamples} and establish that it  holds for a large class of networks and influence structures. Hence, the results of this section are applicable to those general settings.

To facilitate the analysis, we next introduce additional notation.
We use
$\Phi$ to denote the cumulative density function of the standard normal distribution.
We define $\bar{V}_O := \{0\} \cup V_O$,
and index the entries of vectors in $\mathbb{R}^{|V_O|}$ and
$\mathbb{R}^{|\bar{V}_O|}$ by the elements of $V_O$ and $\bar{V}_O$ respectively.
For instance, $(1;p_O^{(t)})_j$ is equal to
$p_j^{(t)}$ if $j\in V_O$, and is equal to $1$ otherwise, where $(1;p_O^{(t)}) \in \mathbb{R}^{|\bar{V}_O|}$. We denote by $\mathbf{1}_{kj}$
with $k,j\in \bar{V}_O$ an indicator variable that takes the value one if $k=j$ and zero otherwise.
We
denote by $\| {v}\|_p$ (for $p\in \{1,2,\infty\}$)
	the
	$\ell_p$ norm of a vector $ v$, i.e.,
	$\| {v}\|_p = (\sum_i |v_i|^p)^{1/p}$ (with the convention $\| { v}\|_\infty= \max_i |v_i|$).
	Similarly, for a matrix $A$, $\| A \|_p$ denotes the induced matrix $p$-norm, i.e.,    $\| A \|_p := \sup_{{x}\neq 0} \frac{\|A{x}\|_p}{\|{x}\|_p}$.
	Observe that for $p=\infty$ we get
	the maximum absolute row sum
	($||A||_\infty = \max_i \sum_{j} |A_{ij}|$), and for $p=1$ we get
	the maximum absolute  column sum
	($||A||_1 = \max_j \sum_i |A_{ij}|$) of the relevant matrix.

\subsection{Approximate Sparsity} \label{subse:approximateSparsity}
Before we proceed with the details of our estimation approach, we formalize our approximate sparsity notion.
\begin{definition} \label{def:appxSparse}
	We say that a matrix is \emph{$s$-sparse} if it has at most $s$ nonzero entries in each row and column.
Moreover, 	
	we say that the
\emph{matrix
	$H^{-1}$
	admits
	 an
	 $(s, r_1)$-sparse approximation} if
	there exists an $s$-sparse
	 matrix $\bar{W}$
	 such that
	 \[
	 \max{\bigg \{ }
	 \| H^{-1}-\bar{W} \|_1,\| H^{-1}-\bar{W} \|_\infty
	 {\bigg \} }
	  \leq r_{1}.
	  \]

\end{definition}

Motivated by Definition~\ref{def:appxSparse}, we say that
a network
admits an
$(s,r_1)$-sparse approximation if the associated $H^{-1}$  matrix does so.
A given  network can admit different
sparse approximations with different parameters
$(s, r_1)$.
There is a clear trade-off between such approximations:
more sparse approximations will lead to higher approximation errors.
To achieve the best tradeoff in the estimation,
it may be appropriate to consider less/more sparse
approximations (i.e., larger/smaller~$s$)  depending on the  number of available observations.
In what follows, we  focus on
$(s_n, r_{1n})$-sparse approximation
	of networks, when
	there are $n$ observations $\{({y}^{(t)}_O,{p}^{(t)}_O)\}_{t=1}^n$.
We
provide the convergence rates of our estimator (in various norms) in terms of
$s_n$ and  $r_{1n}$.

\subsection{The Estimation Algorithm} \label{subse:estimationAlgorithm}

\begin{algorithm}[tbp]
	\begin{itemize}[leftmargin=*]
		\item[] {\bf Input.} The data $\{ ({p}^{(t)}_O, {y}_O^{(t)}) :
		t\in[n]\}$ and the thresholds $\{ \mu_{kj}\}_{ k, j \in V_O}$.
		\item[] {\bf Initialize.}
		 $M_n^2= 1\vee \max_{k\in V_O}\frac{1}{n}\sum_{t=1}^n(p_k^{(t)})^4$, $\tau = \frac{1}{4M_n}$, and $\lambda = \frac{1}{\sqrt{n}}  \Phi^{-1}\left(1-\frac{1}{3n|V_O|^2}\right)$.

		\item[]  {\bf Step 1.}
		Compute an initial estimate $(\hat v,\hat W)$ by solving the following optimization problem:
		\begin{equation}\label{def:lasso}
		\begin{array}{rl}
		\displaystyle (\hat W, \hat v, \hat z) \in \arg\min_{\tilde W,\tilde v,z} \quad &
		\sum_{k\in V_O} \| (\tilde v_k, \tilde W_{k,\cdot})\|_{1} + \tau z_k\\
\mathrm{s.t.} \quad & \left| \frac{1}{n}\sum_{t=1}^{n} \{y_k^{(t)} - (\tilde v_k-		\tWkdot p_O^{(t)})\}(1;p_O^{(t)})_j\right| \leq \lambda z_k, \ \
k\in V_O, j\in
\bar{V}_O,\\
		&  \left\{ \frac{1}{n}\sum_{t=1}^{n} \{y_k^{(t)} - (\tilde v_k-\tWkdot  p_O^{(t)})\}^2(1;p_O^{(t)})_j^2\right\}^{1/2} \leq z_k, \ \ k\in V_O, j\in
		\bar{V}_O.\\
		\end{array}
		\end{equation}

	\item[] {\bf Step 2.} For the design matrix $\hat\Sigma:= \frac{1}{n}\sum_{t=1}^n(1;p_O^{(t)})(1;p_O^{(t)})^T$, compute a debiasing matrix $\hat \Psi$ by solving
	\begin{equation}\label{def:lasso:ortho}
	\begin{array}{rl}
	\displaystyle  (\hat z, \hat \Psi) \in \arg\min_{z,\Psi} \quad  &  \sum_{k\in \bar{V}_O} \left(\frac{1}{n}\sum_{t=1}^n|\Pkdot
 (1;p_O^{(t)})|^4\right)^{1/4} +  z_k\\
	\mathrm{s.t.} \quad 	& |	\Pkdot  \hat\Sigma_{\cdot,j} - \mathbf{1}_{kj} | \leq \lambda z_k, \ \ k,j\in \bar{V}_O, \\
	&  \left\{ \frac{1}{n}\sum_{t=1}^{n} \{ \Pkdot(1;p_O^{(t)}) (1;p_O^{(t)})_j - \mathbf{1}_{kj}\}^2\right\}^{1/2} \leq z_k, \ \ k,j \in \bar{V}_O.\\
	\end{array}\end{equation}

	\item[] {\bf Step 3.} Compute
	\begin{equation}\label{def:ortho:est}	
	(-\check v^T; \check W^T)  = (-\hat v^T; \hat W^T) -  \hat \Psi
	\left\{\frac{1}{n}\sum_{t=1}^n (1;p_O^{(t)}) \{y_O^{(t)}-(\hat v - \hat W p_O^{(t)})\}^T \right\}.
	\end{equation}

	\item[] {{\bf Step 4.} Compute the thresholded estimator
$$ \check W^\mu = ( \check W_{kj} 1\{|\check W_{kj}|> \mu_{kj} \})_{k,j\in V_O}. $$

		Terminate and return
		$\check W^\mu$, $\check{{v}}$ respectively as the
		estimates of $H^{-1}$ and ${v}_O$.}
		
	\end{itemize}
	\caption{Estimation  of $H^{-1}$.}
	\label{alg:Alg1}
\end{algorithm}

  Our estimator is presented in Algorithm 1.
This algorithm builds on different ideas in the high-dimensional statistics literature.
It can be viewed as a thresholded bias-corrected Dantzig selector  estimator whose penalty parameter is pivotal.\footnote{That is, it does not depend on unknown quantities such as the variance of the noise.}
Under (approximate) sparsity assumptions, high-dimensional models are estimable with the introduction of regularization (in this case the $\ell_1$-penalty).
However, the regularization requires carefully setting penalty parameters and typically yields estimators that are consistent but not asymptotically normal. Algorithm 1 addresses both issues through the use of self-normalized moderate deviation theory.

 Step 1 of the algorithm obtains a preliminary estimate $(\hat v,\hat W)$ of $(v_O,H^{-1})$ based on a pivotal version of the Dantzig selector estimator; see  \cite{CandesTao2007,BickelRitovTsybakov2009,belloni2017pivotal}.
This preliminary estimate is not necessarily asymptotically normal, which necessitates the subsequent debiasing step (Step 3).

Step 2 of the algorithm solves an auxiliary regularized estimation problem (again with pivotal choices of the penalty parameter) to compute a pseudo-inverse of the (empirical) covariance matrix $\hat \Sigma$ (which is rank-deficient due to the high dimensionality). The variant used in Algorithm~\ref{alg:Alg1} is similar to the formulations in \cite{javanmard2014confidence}
and \cite{zhu2016significance}
but involves significant differences.
First, in addition to handling the network setting, the specific form of (\ref{def:lasso:ortho}) used here is new. Notably,
	the optimization formulation  in (\ref{def:lasso:ortho}) is always feasible (unlike the corresponding problem in \cite{javanmard2014confidence}).
 Second, it exploits self-normalization to achieve pivotal choices of the penalty parameter $\lambda$. Third, we use a new objective function and minimize a function of  the average (empirical) fourth moment of $\Psi_{k,\cdot}(1;p_O^{(t)})$.
Leveraging the optimality of
$\hat\Psi$ in
(\ref{def:lasso:ortho}), this
novel criterion
 leads to bounds on the higher order empirical moments of
$\bar\Psi_{k,\cdot}(1;p_O^{(t)})$, where
$\bar \Psi:=E[(1;p_O^{(t)})(1;p_O^{(t)})^T]^{-1}$.
As we shall see in the next section, such bounds
 in turn allow us to
 achieve desired rates of convergence
 for a rich class of data-generating processes where
 it is not required for
 (i)  shocks to be Gaussian, and
 (ii) $\hat \Psi_{k,\cdot}$ to converge to $\bar \Psi_{k,\cdot}$.\footnote{It is worth pointing out that if  $E[(1;p) (1;p)^T]$ is known, then its inverse can readily be used in place of $\hat \Psi$ thereby simplifying the estimator.
Our estimator works even in the absence of this information.
}

Step 3 uses the pseudo-inverse $\hat \Psi$ of $\hat\Sigma$ computed in Step 2 to reduce the bias in the preliminary estimator $\hat W$ obtained in Step 1, and leads to the debiased estimator $(\check v,\check W)$. (This can be seen as a Newton step from $(\hat v,\hat W)$.)  In the high-dimensional case such ideas have recently been  used by different authors with different variants and assumptions; see, e.g., \cite{BelloniChernozhukovHansen2011,c.h.zhang:s.zhang,GRD2014,javanmard2014confidence,BCK-LAD}. Similar ideas can be traced back to \cite{neyman1965use} and \cite{Neyman1979} in the fixed dimensional case with the use of the so-called orthogonal moment conditions to reduce the impact of estimation errors of nuisance parameters.

Finally, Step 4 thresholds the intermediate estimator $\check W$ to obtain $\check W^\mu$. The motivation to consider such a thresholded estimator is somewhat subtle,
and has to do with the convergence rates that can be obtained by different estimators. Intuitively, our approximate sparsity notion encodes sparsity across a collection of linear models (each of which captures the dependence of the outcome of a single agent on the covariates). It turns out that under such sparsity assumptions (and an appropriate choice of thresholds), the thresholded estimator  enjoys good rates of convergence for both the rows and columns of $H^{-1}$ (see Theorem \ref{thm:OrthoHinv}),
thereby allowing us to control   errors  not only within each model (rows of $H^{-1}$), but also    across  models (columns of $H^{-1}$).  This,  in turn, ensures consistency  of the estimates in the matrix $2$-norm. Such guarantees do not hold for the  intermediate estimators obtained in Algorithm 1.
For instance, it can be seen that Step~1 of the proposed Algorithm 1, as well as the lasso estimator and its variants, decouple over the rows of $H^{-1}$, and they can be used to obtain convergence rates for the estimation of rows of $H^{-1}$.	 However, since $H^{-1}$ is not necessarily symmetric, in the high-dimensional setting we consider, such estimates do not provide meaningful guarantees  for  column estimates of $H^{-1}$. Similarly, it can be shown that the 	  estimator $\check W$ obtained in Step 3 has good rates of convergence in the maximum entry-wise error but it  need not be a consistent estimator of $H^{-1}$ in matrix  $\infty$-norm (or the $2$-norm).

{The importance of achieving good rates of convergence for both the rows and columns is justified
in Section \ref{se:applications}.
In that section, we show that
in decision problems  where the payoff of the decision maker naturally depends on the network structure, the optimal decisions and payoffs
depend both on $H^{-1}$ and $H^{-T}$.
Thus, controlling the estimation errors for both the rows and columns of	$H^{-1}-\check{W}^\mu$ is important
for constructing
approximately
(and asymptotically)
optimal decisions. Indeed, we leverage our approach and bounds on
row and column errors
to obtain  asymptotically optimal
targeting decisions
for the applications studied in
   Section \ref{se:applications}.}

\subsection{Convergence Rates}\label{subse:estimationBounds}

In order to provide the convergence rates of our algorithm, we
require the underlying covariate
and error processes
to be ``well-behaved.''
In particular, we impose the following assumption:
\begin{assumption}\label{assumption:basicRandom}
	Suppose that the network  admits an
	$(s_n,r_{1n})$-sparse
	approximation.
	Let $c,C,C'>0$ be  constants such that $c<C$,
	and let
	$M_\varepsilon, M_\Psi$ be parameters satisfying
	$M_\varepsilon, M_\Psi \geq 1 $.
	The following conditions hold:
	{
		\begin{enumerate}[label=\roman*.]
			\item
			The observed data $\{ ({p}^{(t)}_O, {y}_O^{(t)}) : t\in [n]\}$ are i.i.d. random vectors
			that satisfy \eqref{eq:def:noise1}.
			Moreover, the  shock term satisfies ${E}[{\varepsilon}^{(t)}_O \mid  {p}^{(t)}_O ] = 0$ for every $t\in[n]$.
			\label{ass3.1}
			\item
			For every $t\in[n]$ we have
			$\min_{j\in V_O} E[(\varepsilon_j^{(t)})^2 \mid p_O^{(t)}] \geq c$, $\max_{j\in V_O}E[|\varepsilon_j^{(t)}|^4\mid p_O^{(t)}] \leq C$,
			$E[\max_{t\in[n],j\in V_O} |\varepsilon_j^{(t)}|^4  \mid \{p_O^{(t)}\}]\leq M_\varepsilon$.
			\label{ass3.2}
			
			\item
			The matrix $\bar \Psi:=E[(1;p_O^{(t)})(1;p_O^{(t)})^T]^{-1}$ is such that
			$$\min_{k,j\in \bar V_O}E[ \{ \Pk(1;p_O^{(t)})(1;p_O^{(t)})_j - \mathbf{1}_{kj}\}^2] \geq c, \mbox{ and } \max_{k,j \in \bar V_O} E[ | \Pk(1;p_O^{(t)})(1;p_O^{(t)})_j|^4] \leq C.$$
			Moreover,
			$E[ \max_{t\in [n]; j,k\in \bar V_O} |  \Pk(1;p_O^{(t)})(1;p_O^{(t)})_j|^4 ] \leq M_\Psi$.
			Finally, we assume that the eigenvalues of $\bar \Psi$ are upper bounded by a constant.
			\label{ass3.3}

			\item
			We have $ s_n^2(\log |V_O|) (\log n)^3  = o(n)	$,
			$M_\varepsilon M_\Psi   \log |V_O| = o(n)$,
			and
			$ C' \log |V_O| \geq  \log n$.
			\label{ass3.5}
		\end{enumerate}
	}

\end{assumption}

{Assumption \ref{assumption:basicRandom}\ref{ass3.1}
	states that the shock term $\varepsilon_O^{(t)}$ is a zero mean conditional on the covariates  in period $t$.
	Assumptions \ref{assumption:basicRandom}\ref{ass3.2} and \ref{ass3.3} are mild moment conditions on the  shocks and covariates. For example, conditional on covariates,  we require the 	 shocks'  fourth moments to be bounded from above,
and  their second moments to be bounded from below.
	Similarly, the eigenvalues
	of matrices constructed from
	expectations of outerproducts of covariate vectors are well-behaved.
	Such moment conditions readily hold for sub-Gaussian and subexponential distributions as well as more heavy-tailed distributions. These assumptions are commonly employed  in high-dimensional statistics, with a general covariate/observation structure, and are adapted to our setting (see, e.g., \cite{BickelRitovTsybakov2009,belloni2017pivotal,belloni2017simultaneous}).
	Assumption \ref{assumption:basicRandom}\ref{ass3.5} imposes requirements on how the number of agents and the sample size can relate. In particular, we allow for a high-dimensional setting where $|V_O| \gg n$.}

	We proceed with our first result on the estimator provided in Algorithm 1.

  \begin{theorem}\label{thm:OrthoHinv}
  	Under  Assumption \ref{assumption:basicRandom}
	with probability at least $1-o(1)$  the following statements hold:
\begin{enumerate}[label=\roman*.]
	\item
Uniformly over  	$k\in V_O$ we have
\[
	\sqrt{n}\{ (-\check v_k, \check W_{k,\cdot})^T - (-v_k, H^{-1}_{k,\cdot})^T\} =  -
	 \hat \Psi\frac{1}{\sqrt{n}}\sum_{t=1}^n \varepsilon_{k}^{(t)}(1;{p}_O^{(t)})  +
R^k_n
\]
 where $\|R^k_n\|_\infty = O(n^{-1/2}s_n\log |V_O| + r_{1n}\sqrt{\log |V_O|})$.
\label{thm1_1}

	\item
	If $\mu_{kj} \geq 2 |\check W_{kj} - (H^{-1})_{kj} |$ and $\mu_{kj}\leq C_1 \sqrt{\log|V_O|/n}$ for all $k,j \in V_O$, then the thresholded estimator
	$\check W^\mu $ satisfies
	\[\| \check W^\mu-H^{-1}\|_\infty  \leq s_n C_2 \sqrt{\frac{ \log |V_O|}{n} } + 3r_{1n} \ \mbox{and} \
	\|
	\check W^\mu-H^{-1}
	\|_1 \leq
	s_n C_2 \sqrt{\frac{ \log |V_O|}{n}  } + 3r_{1n},\]
	for some constants $C_1,C_2 >0$.
\label{thm1_2}

\end{enumerate}

  \end{theorem}

The first part of Theorem \ref{thm:OrthoHinv}  provides an (approximate) linear representation of the (intermediate) estimator $\check W$ of Step 3. With high probability the estimation error is a zero-mean term plus an approximation error $R^k_n$, which vanishes
provided that $r_{1n}\rightarrow 0$ (since  $s_n/\sqrt{n}\rightarrow 0$ by Assumption \ref{assumption:basicRandom}).
This result is key to establishing the relevant rates of convergence (Theorem \ref{cor:Thm1} below) and also a distributional limit that allows the construction of valid confidence intervals (Theorem \ref{thm:inferenceAlg1} in the Appendix). Therefore we will build substantially on Theorem \ref{thm:OrthoHinv}\ref{thm1_1} in what follows.

	{The second part of  Theorem \ref{thm:OrthoHinv} pertains to the thresholded estimator. It states that if the thresholds are chosen to
be sufficiently larger than the entry-wise estimation errors (at the end of Step~3),
	the thresholded estimator will achieve good rates of convergence both for the rows and for the columns of $H^{-1}$.
		To see why
		the
		thresholded estimator  enjoys good rates of convergence, first observe that
		Assumption \ref{assumption:basicRandom}
		guarantees that
		$H^{-1}$ can be approximated well with a matrix that has a  small number of nonzero entries in each row/column.
		Fix a row,
		and let $T$ denote the entries in this row that take nonzero values in this approximation.
		Suppose thresholds that satisfy the conditions of the theorem are available, and consider the errors
		the thresholded estimator makes
		in the entries that belong to $T$ vs. $T^c$.
		Since $T$ has small cardinality, it can be shown that the absolute sum of the errors for the elements in $T$ cannot be large.
		Similarly, due to approximate sparsity,
		many entries of $H^{-1}$ that belong to
		$T^c$
		are already small.
		Thus,
		the definition of the threshold implies that
		only a small fraction of the corresponding entries  of $\check W$ are above the threshold, and the total error the thresholded estimator incurs for these entries is also small.
		The remaining entries in $T^c$ are below the threshold, and  take the value of zero in the thresholded estimator. Approximate sparsity ensures that
		the corresponding entries of $H^{-1}$ are also small, and hence
		 the total error  due to these entries is small. These observations   can be leveraged to uniformly bound the $1$-norm error in each row of $H^{-1}$. A similar argument also bounds the errors in the columns of $H^{-1}$, and combining these the desirable convergence rates in the theorem, in terms of matrix $1$ and $\infty$ norms,  can be obtained.
		Without thresholding, the last error component would in general be large, and it would not be possible to provide similar guarantees in these matrix norms.}

	Of course, the choice of thresholds $\{\mu_{kj}\}_{ k,j\in V_O}$ is key for this result.
	We deliberately state the second part of  Theorem \ref{thm:OrthoHinv} to allow for different choices of thresholds. A particular choice of $\{\mu_{kj}\}$ can be obtained using analytic bounds, based on self-normalized moderate deviation theory, that are slightly conservative but computationally trivial, or using a bootstrap procedure that exploits the correlation structure
	and still allows for $|V_O|\gg n$,
	but  is computationally more demanding and requires stronger conditions.
	For concreteness we provide the results based on the self-normalization ideas in Theorem~\ref{cor:Thm1} and refer the interested reader to Theorem~\ref{thm:inferenceAlg1} in the Appendix for the result associated with the bootstrap procedure.
Theorem \ref{cor:Thm1} is based on the following thresholds:
	\begin{equation}\label{eq:thresholds}
	\mu_{kj} = 2 (1+\log^{-1}n)
	\frac{\hat\sigma_{kj}}{\sqrt{n}}  \Phi^{-1}\left(1-\frac{1}{3n|V_O|^2}\right)
	= 2(1+\log^{-1}n) \hat\sigma_{kj}\lambda,
	\end{equation}%
where $\hat\sigma_{kj}^2 = \frac{1}{n}\sum_{t=1}^n\{\hat\Psi_{k,\cdot}(1;p_O^{(t)})(y_j^{(t)}-(\hat v_j-\hat W_{j,\cdot}p_O^{(t)}))\}^2$.
\begin{theorem}\label{cor:Thm1}
	Suppose that   Assumption \ref{assumption:basicRandom} holds.
Then, with probability at least $1-o(1)$ the intermediate estimator based on Algorithm 1 satisfies
	$$\max_{k,j \in V_O} |(\check W - H^{-1})_{kj} | \leq C_1 \sqrt{\frac{\log |V_O|}{n}} \ \ \mbox{and} \ \ \max_{k\in V_O} |\check v_k - v_k | \leq C_1 \sqrt{\frac{\log |V_O|}{n}},$$
	for some constant $C_1 > 0$.
Moreover, suppose that  $r_{1n}= o(1/\log n)$.
 Then the thresholds in \eqref{eq:thresholds},
 with probability at least $1-o(1)$, yield
\begin{equation} \label{eq:thresholdConstraints}
 C_2 \sqrt{\log|V_O|/n} \geq
\mu_{kj} \geq 2|(\check W - H^{-1})_{kj}| \ \  \mbox{simultaneously over} \ k,j\in V_O,
\end{equation}
 for some constant $C_2 >0$.
Finally, with probability
$1-o(1)$
the  thresholded estimator based on these thresholds  satisfies
\begin{equation} \label{eq:Thm2Rates0}
\max_{k,j \in V_O} |(\check W^\mu - H^{-1})_{kj} | \leq C_3 \sqrt{\frac{\log |V_O|}{n}}, \mbox{as well~as}
\end{equation}
\begin{equation} \label{eq:Thm2Rates}
\| \check W^\mu-H^{-1}\|_\infty  \leq s_n C_3 \sqrt{\frac{ \log |V_O|}{n} } + 3r_{1n} \ \mbox{and} \
\|
\check W^\mu-H^{-1}
\|_1 \leq
s_n C_2 \sqrt{\frac{ \log |V_O|}{n}  } + 3r_{1n},
\end{equation}
	for some constant $C_3 > 0$.
\end{theorem}

{Theorem \ref{cor:Thm1} implies that Algorithm 1 provides consistent estimates of the entries of $H^{-1}$, uniformly over all entries. Having
	estimates of the entries of $H^{-1}$ is valuable  for understanding the impact of
	the covariate of an agent on another  agent's outcome (accounting for the network externalities through latent agents). Moreover, under the approximate sparsity assumptions that imply that
		$r_{1n}= o(1/\log n)$
	(see Section \ref{se:estimationExamples}), Theorem \ref{cor:Thm1} also shows that there is a suitable choice of thresholding parameters that allows for the estimator $\check W^\mu$ to have desirable rates of convergence for the rows and columns. The proposed thresholds are derived using self-normalized moderate deviation theory. This allows us to handle non-Gaussian shocks  as well as the high dimensionality of the estimand matrix.}

{
\begin{remark}[Bootstrap-based thresholds]
Theorem \ref{cor:Thm1} constructs
thresholds $\{\mu_{kj}\}_{ k,j\in V_O},$
using  self-normalized moderate deviation theory and the union bound, i.e., a Bonferroni correction. In some cases this could be conservative and it is  of interest to pursue a less conservative choice. To accomplish this one needs to account for the correlation structure. Under stronger regularity conditions, this can be done through the use of a multiplier bootstrap procedure conditional on the data. In particular, for each $k,j\in V_O$ define
$$ T_{kj} := \frac{1}{\sqrt{n}}\sum_{t=1}^n \hat\xi^{(t)} \frac{\hat \Psi_{k,\cdot}(1;p_O^{(t)})\{ y_j^{(t)} - (\hat v_j - \hat W_{j,\cdot} p_O^{(t)})\}}{\hat\sigma_{kj}}, $$
where $\hat\sigma_{kj}^2 = \frac{1}{n}\sum_{t=1}^n \{\hat \Psi_{k,\cdot}(1;{p}_O^{(t)})[ y_j^{(t)} - (\hat v_j - \hat W_{j,\cdot} p_O^{(t)})]\}^2$, $\hat\xi^{(t)}$ are i.i.d. standard normal random variables
independent of the data.
 The associated critical value we are interested in is
$$ {\rm cv}^*_{(1-\alpha)} = \mbox{conditional} \  (1-\alpha)\mbox{-quantile of} \ \max_{k,j\in V_O} |T_{kj} | \ \ \mbox{given the data},$$
which can be computed by simulation (by redrawing the Gaussian multipliers). Then
the thresholds can be set   $\mu_{kj} = 2 {\rm cv}^*_{(1-\alpha)} \hat\sigma_{kj}/\sqrt{n}$. Such a bootstrap procedure also leads to the construction of simultaneous confidence intervals.  We refer the interested reader to Section \ref{Sec:SimultaneousBootstrap} in the Appendix for a more detailed discussion.
In this appendix, using  recent central limit theorems for high-dimensional vectors (where $|V_O|^2 \gg n$ is allowed; see, e.g., \citet{chernozhukov2012gaussian,chernozhukov2013gaussian}),
the validity of this procedure is established.
	\hfill \Halmos
\end{remark}

\begin{remark}[Time dependence]
In this work we focus on the impact of latent agents
on the remaining agents
 and throughout the paper we assume i.i.d. observations. The key technical tools we rely on, namely, self-normalized moderate deviation theory and  high-dimensional central limit theorems, have been derived under this assumption.
However, recent works have been generalizing these tools to allow for time dependence as well; see \cite{chen2016self} for results of self-normalized moderate deviation theory and  \cite{chernozhukov2013testing,zhang2014bootstrapping,zhang2017gaussian,BelloniOliveira2018} for high-dimensional central limit theorems under various types of dependence. Therefore it is plausible that most of these tools can be extended to allow for time dependence under more stringent conditions both on the moments and on the growth of $V_O$ and $s_n$ relative to $n$. Although it is beyond the scope of the present work we view this endeavor as a potentially interesting future research direction.
	\hfill \Halmos
\end{remark}}

{
\begin{remark}[Handling endogeneity] \label{rem:Endogenous}
In some applications it is of interest to also allow for endogenous covariates, i.e., $E[p_O\epsilon_O^T] \neq 0$. In such cases it is well known that the moment condition (\ref{eq:moment}) no longer holds and in turn the proposed procedure does not lead to consistent estimates of the matrix $H^{-1}$. Nonetheless the tools proposed here can still be useful when suitable instrumental variables are available; i.e., for each $k \in V_O$, we observe a random vector $z^{k(t)}$ such that $E[z^{k(t)}\epsilon_k^{(t)}] = 0$ and $E[z^{k(t)}(p_O^{(t)})^T]$ is full rank and well-behaved. These instruments  yield a similar moment condition
\begin{equation}\label{eq:momentZ}
{E}[\{y_O^{(t)} - (v_O - H^{-1}p_O^{(t)})\}(1;z^{k(t)})^T]=0,
\end{equation}
and Algorithm 1 can be adjusted accordingly.  This generalization is of interest as it allows for covering a different set of applications. However, its analysis poses interesting technical challenges. The analysis of that new estimator would combine the analysis developed here and the analysis of high-dimensional linear instrumental variables developed in \cite{belloni2017simultaneous}. This is being pursued in a companion work \citet{ata2018IVnetwork}.
	\hfill \Halmos
\end{remark}
}

{
\begin{remark}[Variation in Covariates for Latent Agents] \label{rem:latent}
	{In this paper, we consider the case in which the covariates that the latent agents are exposed to are fixed.
		As discussed in Section \ref{se:applications}, this is natural in the advertising/pricing applications we focus on.
That said, our results could be obtained under weaker conditions.
Specifically,	suppose that the covariates of the latent agents evolve with respect to a  stochastic process $\{p_L^{(t)}\}$.
	Then,
	we can use a similar characterization to Lemma~\ref{lem:firstOrderO} to express observable agents' outcome as
	${y}_O^{(t)}=\tilde{v}_O-H^{-1} {p}_O^{(t)}+\tilde{ \varepsilon}_O^{(t)}$, where
	$\tilde{v}_O := H^{-1}  {a}_O-H^{-1} S_{OL} ( {a}_L - E[{p}_L])$
	(analogous to \eqref{eq:defVo}) and
	  $$\tilde{ \varepsilon}_O^{(t)}:=
	 H^{-1} \xi_O^{(t)}-H^{-1} S_{OL}  ({\xi}^{(t)}_L
	 -p_L^{(t)}+  E[{p}_L^{(t)}])
	 .$$
	An inspection of the proofs reveals that
	it suffices to have
	 $E[\tilde\varepsilon_{O}\mid p_O ] = 0$
	 for the arguments
	 (and the results of this section) to go through
	(after replacing
$v_O$ with $\tilde{v}_O$).
For example, this is achieved if the variation in $p_O$ is introduced in a randomized experiment, which is possible in many online platforms.
In observational studies, additional considerations may be needed to justify this condition,
since
 it imposes a restriction on the relation between $p_L$ and $p_O$, e.g., $E[ p_L - E[p_L] | p_O ] = 0$.
 When such a condition does not hold, it becomes a source of endogeneity, in which case the comments in Remark  \ref{rem:Endogenous} become relevant.
 } \hfill\Halmos
\end{remark}		
}

\section{Applications} \label{se:applications}
In this section, we study applications of our  estimation framework to targeted
advertising (Section \ref{subse:estimationAdvertising})  and
pricing (Section \ref{subse:estimationPrices}) problems.
In both cases, we focus on settings
where agents in a social  network consume a
divisible product that exhibits positive network externalities,
which are of a local nature.
The edge weights
$\{G_{ij}\}$ summarize these externalities, and in particular
 $G_{ij} \geq 0$ captures how much the consumption of an agent $j$  influences her neighbor $i$.
A subset of the agents are observable, e.g., they participate in an online platform, which makes data on their past decisions available.
An advertiser/seller decides on agent-specific advertisement levels/prices for the product to influence (observable) agents' purchase decisions.
Her
payoff depends on the induced sales.
We establish that using the estimation framework of the previous section with historical data on observable agents, near optimal advertisement levels/prices can be obtained.

In our applications, the relationship between the agents' outcomes and covariates  take the form presented in Section \ref{se:model}.
We conduct our analysis under the following assumption:
\begin{assumption} \label{ass:DiagDominance}
	There exists some
	$\zeta>0$
	such that
	for each $i\in V$, we have
	$\lambda_i  \geq  \sum_j G_{ij}  + \zeta 	$
	and
	$	\lambda_i \geq  \sum_j G_{ji}  + \zeta $.	
\end{assumption}
This assumption guarantees that
the matrix $M=\Lambda-G$ is strictly row and column diagonally dominant.
Here, $\zeta >0$ is a parameter that captures how large the diagonal entries are relative to off-diagonal entries. The
positivity of this parameter implies that  the eigenvalues of $M$ are bounded away from zero.
Recall that our earlier analysis assumed the boundedness of the absolute row/column sums of $M^{-1}$.
This condition is also readily implied by Assumption \ref{ass:DiagDominance}  (see  Lemma \ref{lem:matBounds}).
Qualitatively,  Assumption \ref{ass:DiagDominance} implies that the network externality
that an agent exerts on the rest of the network (or vice versa) is not too large.\footnote{Variants of this assumption have appeared in pricing in social networks literature to ensure that equilibrium solutions are interior and induced pricing problems are concave. See e.g., 	\cite{ballester2006s,candogan2012optimal,fainmesser2015pricing,zhou2015key}.}

\subsection{Targeted advertising} \label{subse:estimationAdvertising}

We first focus on
the problem of
an
advertiser who advertises a product that exhibits network externalities through an online (social networking) platform.
Specifically, we study a model where the consumption  $y_i$ of  agent $i$ is given by
\begin{equation} \label{eq:advertisingModel0}
	y_i = a_i+  z_i f( \tilde p_i)  +    \sum_j  G_{ij} y_j + 	\xi_i.
\end{equation}
Here,  $   \sum_j  G_{ij} y_j $ captures network externalities.
The agent-specific parameter
$a_i \geq 0$  represents agent $i$'s affinity for the product which captures her (mean) consumption in the absence of any network or advertising effects.
The parameter $\tilde p_i \geq 0$ represents
exposure
level
of agent $i$ to  ads on the online platform.
The coefficient $z_i \geq 0$ captures how responsive customer $i$ is to advertising on the platform.
In different settings it has been documented that advertising exhibits diminishing marginal returns (see, e.g., \citet[][p. 267]{lilien1995marketing} and \citet{manchanda2006effect}).
Consistently with this $f$ is a concave advertising response function that  captures how advertising translates into additional consumption.
More concretely, for our analysis in this section we assume $f(\tilde p_i) = \sqrt{\tilde p_i}$ (but we emphasize that
this is for the ease of exposition and
the
analysis and
results can be extended to other concave response functions).
Finally, $\{\xi_i\}$ represent zero-mean shocks to agents' consumption decisions.

As before, we assume that the parameters of the model are bounded, and in particular we require that   $0\leq a_i < \bar a$, and $0< \underline{z} \leq z_i<\bar z$. Here the positive lower bound on $z_i$ is imposed to rule out settings in which advertising has no impact on consumption, since in this case the advertising problem becomes trivial.
We also make  the following assumption, which
ensures  that  all agents' consumptions in
\eqref{eq:advertisingModel0}  are
nonnegative:
\begin{assumption}\label{ass:OutsideOpt0}
	$\mathbb{P}(a_i + \xi_i \geq  0 )=1$
	for all $i\in V$.
\end{assumption}	

We assume that unit ad exposure on the platform can be achieved at a cost of ${\chi}/{2} > 0$.
Only a subset of all of the agents
$V_O \subset V$
participate in the online platform.
The advertiser
targets these  agents with different ad exposure levels $\{\tilde p_i\}_{i\in V_O}$.
For a latent agent $i\notin V_O$, we set $\tilde{p}_i=0$ (since latent agents do not participate in the platform, and hence cannot be exposed to ads there).
The cost of choosing exposure levels $\{\tilde p_i \}_{i\in V}$ to the advertiser is given by
$\frac{\chi}{2} \sum_{i\in V} \tilde{p}_i= \frac{\chi}{2} \sum_{i\in V_O} \tilde{p}_i$.

Note that advertising the product on the platform
leads to an increase in observable agents' consumptions through two different channels.
The first one is the direct effect of advertising, which motivates the observable agents to consume more.
Through network externalities this triggers latent agents to consume more, which then leads to a further (indirect) increase in the consumptions of observable agents.

Let
${y}_O(\tilde{{p}}_O)$ denote the
consumption  levels of the observable agents
under exposure levels $\tilde p_O$.
We assume that the advertiser  enjoys a unit payoff for each unit consumed by the observable agents. Her objective is to maximize total consumption
by these agents
 minus the  cost of advertising. Specifically, for given $p_O$ the expected total payoff of the advertiser is given by:
\[
\Pi(\tilde p_O) = E [e^T y_O(\tilde p_O)] -  \frac{\chi}{2}  \sum_{i\in V_O} \tilde p_i.
\]

It is more convenient to formulate the problem of the advertiser after a change of variables: $p_i = \sqrt{\tilde p_i}$.
With some abuse of terminology in what follows we refer to $p_i$ as the advertising intensity for agent $i$.
After this change of variables,
the dependence of agents' consumptions on advertising intensities takes the following form:
\begin{equation} \label{eq:advertisingModel}
	y_i = a_i+  z_i  p_i  +    \sum_j  G_{ij} y_j + 	\xi_i.
\end{equation}
Slightly abusing the earlier notation,
we denote by $y_O(p_O)$ the  consumption levels
of observable agents
under advertising intensities $p_O$, and
restate the induced payoffs as follows:
\begin{equation} \label{eq:profitAd0}
\Pi( p_O) = E [e^T y_O(p_O)] -  \frac{\chi}{2}  \sum_{i\in V_O}  p_i^2.	
\end{equation}
Note that this payoff form is consistent with
the convex quadratic cost assumptions used in the advertising literature, e.g., see
\cite{slade1995product,dube2005differences}.
The problem of the advertiser is to choose
advertising intensities  $p_i\in [0, \bar p]$
for $i\in V_O$
to maximize her expected payoffs.\footnote{The results of this section go through if $p_i$ is allowed to be unbounded.}

We assume that the advertiser does not know the influence structure $G$, or the parameters $\{a_i\}, \{z_i\}$.
Thus, it is not possible to directly solve for the advertising intensities that maximize \eqref{eq:profitAd0}.
However, we assume that the advertiser has historical data $\{y_O^{(t)}, p_O^{(t)}\}_{t\in[n]}$ on past consumption/advertising levels
 for observable agents.
In the remainder of this subsection, we   argue that near-optimal advertising levels can be obtained by first estimating the underlying parameters using the available data, and then  using these estimates to construct the advertising levels.

To this end, we first map the model introduced in this section back to the   model introduced in Section \ref{se:model}, by setting
$\lambda_i=1$ for $i\in V$, $p_0=0$, and $M=\Lambda-  G=I- G$, where $I$ is the identity matrix.
We rearrange the terms in
\eqref{eq:advertisingModel} and restate it
(by using matrix notation and making time dependence explicit)
as follows:
\begin{equation} \label{eq:firstOrder_v2}
	{y}^{(t)}= M^{-1} ({a} + {\xi}^{(t)} + Z {p}^{(t)}).
\end{equation}
Here $Z$ is a diagonal matrix whose diagonal entries are given by $z_i$ for $i\in V_O$ (and zero for the remaining agents).
This equation is almost identical to \eqref{eq:firstOrder}, the only difference being the $Z$ matrix.
Thus, an analogous result to Lemma \ref{lem:firstOrderO} holds, and
yields:
\begin{equation}\label{eq:yArr_obs}
	{y}_O
	=  {v}_O- \tilde{H}^{-1} {p}_O +\varepsilon_O^{(t)},
\end{equation}
where
$\tilde{H}^{-1}$ is a row-scaled version of $H^{-1}$, i.e., $\tilde H^{-1}=	-H^{-1} Z_{OO} $, and
we redefine $v_O$ as
\begin{equation}\label{eq:defVo2}
	{v}_O := H^{-1}  {a}_O-H^{-1} S_{OL} {a}_L ,
\end{equation}
since $p_i=0$ for $i\in V_L$.
The remaining variables are defined as in \eqref{eq:sMatDef} and \eqref{eq:defVo}.
Note that since $|z_i|\leq \bar z$, it follows that $\tilde H^{-1}$ satisfies the approximate sparsity conditions introduced in Section~\ref{se:estimation} if and only if $H^{-1}$ does so. This observation together with the fact that
\eqref{eq:yArr_obs} has the same form as in \eqref{lem:firstOrderO},
implies that the algorithm of Section \ref{se:estimation} can be used to estimate $v_O$ and $\tilde H^{-1}$, achieving the error rates provided
in Theorems
\ref{thm:OrthoHinv}
and
\ref{cor:Thm1}
for a given approximately sparse network.

Using this notation in \eqref{eq:profitAd0}, the  expected payoff of the advertiser
for advertising intensities $p_O$ can be more explicitly expressed as:
\[
\Pi(p_O)
=
e^T( {v}_O- \tilde H^{-1} {p}_O  )- \frac{\chi}{2} \sum_{i\in V_O} p_i^2
.
\]
Let $p_O^\star$ denote the advertising intensities that  maximize the advertiser's expected payoff, i.e., $p_O^\star$ solves:
\begin{equation}\label{eq:adOpt}
	\begin{aligned}
		\max_{ 0\leq  {p}_O \leq \bar{p} \cdot {e}_O} \qquad & e^T( {v}_O-\tilde H^{-1} {p}_O  )- \frac{\chi}{2} \sum_{i\in V_O} p_i^2.
	\end{aligned}
\end{equation}
Observe that the payoff function is concave in $p_O$ and
$
\nabla \Pi(p_O) = -\tilde H^{-T } e  - \chi p_O $.
Thus, if the constraints are not binding, the optimal advertising intensities are given by
\begin{equation}\label{eq:optAdv}
p_O^\star= -\frac{1}{\chi}  \tilde H^{-T} e.
\end{equation}

Suppose that the advertiser chooses advertising levels  ${{p}}_O$ instead of ${p}_O^\star$.
Her payoff loss from using the former  advertising levels can be measured as follows:
\begin{equation}\label{eq:profitLoss}
	R({{p}}_O) := \frac{\Pi({{p}}^\star_O) -\Pi({{p}}_O)}{\Pi({{p}}_O^\star)}.
\end{equation}
Here, the numerator gives the absolute payoff difference under the optimal decisions  and advertising intensities ${{p}}_O$.
The denominator is the payoff under optimal decisions. The ratio measures the payoff loss from using prices ${{p}}_O$.

We next show that the  advertiser can approximately maximize her payoff by first using our estimator to estimate $ \tilde H^{-1}$ and then computing the optimal $p_i$ based on this estimate using \eqref{eq:optAdv}.
\begin{theorem}\label{thm:optimalAdv}
	Suppose that the influence structure admits an $(s_n,r_{1n})$-sparse approximation,  Assumption \ref{assumption:basicRandom} holds, and
	$ 0<{p}_O^\star < \bar p \cdot {e}_O$.
	Suppose further that
	$s_n \sqrt{\log(|V_O|)/n} +  r_{1n} = o(1)$.
	Let $\check{W}^\mu$ denote the thresholded estimator
	(of $\tilde{H}^{-1}$) given
	in Theorem \ref{thm:OrthoHinv}, and define
	\begin{equation} \label{eq:priceThm3adv}
		\hat {{p}}_O =  \left( - \frac{1}{\chi} (\check{W}^\mu )^T e \wedge \bar p \cdot e_O \right)_+
	\end{equation}
	Then,
	for some constant $C_1>0$,
	with probability
	$1-o(1)$ we have
	$R(\hat{{p}}_O)\leq
	C_1 \left(
	s_n^2  \frac{\log(|V_O|)}{n} +  r_{1n}^2 \right)$.

\end{theorem}
This   result  relies on bounding the loss in the advertiser's payoff due to using solution $\hat p_O$ as opposed to $p_O^\star$ in terms of $\| \hat p_O - p_O^\star\|_2$.
When the gap between $\tilde H^{-1}$ and its estimate is small with respect to the  $2$-norm, we show that the latter quantity is also small.
Theorem \ref{thm:OrthoHinv} implies that the
aforementioned gap
is small with respect to the $1$ and $\infty$-norms,
which in turn enables bounding the errors with respect to the $2$-norm.
Leveraging this observation, we obtain the bound in Theorem \ref{thm:optimalAdv}.

 This result implies that when the number of samples is at least logarithmic in the number of observable agents, for approximately sparse networks (e.g., those discussed in Section \ref{se:estimationExamples}), small payoff loss can be guaranteed.
 Moreover,
 the targeting decisions our approach yield are asymptotically optimal, i.e.,
 as the number of samples goes to infinity $R(\hat p_O)$ goes to~zero.

\subsection{Obtaining Approximately Optimal Prices} \label{subse:estimationPrices}

We next study the problem of a seller who offers targeted promotional prices for a product that exhibits network externalities.
We start by explicitly defining   agents' payoffs.

The payoff function of agent $i\in V$
has the same structure every period, and
consists of an individual consumption  term, a network externality term, and a payment term.
Suppose  that in period $t$,
agent $i$ consumes  $y_i\geq 0$ units of the product at  unit price  $p_i$, and the remaining agents consume $y_{-i}\geq 0$ units of the product. Then, the payoff of agent $i$ is given by\footnote{It is standard
to provide microfoundation for linear econometric   models of the type we introduced in Section \ref{se:model} using  quadratic payoff functions
(see, e.g.,
\cite{calvo2009peer,blume2015linear,topa2015neighborhood}).
In a similar spirit, the
payoff function provided here offers a microfoundation for the econometric model studied in the remainder of this section.
}
\begin{equation}\label{eq:basicUtil}
	u_i^{(t)}(y_i,y_{-i}, p_i)=
	\underbrace{(a_i+\xi_i^{(t)})    y_i-b_i y_i^2}_{\mbox{individual consumption}} + \underbrace{\sum_{j} G_{ij} y_i y_j}_{\mbox{network externality}} - \underbrace{p_i y_i}_{\mbox{{payment}}}.
\end{equation}
The first term $(a_i+\xi_i^{(t)}) y_i-b_i y_i^2$ in the payoff function
determines the value the agent derives from her own consumption of the product. We assume that $a_i,b_i>0$ so that this term is concave, and agents' marginal payoffs are decreasing in their own consumption.
We also assume that
$\bar a \geq a_i$, $\bar{b} \geq b_i$ for all $i$ and some $\bar a, \bar b \in \mathbb{R}$.
Here $\xi_i^{(t)}$ denotes an {(idiosyncratic)} taste shock for agent $i$ at time $t$, which
impacts her marginal value and consumption. As before we assume that $\{\xi_i^{(t)}\}_{i,t}$ have zero mean, and are independent over time,  but are possibly correlated across agents.
Note that if there are no taste shocks (i.e.,  $\xi_i^{(t)}=0$ for all $i$), the payoffs reduce to those considered in \citet{candogan2012optimal}.
The  term $\sum_{j} G_{ij} y_i y_j$ captures the positive externality that the consumption of her neighbors imposes on agent $i$. The positive externality increases with the consumption ($y_i$) of agent $i$, as well as with that of her  neighbors in the underlying network ($y_j$ for $j$ such that $(i,j)\in E$).
The last term captures the cost incurred by agent $i$ for consuming $y_i$ units of the product at unit price $p_i$.

In every period,
given a vector of prices ${p}$,  each agent chooses the   consumption level that maximizes her payoff.
Since agents' payoffs depend on each other's consumption decisions,
the consumption levels are determined at a
corresponding consumption equilibrium:
\begin{definition}[Consumption equilibrium]
	For a given vector of prices ${p}$, a vector ${y}\geq 0$ is a consumption equilibrium
	in period $t$
	if, for all $i\in V$,
	\begin{equation} \label{eq:optConsumption}
		y_i \in \arg \max_{z\geq 0} u_i^{(t)}(z,y_{-i}, p_i).
	\end{equation}
\end{definition}
Note that a consumption equilibrium corresponds to the Nash equilibrium of the normal form game with a set of agents $V$, a strategy set $[0,\infty)$ for each agent $i$, and payoffs given as in \eqref{eq:basicUtil}.\footnote{This equilibrium concept implicitly assumes that agents know the underlying network structure, and each other's payoff functions (including taste shocks, and prices).
	That being said, in order to determine her optimal consumption level in \eqref{eq:optConsumption}, agent $i$ needs to observe only her neighbors' consumption levels.
	Moreover, it can be shown that for any set of prices/taste shocks, the induced game among agents is supermodular, and agents' best-responses converge to a consumption equilibrium  \citep[see, e.g.,][]{candogan2012optimal}.
}

Hereafter, we denote by ${{p}}^{(t)}$ the (lowest) prices available to the agents in period $t$, and by ${y}^{(t)}$ the induced equilibrium consumption levels.\footnote{Here,
	we assume that
	the agents choose their consumption levels in a period,  based only on the prices offered in that period. Prices/consumption could vary over time due to a variety of factors ranging from inventory imbalances to seasonality. }
The former corresponds to the covariates and the latter the outcomes in the abstract setting discussed in Section \ref{se:model}.
If in a consumption equilibrium agents' consumption amounts are strictly positive, then
by the first-order optimality conditions in \eqref{eq:basicUtil}, it can be seen that in period $t$ the equilibrium consumption levels are given
as in \eqref{eq:firstOrder}
after letting
$\lambda_i=2b_i$ (and $\bar \lambda= 2\bar b$), for all $i\in V$.
Hence, each agent's consumption depends linearly on the price offered to her, as well as to the other agents.

A subset of agents participate in an online (social networking) platform, and a seller (hereafter the platform seller), offers targeted promotional prices to these agents.
The platform seller has access to historical data on the past prices as well as the consumption decisions of these (observable) agents.
Both observable and latent agents can purchase the product at
price $\bar p>0$ from a different channel (hereafter outside sellers), and this price  is public knowledge.
This constitutes an outside option for the observable agents. We assume that the prices offered by the
platform seller
to observable agents are  nonnegative and
weakly lower than $\bar p$.
If this were not the case, then agent $i \in V_O$ would prefer to purchase the product from the outside sellers. Hence, a revenue-maximizing platform always finds it optimal to offer prices weakly lower than
$\bar p$.
 Thus, for any $i\in V_O$, we ignore the outside option $\bar p$, and focus only on the price offered by the platform.
We assume that the outside option $\bar{p}$ is not time varying (see Remark \ref{rem:static}), whereas the prices offered by the platform to the observable agents possibly are.
In other words,
${p}^{(t)}$
is such that $p_i^{(t)}=p_0=\bar{p}$ for  $i\in V_L$ and
 $0\leq p_i^{(t)}\leq \bar{p}$  for $i\in V_O$.

We conduct our analysis under the following assumption:
\begin{assumption}\label{ass:OutsideOpt}
	The outside option
	$\bar p$
	is such that $\mathbb{P}(a_i + \xi_i^{(t)} > \bar p )=1$
	for all $i\in V$ and $t\in \mathbb{Z}_{++}$.
\end{assumption}
Note that this assumption requires negative shocks to be bounded, but allows for unbounded positive shocks.
If no agent consumes the product, the marginal utility of agent $i$ is given by $a_i+ \xi_i^{(t)} - p_i^{(t)}$.
Since for every $i$, $p_i^{(t)}$ is weakly lower than $\bar p$,  this
assumption
ensures that the marginal utilities are positive
for any realization of taste shocks.
As formally established in Lemma
\ref{lem:interiorCons} (Appendix \ref{app:NetworkInsight})
this implies that
all agents consume positive amounts of the product.
Hence this assumption plays a similar role to Assumption \ref{ass:OutsideOpt0} in the previous section, and ensures the linear dependence of agents' outcomes on their covariates (as in \eqref{eq:firstOrder}).

Since agents' consumption decisions are expressed as in \eqref{eq:firstOrder},
the dependence of the consumption decisions of observable agents
on the prices offered to them can be given as in
Lemma \ref{lem:firstOrderO}.
A central question we investigate  is how the platform seller should
offer targeted price discounts  to maximize her expected revenues.
If the network were known (and recalling that the platform seller sells only to the observable agents), we  would focus on the following optimization problem:
\begin{equation}\label{eq:sellerOpt}
	\begin{aligned}
		\max_{ 0\leq  {p}_O \leq \bar{p} \cdot {e}_O,~ {y}} \qquad & {E}_{\xi}[\langle {p}_O, {y}_O \rangle] \\ %
		s.t. \qquad & y_i\in \arg\max_{z\geq 0} u_i(z,y_{-i}, p_i),  \quad  i\in V,
	\end{aligned}
\end{equation}
where the expectation is taken over taste shocks.
The constraint reflects that each agent consumes the payoff-maximizing amount, given the prices and the remaining agents' consumption levels.  The optimal price vector for problem \eqref{eq:sellerOpt} is denoted by ${p}_O^\star$.

It was established in \citet{candogan2012optimal} that
when all agents are observable, the optimal prices set by the platform seller are independent of the network structure, whenever the underlying influence structure is symmetric (and $a_i=\tilde a, b_i=\tilde b$ for all $i\in V$). Interestingly, in Appendix~\ref{app:NetworkInsight}
we show that this is no longer the case when there are latent agents.
In this case,
the platform seller finds it optimal to
 increase the prices offered to observable agents, proportional to how much they are \emph{influenced by} the ``central'' latent agents.
Intuitively,
this is the case since
such observable agents have a strong incentive to consume the product (due to the positive  influence of the latent agents on them), and the platform seller can
improve her profits by charging higher prices to those agents. We detail and formally discuss these points in Appendix~\ref{app:NetworkInsight}.

{Following a similar approach to the previous section, we
denote the
consumption  levels of the observable agents in the  consumption equilibrium induced by some price vector ${{p}}_O$ by ${y}_O({{p}}_O)$.
We denote the corresponding expected revenues
by $\Pi({{p}}_O)$, i.e.,
$\Pi({{p}}_O)={E}[\langle {{p}}_O,
	{y}_O({{p}}_O) \rangle]$.
If the platform seller uses  price vector ${{p}}_O$ instead of ${p}_O^\star$, we capture the induced revenue loss as in \eqref{eq:profitLoss}.
}

Recall that the prices used by the platform seller are nonnegative and less than $\bar p$.
For a given vector of prices ${ p}_O$, we denote by
$({ p}_O \wedge {\bar{p}} \cdot {e}_O)_+$
the vector obtained by capping these prices at $\bar p$ and projecting them to nonnegative reals, i.e.,
$[({ p}_O \wedge {\bar{p}} \cdot {e}_O)_+]_i = \min\{\bar{p}, \max\{p_i,0\}\}$ for all $i\in V_O$.
Using this notation we now state
the main result of this section.

\begin{theorem}\label{thm:optimalprices}
	Suppose that the influence structure admits an $(s_n,r_{1n})$-sparse approximation,  Assumption \ref{assumption:basicRandom} holds, and
	$ 0<{p}_O^\star < \bar p \cdot {e}_O$.
	Suppose further that
	$s_n \sqrt{\log(|V_O|)/n} +  r_{1n} = o(1)$.
	Let $\check{W}^\mu$ denote the thresholded estimator in Theorem \ref{thm:OrthoHinv}, and define
	\begin{equation} \label{eq:priceThm3}
	\hat {{p}}_O =\left( [( \check{W}^\mu +  (\check{W}^\mu)^T)^{-1}
	\check{{v}}_O]\wedge \bar{p} \cdot {e}_O \right)_+.
	\end{equation}
	Then,
	for some constant $C_1>0$,
	with probability
	$1-o(1)$ we have
			$R(\hat{{p}}_O)\leq
	C_1 \left(
	s_n^2  \frac{\log(|V_O|)}{n} +  r_{1n}^2 \right)$.

\end{theorem}

Intuitively, if
$W^\mu$ constitutes a good estimate of $H^{-1}$, then we can exploit this approximation
to
 (solve \eqref{eq:sellerOpt} and)
 construct
 approximately optimal prices.
Theorem \ref{thm:optimalprices} formalizes this intuition.
In particular,
this result establishes that
a seller can
 leverage the algorithm of the previous section to estimate $H^{-1}$,  and  compute prices as in \eqref{eq:priceThm3}
 using these estimates.
 Moreover, it is possible to quantify the revenue loss from using these approximately optimal prices as opposed to the optimal prices.
Our theorem sheds light on how this loss diminishes as a function of the number of available observations.
In settings where the underlying influence structure is approximately sparse (with small $s_n$ and $r_{1n}$), our approach is particularly powerful, and guarantees small revenue loss
and prices that are asymptotically optimal.

{\begin{remark}[Choosing prices adaptively]
Here,
	we do not model the platform seller as an entity that strategically experiments with prices  over time to choose the prices that should be offered to different agents.
	Instead, we assume that there is historical data on the prices used in the past and the induced consumption, and we explore how such data might be leveraged to design future prices.
There is a corresponding online decision problem
where the platform
has no historical data but
adaptively learns optimal prices.
Since the network  is large
the possible externality structures are rich, and this naturally induces an online decision  problem in the high-dimensional regime.
This is a growing area of research (see, e.g., \cite{bastani2015online}), and applications of similar ideas to network pricing problems remain to be interesting research directions.
Note that in our setting
 the
mapping ($H^{-1}$) from the prices to the induced consumption is not sparse, which  makes the  learning problem challenging.
Under approximate sparsity assumptions similar to ours (which are satisfied by various networks as discussed in Section~\ref{se:estimationExamples}),
we suspect that
 it may be possible to develop online learning algorithms with desirable  guarantees   as well.
\hfill \halmos
\end{remark}
}

{\begin{remark}[Static outside price] \label{rem:static}
In this section, we assumed that the price of the product through the alternative channel is fixed at $\bar p$, and the prices offered through the platform are weakly lower.
Fixed outside price  (that is independent of the price offered through the platform)
is natural in some settings.	For instance, some manufacturers employ
minimum advertised pricing or
resale price maintenance policies (see, e.g., \cite{elzinga2008economics,israeli2016minimum,bazhanov2019resale}), where authorized resellers are effectively restricted to offering the product at the manufacturer suggested retail price (or a price close to it).
The products are often not discounted
when purchased directly through the manufacturer (or authorized resellers), though other sellers can still offer the product with some discounts.
Viewing the outside sellers as the manufacturer/authorized resellers, while the platform seller as a 3rd party seller yields the structure in our model.

More interestingly, such prices could also emerge as equilibrium prices in natural settings.
For instance, suppose that  the outside sellers
have access to the real-time information
on the prices used  on the network, and best respond to these prices.
This can be modeled as a  Bertrand competition (with network externalities) among the platform seller and the outside sellers, e.g., brick and mortar stores.
Many online retailers have lower operating costs, as they do not incur costs due to running brick and mortar stores.
Let us assume lower marginal cost for the platform seller,
	and, higher (and identical) ones for the outside sellers.
At the induced equilibrium
(i)
	the prices of the outside sellers would be equal to  their marginal costs, and (ii)
the platform seller's prices would be lower, consistently with our model. \hfill \halmos
	\end{remark}	
}

\begin{remark}[Alternative approaches for approximately optimal prices]
	An alternative approach for constructing approximately optimal prices involves first estimating $H$ (as opposed to $H^{-1}$), and then using the prices that would be optimal if $H$ were equal to this estimate (see Lemma \ref{lem:optPrices} for the dependence of $p_O^\star$ on $H$).
	However, estimating  $H$ presents similar challenges to those of estimating $H^{-1}$.
First, note that in the presence of latent agents the matrix $H$ is not sparse in general.
	To see this,  note on the one hand  that the consumption of observable agents has \emph{direct} influence on the consumption of their observable neighbors, which can be represented by a sparse matrix when the underlying network is sparse.  On the other hand, observable agents also influence other observable agents \emph{indirectly} through the influence they exert on the latent agents (who in turn influence other observable agents). This latter influence structure  is not sparse in general. 	 Hence, the  matrix $H$ that captures the aggregate influence that the observable agents exert on each other need not be sparse.
	Second,  to construct approximately optimal prices through the estimates of the $H$ matrix, it is necessary to obtain  small  estimation errors for both rows and columns.
	When estimating $H^{-1}$,
	this was accomplished by our algorithm in Section~\ref{se:estimation}.
	We explore how ideas such as approximate sparsity can be exploited to obtain estimators
	for the aggregate influence structure $H$
	with desirable guarantees on row/column errors in our companion paper \citet{ata2018IVnetwork}. \hfill \halmos
\end{remark}

{
\begin{remark}[Relaxing the assumptions]
	In this and the previous section, we assumed that externalities are positive, i.e., $G_{ij} \geq 0$, and required that the $a_i$ parameter is not small (through
	Assumptions \ref{ass:OutsideOpt0} and \ref{ass:OutsideOpt}).
	These assumptions were made to ensure
	the nonnegativity of the induced consumption levels, and hence
	the linear dependence of outcomes on the covariates. It is worth highlighting that weaker assumptions that ensure such linearity are sufficient for the results to go through. For instance, limited negative externalities (together with bounded covariates) can be allowed.  \hfill \halmos
\end{remark}
}

\subsection{Numerical Example}
\begin{figure}
	\centering
	\includegraphics[width=0.6\linewidth]{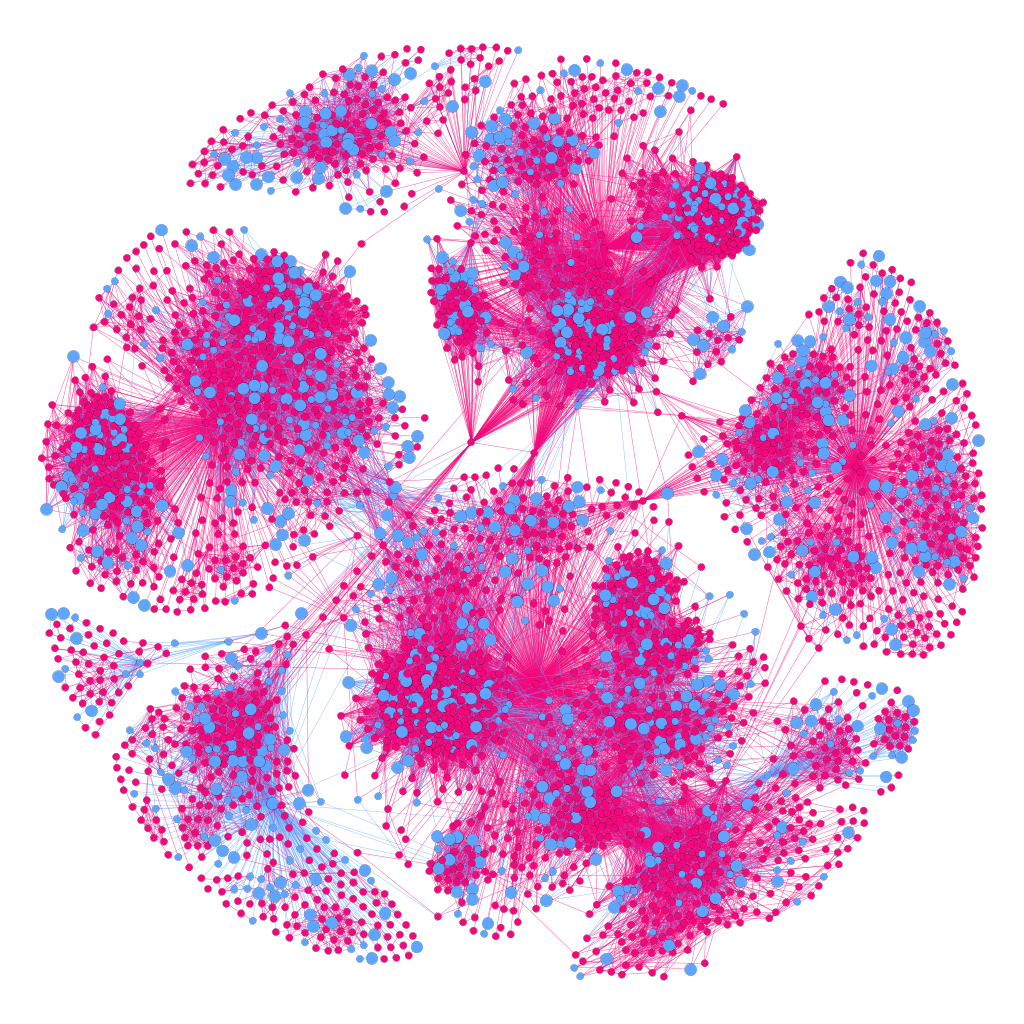}
	\caption[]{Facebook subnetwork from \cite{snapnets}. Observable agents are highlighted in blue, and the remaining agents are latent.}
	\label{fig:facebookv4}
\end{figure}

In this section we illustrate our applications by focusing on a nontrivial network structure.
Specifically,
we focus on an induced subnetwork of the Facebook network, provided by \cite{snapnets}.
The subnetwork consists of 4,039 nodes (agents) and 88,234 edges.
The degrees of the agents
(which vary between 1 and 1,045) and the
connection structure are quite heterogeneous, as can be seen from  Figure \ref{fig:facebookv4}.

We focus on the advertising application
(where $\lambda_i=1$ for all $i$)
and assume that the edge  weights are given by $G_{ij} \in \{0,1/2b\}$ for some parameter $b$.
We set this parameter equal to
$100$, and note that in this case Assumption \ref{ass:DiagDominance} does not hold.
In particular,
due to the large degrees of some nodes,
for many rows (or columns) of $M$ the absolute sum of off-diagonal entries exceeds~$1$.
Our objective is to illustrate the applicability and the performance of our estimator even in settings where some assumptions made for our asymptotic results no longer hold.
We uniformly at random choose $1000$ nodes, and assume that they are observable  while the remaining nodes are latent (see Figure \ref{fig:facebookv4}).
We assume that $a_i=1$ and
$z_i=10$
for all agents, and we assume that these parameters as well as the network structure are unknown to the advertiser.

We assume that
$n=50$ samples of
advertising decisions/outcomes, denoted by
$\{(y_O^{(t)},p_O^{(t)} )\}_{t\in [n]}$
are available for  observable agents.
For $i\in V_O$, we draw each $p_i^{(t)}$ independently from $U[0,1]$, while we set
$p_i^{(t)}=0$ for $i\in V_L$.
Similarly, the taste shocks
$\{\xi_i^{(t)}\}$
are also independent (from each other and $\{p_i^{(t)}\}$), and drawn from
$U[-1,1]$.
As before, we assume that the problem of the advertiser is to choose  advertising  intensities $\{p_i\}_{i\in V_O}$ to maximize
$ \Pi(p_O) = E [e^T y_O(p_O)] -  \frac{\chi}{2}  \sum_{i\in V_O} p_i^2$.
For our numerical studies we assume that $\chi=1$.

We do not restrict the maximum advertising intensity, i.e., we set $\bar p=\infty$.
Let $p_O^\star$ be
the optimal solution to \eqref{eq:adOpt}.
We observe that this vector has strictly positive  entries
and is given as in \eqref{eq:optAdv}.
Recall that $p_O^\star$ gives the advertising intensities that maximize the payoff of the advertiser, if the
parameters of the underlying system were known.
Since she does not know these parameters, we assume that the
advertiser follows  the approach discussed in Section  \ref{subse:estimationAdvertising}.
In particular,
she uses the available data to obtain an estimate of
$\tilde{H}^{-1}$, and then constructs $\hat{p}_O$
as in
Theorem~\ref{thm:optimalAdv} using these estimates.

{To estimate $\tilde{H}^{-1}$,
	we apply Algorithm  \ref{alg:Alg1} with three standard modifications
(see, e.g., \cite{tibshirani1996regression})
 which boost   performance when small number of samples are available.
In particular,
(i) we use demeaned and standardized covariates, (ii) we do not penalize the intercept, and (iii) we cross validate some model parameters.
We detail these modifications next.}

Let
$\bar{p}_i = \frac{1}{n}\sum_{t\in [n]} p_i^{(t)}$, and
$\sigma_i =\sqrt{ \frac{1}{n} \sum_t ( p_i^{(t)} - \bar{p}_i^{(t)} )^2}$.
We compute these quantities and construct
the demeaned and scaled covariates
$d{p}_i^{(t)} = ( p_i^{(t)} - \bar{p}_i)/\sigma_i$ (hereafter we use the prefix $d$ to denote quantities associated with the demeaned/scaled system).
Note that using these variables
\eqref{eq:yArr_obs} can be rewritten as:
\begin{equation} \label{eq:newLinSys}
	{y}_O^{(t)} =
	d{v}_O  -d{H}^{-1} d{p}_O^{(t)} +\varepsilon_O^{(t)},
\end{equation}	
where
\begin{equation}\label{eq:scalingEq}
	\mbox{$d{v}_O=({v}_O-\tilde H^{-1} \bar p_O) $, \quad
		$d{H}^{-1} =\tilde H^{-1}S$	},
\end{equation}
and $S $ is a diagonal matrix with diagonal entries $\{\sigma_{i}\}_{i\in V_O}$.
We use Algorithm  \ref{alg:Alg1}  to estimate the linear system in \eqref{eq:newLinSys} as opposed to the one in \eqref{eq:yArr_obs}.
That is, we use the algorithm with input $\{(y_O^{(t)}, dp_O^{(t)} )\}_{t\in [n]}$, and
obtain
the estimates
$d{\check{v}}_O$ (and $d\check{W}^\mu$)
of
$d{v}_O$ (and $d{H}^{-1}$).
Then we ``rescale'' these estimates  (using the equations in \eqref{eq:scalingEq} that relate $d{v}_O$ and $d{W}$ to the parameters
$v_O$, $\tilde H^{-1}$, $\bar p_O$
of the model)
to obtain estimates
of $v_O$ and $H^{-1}$.
More precisely, we set
$\check W^\mu = d{W}^\mu S^{-1}$ and
$\check v_O = d\check{v}_O+ \check W^\mu \bar p_O$.
We choose the threshold parameters $\mu$
through bootstrapping
as described in  Remark  2
(and Appendix \ref{Sec:SimultaneousBootstrap}; using the quantile parameter $\alpha=0.05$).
Though, we verify that the alternative choice of these parameters  in \eqref{eq:thresholds} also gives similar results.

Second,
in Step 1 of the algorithm,
we modify the first constraint for $j=0$. Specifically, in this case we replace the aforementioned constraint with
$\left| \frac{1}{n}\sum_{t=1}^{n} \{y_k^{(t)} - (\tilde v_k-		\tWkdot p_O^{(t)})\}\right| \leq 0$.
That is, when $j=0$, we replace the quantity on the right hand side with zero.
This allows us to treat the intercept in our linear system differently, and reduce the errors in its estimation in Step 1.

Third, in Step 1 as opposed to using the $\tau$ parameter given in the algorithm, we identify it via cross-validation (and use the cross-validated parameter value $\tau=2$).
Our asymptotic results are robust to the exact choice of this parameter (e.g., a different constant could be used as opposed to $4$ in its definition
in Algorithm \ref{alg:Alg1}
 while still yielding same asymptotic guarantees), however, different parameter values can lead to different finite sample performance.
We optimize it via cross-validation (over a grid of 100 values containing the value provided in the algorithm).
Note that other parameters of the algorithm (such as $\lambda$) can also be optimized using cross-validation, but
we do not explore this here, since
even without  cross validating those parameters we obtain good performance from our estimators.

For the setting described here, we see that despite having access to a small number of
samples, the advertiser can achieve  a payoff that is very close to the optimal.
Specifically, we obtain that
$\Pi(p_O^\star)\approx 66858$,
$\Pi(\hat p_O)\approx 62326$.
These in turn imply that $R(\hat p_O)\approx 0.067$. That is, using the available data a targeting structure that is within  $6.7\%$ of the optimal can be achieved.
Without estimating the parameters of the underlying system it is not clear how to set the advertising intensities.
It is also worth noting that in this example not advertising at all yields substantially lower aggregate consumption (e.g., about 1500 units vs. more than 60000 units with optimal advertising) and payoff to the advertiser.
\section{Examples of Approximately Sparse Networks} \label{se:estimationExamples}

The results in Section
	\ref{se:estimation}
	rely on approximate sparsity of $H^{-1}$.
	The rates of convergence of the proposed estimators (in Theorem \ref{thm:OrthoHinv} and Theorem \ref{cor:Thm1}) are particularly useful if there exists a sparse approximation to $H^{-1}$ with a few nonzero entries in each row/column that
	leads to small approximation errors, i.e.,
	if we can set  the maximum number of nonzero elements in each row and column ($s_n$) so that  $s_n^2/n$ and the approximation error $r_{1n}$ are small.
	In this section,
	we establish that this indeed is the case for many classes of networks.

Throughout the section we impose Assumption \ref{ass:DiagDominance},  and   analyze different classes of networks.
We start with two special cases: (i) $m$-banded networks and (ii) matrices that exhibit polynomial off-diagonal decay of connection strength.
Then, we consider a fairly general class of networks,
where we require agents' neighborhoods  not to grow too fast.
For all these cases, we establish
$(s_n,r_{1n})$-sparsity of the $H^{-1}$ matrix, and provide results on convergence rates when Algorithm 1 is employed.
The proofs of the results in this section can be found in Appendix~\ref{ref:SecEstEx}.

\subsection{$m$-Banded Networks}
A matrix $A$ is \emph{$m$-banded} if its
nonzero entries are at most $m$ entries away from its diagonal, i.e.,
$A_{ij} = 0$ for $|i-j|>m$.
Motivated by this definition,
we say that a network $\cal G$ is \emph{$m$-banded} if
for some permutation $\ell:V\rightarrow \{1,\dots, |V| \}$,
its adjacency matrix satisfies
\begin{equation} \label{eq:mBanded}
G_{ij}=0 \qquad \mbox{if \ $|\ell(i)-\ell(j)|>m$}.
\end{equation}
In other words, for an $m$-banded network a permutation of  rows/columns of  the corresponding matrices $G$ and $M=\Lambda-G$ are $m$-banded (e.g., consider the matrix $G^\ell$ such that $(G^\ell)_{\ell(i),\ell(j)}:= G_{ij}$ for all $i,j\in V$).	
This definition captures cases where nodes
are embedded at the integer points on the real line, and they  have only
``local'' connections, i.e., connections with $m$ nodes to the left and $m$ nodes to the right.\footnote{We emphasize that in our context the
	nodes that correspond to observable agents need not be consecutive integers on the real line. Nevertheless, it can be seen that if the underlying  network is $m$-banded, then the induced subgraph of observable (similarly latent) agents is also $m$-banded.}

It is known that	if $A$ is an $m$-banded matrix,
then $A^{-1}$ exhibits
``exponential decay'' (see \cite{demko1984decay}), where the rate of  decay is characterized in terms of
the singular values of $A A^T$.
That is,
as we get away from the diagonal the magnitude of the entries decays exponentially.
We next
adapt this result to our setting
	and obtain an exponential decay result on the inverse of $M$.
	In this subsection, we discuss our results, using the shorthand notation
	$\tilde{C}_1:= 	2(m+1) \frac{ \bar{\lambda}(2 \bar{\lambda}  - \zeta)}{\zeta^2 ( \bar{\lambda}-\zeta)}$ and
	$\lambda_1 := \left(\frac{ \bar{\lambda}-\zeta}{ \bar{\lambda}}
	\right)^{1/m}$.
\begin{lemma} \label{lem:m-banded}
	Suppose that the underlying network $\cal G$ is $m$-banded. Then, for any $i,j\in V$ we have
	\begin{equation}\label{eq:m-bandedLem}
	| (M^{-1})_{ij}|
	\leq \tilde{C}_1 \lambda_1^{|\ell(i)-\ell(j)|}.
	\end{equation}
	Similarly,
	for $i,j\in V_O$,
	$|(H^{-1})_{ij}|
	\leq \tilde{C}_1 \lambda_1^{|\ell(i)-\ell(j)|}$.
\end{lemma}

This result suggests a natural $(s_n,r_{1n})$-sparse approximation for $H^{-1}$:
since entries of
$H^{-1}$ decay (exponentially)
in $|\ell(i)-\ell(j)|$
for any  $i,j\in V_O$, for an $(s_n,r_{1n})$-sparse approximation of these matrices we just focus on
$i,j$ for which $|\ell(i)-\ell(j)|$ is small.
In particular,
let  $\bar W\in \mathbb{R}^{|V_O|\times |V_O|} $ be a matrix,  such that for $i,j\in V_O$ we have
\begin{equation}\label{eq:ex1Wbar}
\bar{W}_{ij}=
\begin{cases}
(H^{-1})_{ij} & \mbox{for $|\ell(i)-\ell(j)|\leq \frac{s_n-1}{2} $},\\
0 & \mbox{otherwise.}
\end{cases}
\end{equation}	
Observe that $\bar{W}$ has at most $s_n$ nonzero elements in each row and column by construction.
Our next result
establishes that $\bar{W}$ yields a sparse approximation of $H^{-1}$.

\begin{lemma}\label{lem:decayMbanded}
	Suppose that $\cal{G}$ is $m$-banded. Let
	$\bar{W}$
	be given as in \eqref{eq:ex1Wbar}.
	Then $\bar{W}$
	is an $(s_n,r_{1n})$-sparse approximation
	of  $ H^{-1}$, where
	$s_n=\tilde{C}_2 \max\left\{m,  \log {n}  \right\} $ and
	$r_{1n}=\frac{1}{n}$
	for some constant $\tilde{C}_2>0$.
\end{lemma}
In this result, $\tilde{C}_2$ is a constant that depends on $\tilde{C}_1$ and  $\lambda_1$, and its precise form is provided in the proof of the lemma.
This result shows that for $m$-banded networks, a sparse approximation where
$s_n$ is logarithmic in $n$ and the error term
$r_{1n}$
scales with $1/n$ can be obtained.
Thus for such networks,  Algorithm 1 can be used to obtain a consistent estimate of $H^{-1}$
with desirable rates. Our next result explicitly characterizes the rates of this estimator.

\begin{corollary}
	Suppose that  $\cal G$ is $m$-banded.
	Consider the estimators
	$\check W^\mu$ and $\check W$
	given  in
	Algorithm \ref{alg:Alg1}.
Under the Assumptions of Theorem \ref{thm:OrthoHinv}, with probability $1-o(1)$ we have
	\[
	\max {\bigg \{ }
	\| \check W^\mu-H^{-1}\|_\infty  , \|
	\check W^\mu-H^{-1}
	\|_1
	{\bigg \} }
	\leq
\tilde{C}_3 \max\left\{m,  \log {n}  \right\}
	\sqrt{\frac{\log V_O}{n} },\]
	for some constant $\tilde{C}_3>0$.
	Moreover,
	\[\sqrt{n}\{ (-\check{{v}}_k, \check W_{k,\rowdot})^T - (-v_k, H^{-1}_{k,\rowdot})^T\} =   - \hat \Psi\frac{1}{\sqrt{n}}\sum_{t=1}^n \varepsilon_{k}^{(t)}(1;{p}_O^{(t)}) + R^k_n\]
	where
	$\|R^k_n\|_\infty
	= O\left(\frac{1}{\sqrt{n}}  \max\{m,\log n \}
		\log |V_O|   \right)$.
\end{corollary}
We omit the proof of this result, as it is an immediate corollary of Lemma \ref{lem:decayMbanded},
Theorem \ref{thm:OrthoHinv}, and Theorem \ref{cor:Thm1}.
Recall that in
Theorem \ref{thm:OrthoHinv} and Theorem \ref{cor:Thm1}
we have three sets of estimation results. The first set
focuses on ``asymptotic normality'' and the second one
characterizes the $\ell_1, \ell_\infty$ errors for the corresponding thresholded estimators.
These two sets of results depend on the approximate sparsity of the underlying matrices, and  are reported above for $m$-banded networks.
The third set of results pertains to the entry-wise estimation errors for estimators $\check W$, which is independent of the specifics of the sparse approximation $(s_n,r_{1n})$ and hence is excluded from our corollary.
 In subsequent subsections,
as we exemplify other approximately sparse networks,
for compactness
we  report our results only on $\ell_1, \ell_\infty$ errors. However, we emphasize that similar bounds for the asymptotic normality results can also be readily obtained by using
the bounds of Theorem  \ref{thm:OrthoHinv} and Theorem \ref{cor:Thm1}
together with the approximate sparsity bounds $(s_n,r_{1n})$ of the underlying networks.

\subsection{Polynomial Decay of Connection Strengths}

The $m$-banded
structure discussed in the previous section is very special, and in particular it disallows for long-range connections considered in small-world models (where agents in addition to local connections are allowed to have a few long-range connections, eventually leading to a network with a small diameter;
see
\citet{watts1998collective,kleinberg2000small}).
We next argue that our results on $m$-banded networks carry over when such connections are allowed, provided that their strength is decaying with the distance between agents.

In particular,
we next consider networks
for which the
entries of $G$
satisfy
\begin{equation} \label{eq:decayCond}
G_{ij}\leq \frac{ \tilde C}{(1+|\ell(i)-\ell(j)|)^{\theta}} \qquad \mbox{for some $\theta>1, \tilde C>0$, and all $i,j\in V$},
\end{equation}
where, as before, $\ell:V\rightarrow \{1,\dots,|V|\}$ is some permutation.
Thus, the structure in
\eqref{eq:decayCond}
allows for long-range connections, but requires the connection strength to decay polynomially as the ``distance'' increases.\footnote{Once again, the nodes can be thought of as corresponding to integer points on the real line  and the observable (latent) nodes need not be  contiguous. It can be readily seen that if the network satisfies \eqref{eq:decayCond}, then the induced subnetworks of observable/latent nodes also satisfy this condition.}

Jaffard's theorem \citep{jaffard1990proprietes}
is essential for our analysis in this section.
This theorem implies that the inverse of  a matrix whose entries decay polynomially
with distance from the diagonal
also exhibits similar decay properties.
More precisely,
the aforementioned result
focuses on infinite matrices and establishes that
Banach algebras  of matrices with a polynomial off-diagonal decay property are inverse-closed \citep{grochenig2010noncommutative,grochenig2014norm}.
A quantitative version of this result, with explicit bounds on decay parameters, is given in \citet{grochenig2014norm}.
Even though this result is for infinite matrices, viewing finite matrices as diagonal subblocks of infinite matrices immediately yields the following result, which we state without a proof:
\begin{lemma}\label{lem:jaffardDecayOrig}
	Let $A \in \mathbb{R}^{k\times k}$ be a matrix such that
	$|A_{ij}|\leq \tilde C/(1+ |i-j|)^\theta$.
	Then, $|A^{-1}_{ij}|\leq  \tilde C_1/(1+ |i-j|)^\theta$, where
	$\tilde C_1$ is a constant
	that depends only on $\theta, \tilde C$, and $\|A^{-1} \|_2$.
\end{lemma}

Applying this result
to network matrices that satisfy \eqref{eq:decayCond},
after some algebraic manipulations,
we obtain the following result:	
\begin{lemma}\label{lem:jaffardDecay}
	Suppose that \eqref{eq:decayCond} holds. Then,
	$\max\{ |M^{-1}_{ij}|,|H^{-1}_{ij}|\} \leq
	\frac{\tilde C_1}{(1+|\ell(i)-\ell(j)|)^{\theta}}$
for some constant
$\tilde{C}_1 > 0$.
\end{lemma}
In this result $\tilde{C}_1$
depends on
$\tilde C,\bar{\lambda}, \theta, \zeta$.
This lemma together with the decay property in \eqref{eq:decayCond}
provides means of obtaining sparse approximations of $H^{-1}$.
The high-level idea here is
similar to the one in the context of $m$-banded matrices:
prune entries that correspond to pairs of nodes that are ``far away.''
More precisely,  given number of nonzero elements in each row/column $s_n$,
let
$\bar W$ once again be  given as in
\eqref{eq:ex1Wbar}.
As before, $\bar{W}$ has at most $s_n$ nonzero elements in each row and column by construction.
Our next result establishes that thanks to the polynomial decay property of the entries of $G$,
the error  due to omitting
entries that correspond to faraway pairs of nodes is insignificant.

\begin{lemma}\label{lem:polyDecay}
	Suppose that \eqref{eq:decayCond} holds.
	Let
	$\bar{W}$
	be given as in \eqref{eq:ex1Wbar}.
	Then $\bar{W}$
	is an $(s_n,r_{1n})$-sparse approximation
	of  $H^{-1}$, where
	$s_n =   (n/\log |V_O|)^{1/2\theta}  $ and
	$r_{1n} \leq \frac{\tilde C_1 2^\theta }{\theta-1}
	\left(\frac{\log|V_O|}{n}\right)^{(\theta-1)/2\theta}$.
\end{lemma}

Using this result, we obtain
the following rates of convergence for the  estimator defined in
Algorithm 1:

\begin{corollary} \label{cor:polyDecay}
	Suppose that  \eqref{eq:decayCond} holds.
		Consider the
		thresholded estimator
	$\check W^\mu$
	given  in
	Algorithm \ref{alg:Alg1}.
Under the assumptions of Theorem \ref{thm:OrthoHinv}, with probability $1-o(1)$
we have
	\[ \max{\big \{}\| \check W^\mu-H^{-1}\|_\infty  , \|
	\check W^\mu-H^{-1}
	\|_1  {\big \}}
	\leq
	\left(\tilde{C}_2+\frac{\tilde{C}_3 2^\theta }{\theta-1}  \right)
	\left(\frac{\log|V_O|}{n}\right)^{(\theta-1)/2\theta},
	\]
	for some constants $\tilde{C}_2,\tilde{C}_3 >0$.
\end{corollary}
This result is an immediate corollary of   Lemma
\ref{lem:polyDecay}
 and
Theorem \ref{thm:OrthoHinv}, and hence its proof is omitted.

\begin{remark}[Exponential Decay] Here we restrict attention to polynomial decay of entries of $G$.
	If the decay rate is exponential, i.e.,
	$ |M_{ij}|\leq \tilde C \exp(-\theta|\ell(i)-\ell(j)|)$, then a version of Jaffard's theorem implies that $M^{-1}$ also exhibits exponential decay  \citep{benedetto2006}. In this case,
	by choosing $s_n, r_{1n} $  appropriately,
	it is possible to  obtain faster rates of convergence than those of Corollary \ref{cor:polyDecay}.
	Moreover, in this section (as in the previous one) we implicitly assume that nodes are embedded in a one-dimensional grid.
	Similar polynomial/exponential decay results continue to hold if the nodes are embedded in a $k>1$-dimensional grid,
	and the decay of influence parameters between nodes $i$ and $j$ is expressed in terms of
	the distance between the nodes
	 in each dimension
	(see, e.g., \citet{grochenig2010noncommutative}).
	Thus the results of this subsection readily extend to more general network structures with appropriate decay of connection strength. \hfill \halmos
\end{remark}

{
	\begin{remark}[Small worlds]
In many real-life social networks, agents have many local ties (e.g., friends who live in the same city) and a few long distance ties (e.g., friends who are in other cities or countries an individual lived in the past).
This was the key idea in the seminal work of \cite{kleinberg2000small},
where the network was modeled as a $2$-dimensional grid with few additional long distance edges (and where the probability of a connection decays with distance).
Moreover, it is natural that local links are more influential than the long distance ones (since an individual is more likely to interact with local friends and be influenced by them). In such settings, the induced influence structure satisfies
the assumptions of this section (or their extensions to the $k$-dimensional grid as explained in the previous comment), and
 our framework applies. Richer degree distributions  (e.g., power laws) can also be allowed (e.g., by randomly drawing the  number
of  ties each agent has from an appropriate distribution and randomly wiring them in a way that makes the longer distance connections less likely  as in Kleinberg's model), for as long as the property that local links are more influential is satisfied.
 \hfill \halmos
	\end{remark}
}
\subsection{A General Class of Networks with Bounded Neighborhood Growth}
Our examples thus far have focused on settings where the connections are ``local,'' or their strength decays as
the distance between nodes increases.
While these assumptions simplify the analysis, they are not necessary for employing our estimator in Algorithm \ref{alg:Alg1}. In this section, we introduce a fairly general model of network connections that allows for strong long-distance connections and establish that our algorithm still achieves desirable convergence rates.

We start by stating the assumption we impose  in this section:
\begin{assumption} \label{ass:diagonalizable}
	$M$
	and $M_{LL}$ are
	diagonalizable, i.e., $M= X D X^{-1}$ and
	$M_{LL}= Y D_{LL} Y^{-1}$
	for some diagonal matrices $D$,  $D_{LL}$ and invertible matrices $X$ and $Y$.
	In addition,
	the condition numbers of matrices $X$ and $Y$ are bounded by a constant, i.e.,
	$\kappa(X), \kappa(Y) \leq \bar{\kappa}$
	where $\kappa(A) := ||A||_2 \times ||A^{-1}||_2$.	
\end{assumption}

When
$M= X D X^{-1}$,
the diagonal entries of $D$ correspond to the eigenvalues of $M$, and the columns of $X$ correspond to the right eigenvectors (similarly for $M_{LL}$).
Diagonalizability is a mild condition since the set of diagonalizable matrices is dense in the set of all (square) matrices, and $M$ is diagonalizable, e.g.,
if it has $|V|$ distinct eigenvalues \citep{golub2012matrix}.
The  assumption on the condition number  of $X$ holds, for instance, if $M$ admits an orthogonal set of eigenvectors since in this case $X$ can be chosen to have orthonormal columns and $\kappa(X)=1$.
\footnote{A sufficient condition for
	$M$ to admit an orthogonal set of eigenvectors is for $G$ (and hence $M$) to be symmetric.
	Another condition is to have
	$\lambda_i=\bar{\lambda}$ for all $i$ and
	the underlying weighted directed graph to
	satisfy $\sum_k g_{kj} g_{ki}= \sum_k g_{ik} g_{jk}$.
	If weights are binary, the latter
	can be interpreted as a regularity condition,
	as it guarantees that the total number of common out-neighbors of $i,j$ (i.e., nodes that are influenced by both $i$ and $j$) is equal to the total number of common in-neighbors (i.e., nodes that influence both $i$ and $j$). In matrix notation, this condition can be stated as $G^T G = G G^T$. Hence, it implies that $G$ is a normal matrix, and that its eigenvectors are orthogonal.
	When $\lambda_i=\bar{\lambda}$ for all $i$, this also implies that $M$ has orthogonal eigenvectors.
	Thus, while
	the condition on $\kappa(X)$ is readily satisfied by
	 undirected networks, it is also satisfied by
	 directed structures where $G$ is not symmetric. These conditions focus on the $\kappa(X)=1$ case. However, note that $\bar{\kappa}$ can be any constant, and richer structures are allowed. Furthermore, by allowing $\bar{\kappa}$ to depend on $|V_O|$, similar results to the ones in this subsection can be obtained at the expense of looser bounds.}
In general, this assumption requires the eigenvectors of $M$ corresponding to different eigenvalues to be sufficiently different (i.e., no eigenvector should be well approximated by a linear combination of other eigenvectors).
Assumption \ref{ass:diagonalizable} similarly requires that
$M_{LL}$ be diagonalizable with sufficiently different eigenvectors.

It is known that by using
spectral theory and approximation theory,
elegant approximations of
matrix functions can be obtained
\citep[see, e.g.,][]{demko1984decay,benzi2007decay}.
For completeness, we next outline how
under  Assumption \ref{ass:DiagDominance},
sparse approximation of matrix $M^{-1}$  can be obtained using the aforementioned ideas.

Under Assumption \ref{ass:DiagDominance}, it can be shown that singular values of $M$ belong to the interval
$[\zeta, 2\bar{\lambda}-\zeta]$ (see Lemma \ref{lem:matBounds} in Appendix \ref{subse:MatrixLemma} for a proof).
Let ${\cal D}$ denote the disc
in the complex plane
centered at $\bar{\lambda}$ with radius $\bar{\lambda}-\zeta$ that contains this interval.
Observe that the function $h(x):= x^{-1}$ is analytic in ${\cal D}$.
Let $\pi_k$ denote the set of polynomials of degree at most $k$. Given a subset $K$ of the complex plane
and a function $f$ whose domain contains $K$,
let
$||f||_K := \sup_{z\in K} |f(z)|$, and let
$e_k(K) :=  \inf_{g\in \pi_k} ||f-g||_K$. That is,  $e_k(K)$ corresponds to the error of the best
uniform approximation of $f$ with a polynomial of degree $k$ on set $K$.
We will use the classic results on approximation of $f(\cdot)$ with polynomials (see, e.g., \citet{demko1984decay} and Section 4.3 of \citet{meinardus2012approximation}).\footnote{The best approximating polynomial can also be explicitly obtained.}
\begin{proposition}\label{prop:polyAppx}
	Let $r=(2\bar{\lambda}-\zeta)/\zeta$ and $q= (\sqrt{r}-1)/(\sqrt{r}+1)$. Then,
	$e_k({\cal D}) \leq \frac{(1+r^{1/2})^2}{2\zeta r}q^{k+1}$.
\end{proposition}
Let $g_k^\star$ denote the polynomial that
achieves the best possible approximation error $e_k$ in Proposition~\ref{prop:polyAppx}.
With abuse of notation let $h(M)=M^{-1}$,
and let
$g^\star_k(M)$ be the corresponding  $k$th-order matrix polynomial of $M$.
By diagonalizability of $M$ we have
\begin{equation} \label{eq:polyApprox}
\begin{aligned}
|| h(M) - g_k^\star(M) ||_2
&= || X( h(D)-g_k^\star(D) ) X^{-1} ||_2
\leq ||X^{-1}||_2 || X||_2 || h(D)-g_k^\star(D)||_2\\
&\leq
\kappa (X) \max_{z\in \sigma(M)} |h(z) - g_k^\star(z)|\\
& \leq
\tilde{C}
q^k,
\end{aligned}
\end{equation}
where $\sigma(M)$ denotes the set of singular values of $M$.
Here,   the second inequality follows from spectral theory (see  \citet{demko1984decay}),
the last inequality uses Proposition \ref{prop:polyAppx}, and $\tilde{C}=\bar{\kappa}q\frac{(1+r^{1/2})^2}{2 \zeta r}$ is a constant.

Equation \eqref{eq:polyApprox} suggests that $M^{-1}$ can be approximated through a (matrix) polynomial of $M$.
When this polynomial has a small degree, it induces a sparse approximation of $M^{-1}$.
Moreover,  exploiting the
fact that
$H^{-1}$ can be expressed in terms of a submatrix of $M^{-1}$  (see Lemma~\ref{lem:MInv0}),
we can employ
 this result  to obtain sparse approximations of $H^{-1}$ as well.

To see this, denote by  $\bar W$ the $|V_O| \times |V_O|$ submatrix
of $g^\star_k(M)$ corresponding to observable agents, i.e.,
\begin{equation} \label{eq:ex3barW}
\bar W = [g^\star_k(M)]_{OO}.
\end{equation}
The submatrix of $M^{-1}$ corresponding to observable agents is given by $H^{-1}$. Thus, from the observations above $\bar W$ can be viewed as a sparse approximation of $H^{-1}$. In particular, \eqref{eq:polyApprox} implies that
\begin{equation} \label{eq:polyApproxHinv}
\begin{aligned}
\|  H^{-1} - \bar{W}\|_2 \leq
\| h(M) - g_k^\star(M) \|_2
\leq \tilde{C}
q^k.
\end{aligned}
\end{equation}

The above discussion  suggests that $\bar W$ can serve as an approximation  of $H^{-1}$.
We next  establish that
$\bar W$ indeed  constitutes a sparse approximation and characterize the sparsity parameters $s_n$, $r_{1n}$.
Before we state our result, we introduce a relevant definition.

\begin{definition}
	Let  $\rho(i,j)$ denote the hop distance from node $i$ to $j$, i.e.,
	  the minimum number of edges on a (directed) path from $i$ to $j$:
	$$\rho(i,j) :=\min  \{ k\in \mathbb{Z}_{+}|   i_1=i, i_{k+1}=j,  (i_\ell,i_{\ell+1})\in E \mbox{ for $\ell=1,\dots,k$} \}.$$
	We say that neighborhoods
	of  agents
	exhibit exponential/polynomial growth if  the following conditions hold:
	\begin{itemize}
		\item[(i)] Exponential growth: the number of agents that are at most $k$ hops away from $i$ is bounded by an exponential function with exponent $k$; i.e., there exist constants $C_e, d_e>0$ such that
		$$|\{ j\in V \mid \rho(i,j)  \leq  k \}  \cup \{ j\in V \mid \rho(j,i)  \leq  k \}|  \leq C_e d_e^k, \ \ \ \mbox{for all $i\in V$, $k \in \mathbb{Z}_+$}.$$
		\item[(ii)]	Polynomial growth:
		the number of agents that are at most  $k$ hops away from $i$
		is bounded by a polynomial of $k$; i.e.,  there exist constants $C_p, d_p>0$ such that
		$$|\{ j\in V \mid \rho(i,j)  \leq  k \} \cup \{ j\in V \mid \rho(j,i)  \leq  k \} | \leq C_p k^{d_p}, \ \ \mbox{for all $i\in V$, $k \in \mathbb{Z}_+$}.$$
	\end{itemize}
\end{definition}

\begin{lemma}\label{lem:appxDiagonalizable}
	Suppose that Assumption \ref{ass:diagonalizable} holds.
	Let $\bar{W}$ be defined as in
	\eqref{eq:ex3barW}, for some
	appropriately chosen $k$, and let $q<1$ be defined as in Proposition \ref{prop:polyAppx}.
	\begin{itemize}
		\item[(i)] If the neighborhoods exhibit exponential growth, then
		the matrix $\bar{W}$ is an
		$(s_n,r_{1n})$-sparse approximation
		of  $H^{-1}$, where for
		 $\nu=\log_{d_e/q}(d_e )$
		 and  some constants $\tilde{C}_1,\tilde{C}_2>0$
		 we have
			$$s_n= \tilde{C}_1
		\left(\frac{ \sqrt{n|V_O|}}{\log(|V_O|)} \right)^{\nu}, \ \
		\mbox{and}
		\ \
		r_{1n}=
		\tilde{C}_2
		{|V_O|}^{\nu/2}
		\left( \frac{\log(|V_O|)}{\sqrt{n} } \right)^{1-\nu}.$$

		\item[(ii)] If the neighborhoods exhibit polynomial growth, then
		$\bar{W}$ is an $(s_n,r_{1n})$-sparse approximation
		of  $H^{-1}$, where
		for		 some constants $\tilde{C}_3,\tilde{C}_4>0$
		we have
		$$s_n= \tilde{C}_3
		\left(\frac{n}{\log(|V_O|)}\right)^{1/4},
\ \ \mbox{and} \ \ 		
		r_{1n}=
		\tilde{C}_4 {\sqrt{|V_O|}}
		q^{ \left({ {\left(\frac{n}{\log(|V_O|)}\right)^{\frac{1}{4d_p}}} } \right)}
		.$$
	\end{itemize}

\end{lemma}
The constants
$\tilde{C}_1,\tilde{C}_2,\tilde{C}_3,\tilde{C}_4$
are independent of $n$ and $|V_O|$ but can depend on the parameters
$\tilde{C}, C_e, C_p, d_e, d_p, q, \bar{p}, \zeta$.
Their precise characterizations are given in the proof of the lemma.
Note that this lemma implies that when neighborhood growth is polynomial, it is possible to choose a sparse approximation
that scales with a root of the  number of observations $n$,
while
ensuring that
the approximation error $r_{1n}$ decays exponentially fast.
This observation allows for achieving small approximation errors with  sparse structures.

With exponential neighborhood growth,
achieving small approximation errors requires employing less sparse structures. In this case,
for the approximation error $r_{1n}$  to be $O(1)$, it is necessary to have $n=\Omega(\log(|V_O|)^2 |V_O|^{\nu/(1-\nu)} )$.
On the other hand, in this case,
$s_n=\Omega(|V_O|^{\nu/2(1-\nu)})$.
This implies that in order to ensure a small approximation error it is necessary to have the sparsity parameter scale with a root of $|V_O|$.
Note that when $\nu$ is small, the degree of the root is large and hence it is possible to achieve a small error while still ensuring a significant degree of sparsity.
This is the case, for instance,
when $q$ is small. In terms of the primitives, such cases correspond to settings where $2\bar{\lambda}- \zeta \approx \zeta $ and hence the singular values of $M$ are close to each other.

It is intuitive that with exponential neighborhood growth, small approximation errors necessitate less sparse approximations. This is so because polynomial approximations are more accurate when higher orders of polynomials are employed. But with exponential neighborhood growth, such approximations naturally induce less sparse structures.
When the neighborhood growth is slower than exponential, it is possible to obtain sparser approximations (e.g., as in the case of polynomial neighborhood growth).

Using Lemma \ref{lem:appxDiagonalizable},
we next obtain an immediate corollary (stated without a proof) of
Theorem~\ref{thm:OrthoHinv}
that characterizes the rates of convergence for the thresholded estimator of
Algorithm~1.

\begin{corollary} \label{cor:diagDecay}
	Suppose that
	Assumption \ref{ass:diagonalizable}  holds.
	Consider the thresholded
	estimator
	$\check W^\mu$
	given  in
	Algorithm \ref{alg:Alg1}.
In the exponential neighborhood growth case,
under the assumptions of Theorem~\ref{thm:OrthoHinv}, with probability $1-o(1)$
	we have
	\begin{equation*}
	\begin{aligned}
	\max\{ \| \check W^\mu-H^{-1}\|_\infty  , \|
	\check W^\mu-H^{-1}
	\|_1 \}
	&\leq
	\tilde{C}_5
	\sqrt{\frac{\log |V_O|}{n} }
	\left(\frac{ \sqrt{n|V_O|}}{\log(|V_O|)} \right)^{\nu}
+	\tilde{C}_6
	{|V_O|}^{\nu/2}
	\left( \frac{\log(|V_O|)}{\sqrt{n} } \right)^{1-\nu},	
	\end{aligned}
	\end{equation*}
	for some constants $\tilde{C}_5,\tilde{C}_6 >0$.	
	Similarly, for the polynomial neighborhood growth case, we have
	\begin{equation*}
	\begin{aligned}
	\max\{ \| \check W^\mu-H^{-1}\|_\infty  , \|
	\check W^\mu-H^{-1}
	\|_1 \}
	&\leq
\tilde{C}_7	\sqrt[4]{\frac{\log |V_O|}{n} }
+\tilde{C}_8
	{\sqrt{|V_O|}}
	q^{ \left({ {\left(\frac{n}{\log(|V_O|)}\right)^{\frac{1}{4d_p}}} } \right)}
	\end{aligned}
	\end{equation*}
		for some constants $\tilde{C}_7, \tilde{C}_8>0$.
\end{corollary}

\section{Conclusions} \label{se:conclusions}

We study  the estimation and targeting problems in  social networks in the presence of latent agents.
We focus on a setting where the outcomes of agents depend linearly on the outcomes of their neighbors, and agent-specific covariates.
We assume that data on the outcomes/covariates of only a subset of all agents, referred to as observable agents, is available.
Our results indicate that by using the available data,
it is possible to estimate
a matrix that captures how the
outcome of  each observable agent depends on the covariates of the remaining agents.
Our estimator for this matrix  yields good estimation error guarantees
under an approximate sparsity assumption. This assumption holds for a rich class of networks, making our estimators applicable in interesting high-dimensional settings.
In addition,
as we illustrate through our targeted advertising and pricing applications,
 our estimation framework can be used to improve targeting decisions in social networks.
In particular, we show that  using the
available data to
estimate the aforementioned matrix it is possible to construct advertising/pricing decisions that are asymptotically optimal.

Our work is the first to focus on estimation and targeting problems
in social networks
 in the presence of latent agents.
It opens up a number of interesting research avenues, two of which we highlight here.
First,
consider our advertising/pricing applications, and
suppose that the seller does not have historical data on the consumption decisions of agents.
It would be interesting to study  how the seller can experiment with
targeting decisions   to learn the
impact of different targeting structures on the outcome, and
improve her long-run  payoffs. Such a study would facilitate improved decision making in social networks even in the absence of readily available data, and leverage tools from networks and online convex optimization.
Second,
it may be possible to
infer  the presence of latent agents
and how they are connected to observable ones
from the individual outcomes of
observable agents.
It
would be interesting to study
how and under what conditions
such inferences can be drawn.

\newpage
\bibliographystyle{ormsv080}
\bibliography{optimalPricing}

\newpage
    \begin{APPENDIX}{}
\noindent In Appendix \ref{app:notation}
    	we remind the reader our notation.
    	In Appendices
    	\ref{app:proofs} -- \ref{ref:SecEstEx},
    	we present the proofs of the results stated in Sections \ref{se:model} -- \ref{se:estimationExamples} respectively.
    	In Appendix
    	\ref{Sec:SimultaneousBootstrap}, we provide
    	simultaneous confidence intervals for our estimator, and also present an alternative bootstrap-based approach for choosing the threshold parameters.
Appendix \ref{app:NetworkInsight} sheds light on the structure of the equilibria and optimal prices for the application presented in Section \ref{subse:estimationPrices}.    	
    	Various auxiliary results and technical lemmas that are used in these appendices are presented in Appendix \ref{app:auxResults}.
    	
    	\section{Notation} \label{app:notation}
    	We start by
    	summarizing
 the notation that will be used throughout the appendix.
    	\begin{itemize}
    		\item $E[X]$: Expectation of random vector $X$.
    		\item $\mathbb{E}_n[X] = \frac{1}{n} \sum_{t=1}^n X^{(t)}$, for random vectors $\{X^{(t)}\}_{t\in[n]}$, where $[n]=\{1,\dots,n\}$.
    		\item $\|A \|_p$: Matrix $p$-norm for $p\in \{1,2,\infty\}$. Defined similarly for vectors.
    		\item $\|A\|_{e,\infty}$: Entry-wise maximum absolute entry of $A$, i.e., $\|A\|_{e,\infty}=\max_{i,j} |A_{ij}|$.
    		\item
    		$\| \beta \|_0$:  Number of nonzero entries of a given vector $\beta$.
    		\item $\lambda = \frac{1}{\sqrt{n}}  \Phi^{-1}\left(1-\frac{1}{3n|V_O|^2}\right)$, where $\Phi$
    		denotes the cumulative density function of the standard normal distribution.
    		
    		\item 	 $\bar V_O := \{0 \} \cup V_O$, and for $k,j\in \bar{V}_O$,
    		$\mathbf{1}_{kj}=1$ if $k=j$ and $\mathbf{1}_{kj}=0$ otherwise.
    		We index the entries of vectors in $\mathbb{R}^{|V_O|}$ and
    		$\mathbb{R}^{|\bar{V}_O|}$ by the elements of $V_O$ and $\bar{V}_O$ respectively.
    		For instance,
    		$(1;p_O^{(t)})_j$ is equal to
    		$p_j^{(t)}$ if $j\in V_O$, and is equal to $1$ otherwise, where $(1;p_O^{(t)}) \in \mathbb{R}^{|\bar{V}_O|}$.
    		
    		\item 	$M_n^2= 1\vee \max_{k\in V_O}\frac{1}{n}\sum_{t=1}^n(p_k^{(t)})^4$.
    		
    		\item $\hat\Sigma= \frac{1}{n}\sum_{t=1}^n(1;p_O^{(t)})(1;p_O^{(t)})^T$.

    		\item $\bar \Psi=E[(1;p_O^{(t)})(1;p_O^{(t)})^T]^{-1}$.
    		
    	\end{itemize}

    	\section{Proofs of Sections \ref{se:model}} \label{app:proofs}
    \proof{Proof of Lemma \ref{lem:firstOrderO}.}
 Recall that by \eqref{eq:firstOrder} we have ${y}^{(t)}=M^{-1}({a}+\xi^{(t)} - {p}^{(t)})$.
Using Lemma \ref{lem:MInv0} to express $M^{-1}$ in block format, and focusing on entries of ${y}^{(t)}$ associated with observable agents, the claim follows.
\hfill     \Halmos
    \endproof

\section{Proofs of Section \ref{se:estimation}}
\label{app:estimation}

\subsection{Proof of Theorem \ref{thm:OrthoHinv}}
In this subsection, we establish a more general version of Theorem \ref{thm:OrthoHinv}.
In particular, as opposed to imposing Assumption \ref{assumption:basicRandom}, we impose the following weaker assumption, and show that Algorithm~1 obtains the rates given in the statement of the theorem under this assumption.

\begin{assumption}\label{Assumption:New}
	Suppose that the network admits an $(s_n,r_{1n})$-sparse approximation.
		Let $c,c',C,C'>0$ be  constants such that $c<C$.

\begin{enumerate}[label=\roman*.]
	\item
	The observed data $\{ ({p}^{(t)}_O, {y}_O^{(t)}) : t\in [n]\}$ are i.i.d. random vectors
that satisfy \eqref{eq:def:noise1}.
Moreover, the data satisfy ${E}[({\varepsilon}^{(t)}_O) ({p}^{(t)}_O)^T ] = 0$ for every $t\in[n]$, and
	$$\min_{k\in V_O,j\in \bar V_O}E\bigg[  \{ 	(\varepsilon_{k}^{(t)}) (1;p_O^{(t)})_j \}^2\bigg] \geq c, \mbox{ and } \max_{k\in V_O,j\in \bar V_O} E\bigg[ 	{ \big|  (\varepsilon_{k}^{(t)}) (1;p_O^{(t)})_j }\big| ^4 	\bigg] \leq C.$$
Finally,
 there exists $M_\varepsilon\geq 0$ such that
 $E\big[ \max_{t\in [n]; k\in V_O} |
\varepsilon_{k}^{(t)} |^4\big] \leq M_\varepsilon$ and $n^{-1}M_\varepsilon \log |V_O| =o(1)$.	
	
\label{ass7.1}

\item 			The matrix $\bar \Psi:=E[(1;p_O^{(t)})(1;p_O^{(t)})^T]^{-1}$ is such that
$$\min_{k,j\in \bar V_O}E[ \{ \Pk(1;p_O^{(t)})(1;p_O^{(t)})_j - \mathbf{1}_{kj}\}^2] \geq c, \mbox{ and } \max_{k,j \in \bar V_O} E[ | \Pk(1;p_O^{(t)})(1;p_O^{(t)})_j - \mathbf{1}_{kj}|^4] \leq C.$$
  Moreover, there exists $M_\Psi\geq 0$ such that
$E[ \max_{t\in [n]; j,k\in \bar V_O} |  \Pk(1;p_O^{(t)})(1;p_O^{(t)})_j  - \mathbf{1}_{kj}|^4 ] \leq M_\Psi$  and $n^{-1}M_\Psi\log |V_O| =o(1)$.
{Finally, we assume that the eigenvalues of $\bar \Psi$ are upper bounded by a constant.}
	
	   \label{ass7.2}
	
	   \item
Given the $(s_n,r_{1n})$-sparse approximation of the influence structure,
define the restricted eigenvalue of $\hat{\Sigma}$ as
\begin{equation} \label{eq:kappaCond}
\kappa^2_{\bar c} = \min_{J \subset V_O, |J|\leq s_n}~ \min_{\|{\Delta}_{J^c}\|_1\leq \bar c\|\Delta_J\|_1} s_n\frac{\Delta^T\hat{\Sigma} \Delta}{\|\Delta\|_1^2},
\end{equation}
	    where $\Delta\in \mathbb{R}^{|\bar{V}_O|}$, and for any index set $J\subset |\bar{V}_O|$,
	    $\Delta_J$ corresponds to a vector whose entries consist of the entries of $\Delta$ whose indices belong to $J$ (similarly for $J^c:= \bar{V}_O \setminus J$).
	    For $\bar{c} = 3$,
	    with probability at least $1-o(1)$
	    we have that
	    $\lambda M_n s_n /\kappa^2_{\bar{c}} \leq  1/8$ and
	     $\kappa^2_{\bar c} \geq c'$,
	     where $M_n = \sqrt{ 1\vee \max_{k\in V_O}\frac{1}{n}\sum_{t=1}^n(p_k^{(t)})^4 }$  as given in Algorithm 1.
	   \label{ass7.3}
	
	     \item {$ C' \log V_O \geq  \log n$ and $n^{-1} \log |V_O| = o(1)$.}
	
	     	\label{ass7.4}
	
\end{enumerate}
\end{assumption}
{The main technical
difference of Assumption  \ref{assumption:basicRandom}
from Assumption \ref{Assumption:New}
 is the restricted eigenvalue condition
(see \citet{BickelRitovTsybakov2009}).
 We exploit this condition  in the proof of Theorem
\ref{thm:OrthoHinv}, when bounding estimation errors.
For the sake of exposition,
 we decided to defer
  this technical condition
 to the appendix, and opted to use the slightly stronger Assumption \ref{assumption:basicRandom} in Section \ref{se:estimation}.
We establish in Appendix~\ref{subse:Assumptions} that in fact
Assumption~\ref{Assumption:New} is implied by Assumption \ref{assumption:basicRandom}.  That is, the assumptions stated earlier readily imply the restricted eigenvalue condition. This observation also implies that Theorem \ref{thm:OrthoHinv} applies for a more general class of problem instances.

In order to achieve desirable rates for  estimators, it is often necessary to have
the eigenvalues of the design matrix
bounded away from zero.
In our case, such a condition can be written as
\begin{equation}\label{eq:rEval}
x^T \hat\Sigma x > c_1' \|x\|_2>0
\end{equation}
for some constant $c_1'$ and $x\in \mathbb{R}^{|V_O|}$ with $x\neq 0$.
On the other hand, due to high dimensionality this condition does not hold unless there is abundant data. This can be seen by noting that
for $n < |\bar{V}_O|$
the matrix
$\hat\Sigma= \frac{1}{n}\sum_{t=1}^n(1;p_O^{(t)})(1;p_O^{(t)})^T$
is rank-deficient.
Yet,
despite the rank deficiency,
\eqref{eq:rEval}  can still hold for a subset of values of $x \in R^{|V_O|}$.
In fact, under approximate sparsity conditions,
in order to obtain desirable rates in estimation,
it suffices for \eqref{eq:rEval} to hold
for $x\neq 0$ such that
$\|x_{J^c}\|_1  \leq \bar{c}
\|x_J\|_1$ for some set of indices $J \leq s_n$.
In settings where $s_n \ll n$, the restricted eigenvalue can be bounded away from zero for many common design matrices with high probability as required in Assumption
\ref{Assumption:New}\ref{ass7.3}, which can then be leveraged for characterizing rates of estimators.
}

Before we establish Theorem \ref{thm:OrthoHinv}, we provide three lemmas
that characterize some feasible/optimal solutions of the optimization problems given in Algorithm 1.
In proving these lemmas, as well as  Theorem \ref{thm:OrthoHinv},
for $k\in V_O$
 we use the shorthand notation:
 \begin{equation}\label{eq:deltaShortHand}
\hat \delta^k :=
(-v_k,\Hinvkdot)^T-(-\hat{v}_k,\hWkdot )^T=
(-v_k+\hat v_k, \Hinvkdot-\hWkdot)^T\in \mathbb{R}^{|\bar V_O|}.
 \end{equation}

\begin{lemma}\label{lem:Alg1:DeBias}
	Let $\bar z \in \mathbb{R}^{|\bar{V}_O|}$ be such that
	$\bar{z}_k :=
	\max_{j\in \bar V_O}\sqrt{ \frac{1}{n}\sum_{t=1}^n \{\bar \Psi_k(1;p_O^{(t)})(1;p_O^{(t)})_j- \mathbf{1}_{kj}\}^2}$
	 for $k\in \bar V_O$.	
Under Assumption \ref{Assumption:New}, with probability $1-o(1)$ we have that\\
(i)	$(\bar{z},\bar \Psi)$ is feasible in
	(\ref{def:lasso:ortho});\\
(ii) $(\hat{z},\hat \Psi)$ is such that
	$\max_{k\in \bar V_O} \hat z_k \leq C_1$ for some constant $C_1\geq 0$;\\
(iii) $\sqrt{ \Phk \hat\Sigma  \Phk^T} \leq 1+\hat z_k$ for every $k\in \bar V_O$;\\
(iv) $\sqrt{ \Phk \hat\Sigma  \Phk^T} \geq (1 - \lambda \hat z_k)/\sqrt{\hat\Sigma_{kk}}$ for every $k\in \bar V_O$;\\
(v) {$\frac{1}{n}\sum_{t=1}^n|\Phk (1;p_O^{(t)})|^4 \leq C_2$ for every $k\in \bar V_O$, and some constant $C_2 \geq 0$.}
\end{lemma}
\proof{Proof of Lemma \ref{lem:Alg1:DeBias}.}
To show (i) let $Z^{(t)}_{ij} := \Pii (1;p_O^{(t)})(1;p_O^{(t)})_j - \mathbf{1}_{ij}$ for $t\in[n]$, and $i,j\in \bar{V}_O$.
Denote by
$X^{(t)}$  a  vector of length $|\bar V_O| \times |\bar{V}_O|$ whose entries consist of
$\{Z^{(t)}_{ij}\}_{ i\in \bar{V}_O,j \in \bar{V}_O }$.

Observe that
$
\Pii E[ (1;p_O^{(t)})(1;p_O^{(t)})_j] =  \Pii  (\bar \Psi^{-1})_{\cdot,j} = \mathbf{1}_{ij} $.
Thus, it follows that $E[Z^{(t)}_{ij}]=0$.
Assumption \ref{Assumption:New}\ref{ass7.1} implies that $\{X^{(t)}\}$ are independent vectors.
By Assumption \ref{Assumption:New}\ref{ass7.2}, we have
$E[ (Z^{(t)}_{ij})^2]\geq c$ and $E[ |Z^{(t)}_{ij}|^4]\leq C$
for all $t\in[n]$, $i,j\in \bar{V}_O$.
Note that  by Jensen's inequality we have
\begin{equation}\label{eq:JensenZij}
E[ |Z^{(t)}_{ij}|^\ell]^{4/\ell} \leq
E[ |Z^{(t)}_{ij}|^4] \leq C
\end{equation}
for $\ell\in \{2,3\}$, which in particular implies that
$E[ |Z^{(t)}_{ij}|^3] \leq C_1'$ for some constant $C_1'>0$.
Moreover,
 Assumption \ref{Assumption:New}\ref{ass7.2} also
implies that
	${M}_\Psi \geq E[\max_{t\in[n] } \|X^{(t)}\|_\infty^4 ]$,
where
${M}_\Psi$
is such that
${M}_\Psi {\log |V_O|}/{n} = o(1)$.
Finally,
by Assumption \ref{Assumption:New}\ref{ass7.4}, we have
$ C' \log |V_O| \geq  \log n$ and ${\log |V_O|}/{n}=o(1)$.
Thus, the
conditions of   Lemma \ref{lem:ozanVersion} (with $L=4$)
hold for the constructed $\{Z^{(t)}_{ij} \}$, and
with probability $1-o(1)$ we have
\begin{equation} \label{eq:MDSN_bound3_o}
\begin{aligned}
\lambda  &\geq
\max_{i,j\in \bar{V}_O}
\frac{    \left|\frac{1}{n} \sum_{t=1}^n  Z^{(t)}_{ij}  \right| }{   \sqrt{\frac{1}{n}
		\sum_{t=1}^n \{Z^{(t)}_{ij} \}^2}} \geq
\max_{i\in \bar{V}_O}
\frac{  \max_{j\in \bar{V}_O} \left|\frac{1}{n} \sum_{t=1}^n  Z^{(t)}_{ij}  \right| }{  \max_{j\in \bar{V}_O} \sqrt{\frac{1}{n}
		\sum_{t=1}^n \{Z^{(t)}_{ij} \}^2}}\\
&=
 \max_{i\in \bar{V}_O} \frac{ \max_{j\in\bar{V}_O} \left| \frac{1}{n} \sum_{t=1}^n \{
 	\Pii (1;p_O^{(t)})(1;p_O^{(t)})_j - \mathbf{1}_{ij}\}\right|}{  \max_{j\in\bar{V}_O} \sqrt{\frac{1}{n} \sum_{t=1}^n \{ \Pii (1;p_O^{(t)})(1;p_O^{(t)})_j- \mathbf{1}_{ij}\}^2}} \\
&=
 \max_{i\in{\bar V}_O}\frac{ \max_{j\in\bar{V}_O}  | \Pii \hat{\Sigma}_{\cdot,j} - \mathbf{1}_{ij} | }{\bar{z}_i }
.
\end{aligned}
\end{equation}
Hence, we conclude that
$(\bar{z},\bar \Psi)$ satisfies the
first constraint of \eqref{def:lasso:ortho} with probability $1-o(1)$.
Moreover,
by construction of $\bar{z}$ it follows that
$(\bar{z},\bar \Psi)$ always satisfies the second constraint of \eqref{def:lasso:ortho}.
Thus,
$(\bar{z},\bar \Psi)$ is feasible in \eqref{def:lasso:ortho} with probability at least $1-o(1)$.

To show (ii) suppose that $(\bar{z},\bar \Psi)$ is feasible, which occurs with probability $1-o(1)$ by (i), and consider the optimal solution
$(\hat{z},\hat \Psi)$  of (\ref{def:lasso:ortho}).
Since (\ref{def:lasso:ortho}) decouples over $k\in \bar{V}_O$,
the optimality of the latter implies that
\begin{equation}\label{eq:optVsFeas_o}
\left(\frac{1}{n}\sum_{t=1}^n|\Phk(1;p_O^{(t)})|^4\right)^{1/4} +  \hat z_k \leq  \left(\frac{1}{n}\sum_{t=1}^n|\bar\Psi_{k\cdot}(1;p_O^{(t)})|^4\right)^{1/4}  +  \bar{z}_k.
\end{equation}

Observe that Lemma \ref{lem:ozanVersion} implies that with probability $1-o(1)$,
\begin{equation} \label{eq:PsiBarBound}
\begin{aligned}
\max_{i,j\in \bar{V}_O}
\frac{1}{n}\sum_{t=1}^n|
\Pii (1;p_O^t)(1;p_O^t)_j - \mathbf{1}_{ij}|^4
=
\max_{i,j\in \bar{V}_O}
\frac{1}{n}\sum_{t=1}^n|Z^{(t)}_{ij}|^4 \leq C'_2
\end{aligned}
\end{equation}
for some constant $C'_2 >0$.
This observation, together with the Cauchy--Schwarz inequality, implies that
\begin{equation}
\begin{aligned}
\max_{j,k \in \bar{V}_O} \frac{1}{n} \sum_{t=1}^n | \bar \Psi_k(1;p_O^{(t)})(1;p_O^{(t)})_j- \mathbf{1}_{kj}|^2
=\max_{j,k\in \bar{V}_O,}
\frac{1}{n}\sum_{t=1}^n|Z^{(t)}_{kj}|^2
\leq C'_3
\end{aligned}
\end{equation}
for some constant $C_3'>0$ with probability $1-o(1)$.
By the definition of $\bar{z}_k$, this implies that with probability $1-o(1)$, we have
$\max_{k\in \bar{V}_O}\bar{z}_k \leq \sqrt{C_3'}$.

Note that for any scalars $a_1, a_2$, we have
$(a_1+a_2)^4 \leq \max\{(2a_1)^4,(2a_2)^4 \}\leq 16 (a_1^4 +a_2^4)$.
Setting
$a_1=
\Pii (1;p_O^t)(1;p_O^t)_j - \mathbf{1}_{ij}$, and $a_2=\mathbf{1}_{ij}\leq 1$, this implies that
\begin{equation}
\begin{aligned}
\max_{i,j\in \bar{V}_O}
\frac{1}{n}\sum_{t=1}^n|
\Pii (1;p_O^{(t)})(1;p_O^{(t)})_j - \mathbf{1}_{ij}|^4 &\geq
C_4'
\max_{i,j\in \bar{V}_O}
\frac{1}{n}\sum_{t=1}^n|
\Pii (1;p_O^{(t)})(1;p_O^{(t)})_j|^4  - 1 \\
&\geq
C_4'
\max_{i\in \bar{V}_O}
\frac{1}{n}\sum_{t=1}^n|
\Pii (1;p_O^{(t)})|^4  - 1,
\end{aligned}
\end{equation}
for some constant $C_4'>0$.
Combining this observation with \eqref{eq:PsiBarBound} yields
\begin{equation}
\begin{aligned}
\max_{k\in \bar{V}_O,}
\frac{1}{n}\sum_{t=1}^n|\bar\Psi_{k,\cdot}(1;p_O^{(t)})|^{4}
\leq C'_5
\end{aligned}
\end{equation}
for some constant $C_5'>0$ with probability $1-o(1)$.

Since $(\bar{z},\bar \Psi)$ is feasible
in 	(\ref{def:lasso:ortho})
with probability $1-o(1)$, and both
the quantity $\max_{k\in \bar V_O} \bar z_k$ and $\max_{k\in V_O}\frac{1}{n}\sum_{t=1}^n|\bar\Psi_{k,\cdot}(1;p_O^{(t)})|^4$  are bounded by a constant,
\eqref{eq:optVsFeas_o} implies (ii) and (v).

To show (iii) note that
$$\begin{array}{rl}
\sqrt{ \Phk \hat\Sigma  \Phk^T} & = \sqrt{\frac{1}{n}\sum_{t=1}^n \Phk  (1;p_O^{(t)}) (1;p_O^{(t)})^T \Phk^T } \\
& = \sqrt{\frac{1}{n}\sum_{t=1}^n \{\Phk (1;p_O^{(t)})\}^2}\\
& \leq \max_{j\in \bar V_O} \sqrt{\frac{1}{n}\sum_{t=1}^n \{\Phk (1;p_O^{(t)})(1;p_O^{(t)})_j\}^2}\\
& \leq \max_{j\in \bar V_O} \sqrt{\frac{1}{n}\sum_{t=1}^n \{\Phk (1;p_O^{(t)})(1;p_O^{(t)})_j-\mathbf{1}_{kj}\}^2}+1\\
& \leq 1 + \hat z_k, \\
\end{array}$$
where the first inequality follows since $\max_{j\in \bar V_O} (1;p_O^{(t)})_j \geq 1$, and the second one uses the triangle inequality and $|\mathbf{1}_{kj}|\leq 1$.
The last inequality follows since $(\hat{z},\hat \Psi)$ is feasible in \eqref{def:lasso:ortho}.

Relation (iv) follows from $(\hat{z},\hat \Psi)$ being feasible in \eqref{def:lasso:ortho}. Indeed by the Cauchy--Schwarz inequality we have
\begin{equation}\label{eq:part4Lemma}
\begin{aligned}
 \Phk \hat\Sigma_{\cdot, k} &= \frac{1}{n}\sum_{t=1}^n \Phk (1;p_O^{(t)})(1;p_O^{(t)})^T_k\leq \sqrt{\frac{1}{n}\sum_{t=1}^n|\Phk(1;p_O^{(t)})|^2}\sqrt{\frac{1}{n}\sum_{t=1}^n |(1;p^{(t)}_O)_k|^2}\\
 &= \sqrt{ \Phk \hat\Sigma  \Phk^T} \sqrt{\hat\Sigma_{kk}}.
 \end{aligned}
\end{equation}
On the other hand, by the feasibility of
$(\hat{z},\hat \Psi)$
we have
$
1- \Phk \hat\Sigma_{\cdot k}\leq
| \Phk \hat\Sigma_{\cdot k}- 1| \leq \lambda\hat z_k$.
After rearranging terms this inequality together with
\eqref{eq:part4Lemma} implies   (iv).
\hfill \halmos
\endproof

\begin{lemma}\label{lem:Alg1:Step1}
	Let $\bar z \in \mathbb{R}^{|{V}_O|}$ be such that
		$\bar{z}_k := \max_{j\in \bar V_O}
		\sqrt{ \frac{1}{n}\sum_{t=1}^{n} \{(y_k^{(t)} -  v_k+ H^{-1}_{k,\cdot}p_O^{(t)})\}^2(1;p_O^{(t)})_j^2}$
	for $k\in V_O$.	
Under Assumption \ref{Assumption:New}, with probability $1-o(1)$ we have that
	$(H^{-1}, v_O, \bar{z})$ is feasible in~(\ref{def:lasso}),
	$\max_{k\in  V_O} \bar z_k \leq C_1$,
		$\max_{k\in  V_O} \hat z_k \leq C_1$ for some constant $C_1 \geq 0$.
\end{lemma}
\proof{Proof of Lemma \ref{lem:Alg1:Step1}.}
Let
$Z^{(t)}_{ij} :={(y_i^{(t)} -  v_i+ H^{-1}_{i,\cdot}p_O^{(t)})(1;p_O^{(t)})_j}$ for $t\in[n]$,
$i\in V_O,j\in \bar{V}_O$.
Denote by
$X^{(t)}$  a  vector of length $| V_O| \times |\bar{V}_O|$ whose entries consist of
$\{Z^{(t)}_{ij}\}_{ i\in {V}_O,j \in \bar{V}_O }$.

By Lemma \ref{lem:firstOrderO},
we have
$(y_i^{(t)} -  v_i+ H^{-1}_{i,\cdot}p_O^{(t)}) = \varepsilon_i^{(t)}$ for all $i\in V_O$. Hence,
$Z^{(t)}_{ij}$
can be equivalently expressed as follows:
$Z^{(t)}_{ij} ={\varepsilon_i^{(t)}(1;p_O^{(t)})_j}$.
By Assumption \ref{Assumption:New}\ref{ass7.1} it follows that  $E[Z^{(t)}_{ij}]=0$, and
 $\{X^{(t)}\}$ are independent vectors.
This assumption also implies that
  $E[ (Z^{(t)}_{ij})^2]\geq c$, and
  $E[ |Z^{(t)}_{ij}|^4]\leq C$, which in turn yields (using Jensen's inequality)
  that
  $E[ |Z^{(t)}_{ij}|^\ell]\leq C^{\ell/4}$
 for
 $\ell\in \{2,3\}$ and
 all $t\in[n]$, $i\in V_O,j\in \bar{V}_O$.
 Note that using
Assumption  \ref{Assumption:New}\ref{ass7.1}
(and the fact that covariates are bounded by $\bar p$)
we also obtain
\[
E[\max_{t\in[n] } \|X^{(t)}\|_\infty^2 ] \leq
\bar p^2 E[\max_{t\in[n] } \|\varepsilon_O^{(t)}\|_\infty^2 ]
\leq
\bar p^2 (E[\max_{t\in[n] } \|\varepsilon_O^{(t)}\|_\infty^4 ])^{1/2}
\leq \sqrt{ {M}_\varepsilon } \bar p^2,
\]
 where
 ${M}_\varepsilon$
 is such that
 $ {M}_\varepsilon \frac{\log |V_O|}{n} = o(1)$. This in turn implies that
 $\sqrt{{M}_\varepsilon} \frac{\log |V_O|}{n} = o(1)$, since $\frac{\log |V_O|}{n}=o(1)$.
 Finally,
 by Assumption \ref{Assumption:New}\ref{ass7.4}, we have
 $ C' \log |V_O| \geq  \log n$ and $\frac{\log |V_O|}{n}=o(1)$.
 Thus, the
 conditions of   Lemma \ref{lem:ozanVersion}
 hold
 (with $L=2$)
  for the constructed $\{Z^{(t)}_{ij} \}$, and
 with probability $1-o(1)$ we have
 \begin{equation} \label{eq:MDSN_bound_step1_fin_o}
 \begin{aligned}
 \lambda  &\geq
 \max_{i\in V_O,j\in \bar{V}_O}
 \frac{    \left|\frac{1}{n} \sum_{t=1}^n  Z^{(t)}_{ij}  \right| }{   \sqrt{\frac{1}{n}
 		\sum_{t=1}^n \{Z^{(t)}_{ij} \}^2}}
 \geq
 \max_{i\in {V}_O}
 \frac{  \max_{j\in \bar{V}_O} \left|\frac{1}{n} \sum_{t=1}^n  Z^{(t)}_{ij}  \right| }{  \max_{j\in \bar{V}_O} \sqrt{\frac{1}{n}
 		\sum_{t=1}^n \{Z^{(t)}_{ij} \}^2}}\\
& =
 \max_{i\in {V}_O} \frac{ \max_{j\in\bar{V}_O} \left| \frac{1}{n}
 	\sum_{t=1}^n
 	(y_i^{(t)} -  v_i + H^{-1}_{i,\cdot}p_O^{(t)})(1;p_O^{(t)})_j
 	\right|}{  \max_{j\in\bar{V}_O} \sqrt{\frac{1}{n} \sum_{t=1}^n
 		\{{ (y_i^{(t)} -  v_i + H^{-1}_{i,\cdot}p_O^{(t)})(1;p_O^{(t)})_j} \}^2
 	}} \\
 &=
 \max_{i\in{ V}_O}
\frac{ \max_{j\in\bar{V}_O} \left| \frac{1}{n}
	\sum_{t=1}^n
	(y_i^{(t)} -  v_i + H^{-1}_{i,\cdot}p_O^{(t)})(1;p_O^{(t)})_j
	\right|}
{\bar{z}_i }.
 \end{aligned}
 \end{equation}
Observe that by construction
$(H^{-1}, v_O, \bar{z})$ satisfies the second constraint of
(\ref{def:lasso}).  In addition, \eqref{eq:MDSN_bound_step1_fin_o} implies that the first condition of
(\ref{def:lasso}) is also satisfied with probability $1-o(1)$.
Thus, $(H^{-1}, v_O, \bar{z})$ is feasible in (\ref{def:lasso}) with probability
$1-o(1)$, as claimed.

		We conclude the proof by showing that $\max_{k\in V_O} \hat{z}_k$
and  $\max_{k\in V_O} \bar{z}_k$
are bounded by a constant (with probability $1-o(1)$).
Suppose that $(H^{-1}, v_O, \bar{z})$ is feasible, and consider the optimal solution
$(\hat W, \hat v, \hat {z})$
of \eqref{def:lasso}.
Since \eqref{def:lasso} decouples over $k \in V_O$,
the optimality of the latter implies that
\begin{equation}\label{eq:optVsFeas_o2}
\| (\hat v_k, \hWkdot )\|_{1} + \tau \hat z_k \leq
\| ( v_k,  H^{-1}_{k,\cdot})\|_{1} +   \tau \bar{z}_k,
\end{equation}		
where
$\tau=1/(4M_n)$.
Note that $\max_{k\in V_O} | v_k|=
\|{v}_O\|_\infty $ and
$\max_{k\in V_O} \| H^{-1}_{k,\cdot}\|_1 = \| H^{-1}\|_\infty$
are bounded by constants  {(see Lemmas
	\ref{lem:MInv0},
	 \ref{lem:boundVo} and recall that $M^{-1}$ has bounded absolute row/column sums)}.
Moreover, by definition
$M_n^2= 1\vee \max_{k\in V_O}\frac{1}{n}\sum_{t=1}^n(p_k^{(t)})^4 \leq \bar{p}^4$, and hence
$\tau> c_1'$  for some constant $c_1'>0$.
Thus,
\eqref{eq:optVsFeas_o2} implies that
 to complete the proof
it suffices to show that
$\max_{k\in V_O} \bar{z}_k$ is bounded by a constant.

{Observe that Lemma \ref{lem:ozanVersion} also implies that
\begin{equation}
\begin{aligned}
\max_{j\in \bar V_O ,k\in {V}_O,}
\frac{1}{n}\sum_{t=1}^n|Z^{(t)}_{kj}|^2
=
\max_{j\in \bar V_O ,k\in {V}_O,}
 \frac{1}{n} \sum_{t=1}^n |
(y_k^{(t)} -  v_k + H^{-1}_{k,\cdot}p_O^{(t)})(1;p_O^{(t)})_j|^2
\leq C'_1
\end{aligned}
\end{equation}
for some constant $C_1'>0$ with probability $1-o(1)$.
 	By the definition of $\bar{z}_k$, this implies that with probability $1-o(1)$, we have
$\max_{k\in \bar{V}_O}\bar{z}_k \leq \sqrt{C_1'}$.
Hence, the claim follows.}
\hfill\halmos	
\endproof

\begin{lemma}\label{thm:DS:fast}
	Under Assumption \ref{Assumption:New}
with probability
	$1- o(1)$ for all $k\in V_O$ we have
	$$ \|(\hat v_k, \hWk) - (v_k,
	\Hinvkdot)\|_1 \leq  C_1 r_{1n} + C_1 \lambda s_n/\kappa^2_{\bar c}, $$
	where $\bar c = 3$ and $C_1>0$
	is some constant.
\end{lemma}
\proof{Proof of Lemma \ref{thm:DS:fast}.}
For notational convenience let $W_k = H^{-1}_{k,\cdot}$,
$\hat{W}_k = \hWkdot$, and
$$z_k^2(v',\beta')=\max_{j\in \bar V_O} \frac{1}{n}\sum_{t=1}^{n} (y_k^{(t)} -  v'+ \beta'p_O^{(t)})^2(1;p_O^{(t)})_j^2$$
for $u'\in \mathbb{R}$, $\beta'\in \mathbb{R}^{1\times |V_O|}$.
By Lemma \ref{lem:Alg1:Step1}, under Assumption \ref{Assumption:New} we have that $\{W_k,v_k,z_k(v_k,W_k)\}_k$ is feasible
in (\ref{def:lasso})  with probability at least $1-o(1)$.

Consider the optimal solution  $\{\hat W_k, \hat v_k, \hat z_k\}_k$ of \eqref{def:lasso}.
The definition of $z_k(\cdot,\cdot)$ and the feasibility of this solution imply that
$ z_k(\hat v_k,\hat W_k)\leq \hat z_k$ for all $k$.
Observe that \eqref{def:lasso} decouples over $k$.
Using this observation, the inequality $ z_k(\hat v_k,\hat W_k)\leq \hat z_k$, and
the optimality of $\{\hat W_k, \hat v_k, \hat z_k\}_k$, we obtain that with probability at least $1-o(1)$, for all $k\in V_O$,
the following inequality holds:
\begin{equation}\label{eq:1NormTBound}
\|(\hat v_k,\hat W_k) \|_1 + \tau z_k(\hat v_k,\hat W_k) \leq  \|(\hat v_k,\hat W_k) \|_1 + \tau \hat z_k \leq \|(v_k,W_k)\|_1 + \tau z_k(v_k,W_k),
\end{equation}
where $\tau = 1/(4M_n)$,  $M_n^2= 1\vee \max_{k\in V_O}\frac{1}{n}\sum_{t=1}^n(p_k^{(t)})^4$, as in Algorithm 1.
Rearranging terms, this yields
\begin{equation}\label{eq:preTResult}
\|(\hat v_k,\hat W_k) \|_1    \leq \|(v_k,W_k)\|_1 + \tau \left( z_k(v_k,W_k) -  z_k(\hat v_k,\hat W_k) \right).
\end{equation}

By the definition of $z_k(\cdot,\cdot)$ it follows that
\begin{equation} \label{eq:tkBound}
z_k(v'',\beta'') - z_k(v',\beta') \leq
\max_{j\in \bar{V}_O}
\left|
h_{k,j}(v'',\beta'')
-h_{k,j}(v',\beta')\right|,
\end{equation}
where $\beta',\beta'' \in \mathbb{R}^{1\times |V_O|}$, $v',v''\in \mathbb{R}$, and
$h_{k,j}(v',\beta'):=\frac{1}{\sqrt{n}}
\sqrt{\sum_{t=1}^n (1;p_O^{(t)})_j^2(y_k^{(t)}-v'+\beta'p_O^{(t)})^2}$.
It follows from Lemma \ref{lem:hBound} that
\begin{equation}\label{eq:htkBound}
\left|
h_{k,j}(v'',\beta'')
-h_{k,j}(v',\beta')\right| \leq
M_{k,j}
 \| (v'',\beta'')-(v',\beta') \|_1,
\end{equation}
where $M_{k,j}=
\max_{i\in \bar V_O} {\sqrt{\frac{1}{n} \left(\sum_{t=1}^n  (1;p_O^{(t)})_j^2 (1;p_O^{(t)})_i^2  \right) } }$.
Equations \eqref{eq:tkBound} and \eqref{eq:htkBound} jointly imply that
\begin{equation}\label{eq:tkBeforeFinal}
z_k(v'',\beta'') - z_k(v',\beta') \leq
\max_{j\in \bar{V}_O} M_{k,j}
\| (v'',\beta'')-(v',\beta') \|_1.
\end{equation}
On the other hand,
\begin{equation}
\begin{aligned}
\max_{j\in \bar{V}_O} M_{k,j} &\leq
\max_{i,j\in\bar{V}_O}
{\sqrt{\frac{1}{n} \left(\sum_{t=1}^n  (1;p_O^{(t)})_j^2 (1;p_O^{(t)})_i^2  \right) } }
\end{aligned}
\end{equation}
For $i\in \bar{V}_O$ let $X_i$ denote a vector of length $n$, whose $t$th entry is given by
$(1;p_O^{(t)})_i^2$. Using this to rewrite   the above inequality  we obtain
\begin{equation}
\begin{aligned}
\max_{j\in \bar{V}_O} M_{k,j} &\leq
\max_{i,j\in\bar{V}_O}
{\sqrt{\frac{1}{n}
		\langle X_i, X_j \rangle } }
	\leq
	\max_{i,j\in\bar{V}_O}
	{\sqrt{\frac{1}{n}
			\| X_i\|_2 \cdot \|X_j\|_2 }}
		=
			\max_{j\in\bar{V}_O}
		{\sqrt{\frac{1}{n}
				\| X_j\|_2^2 }}
	\\
&=
\max_{j\in \bar{V}_O}
{\sqrt{\frac{1}{n}
		\langle X_j, X_j \rangle } }
=
\max_{j\in V_O} {\sqrt{\frac{1}{n} \sum_{t=1}^n \left((p_j^{(t)})^4 \right) } }  \vee 1
=M_n,
\end{aligned}
\end{equation}
where
$\langle \cdot,\cdot \rangle$  denotes the inner product between the relevant vectors
and we use the fact that the inner product between the two vectors is smaller than the product of their norms.
Combining this with \eqref{eq:tkBeforeFinal}, we obtain
\begin{equation}
\label{eq:boundTkFinForm}
z_k(v'',\beta'') - z_k(v',\beta') \leq   M_n \| (v'',\beta'')-(v',\beta') \|_1.
\end{equation}
Using this inequality with \eqref{eq:preTResult} (and recalling that $\tau=1/(4M_n)$), we conclude that
\begin{equation}\label{eq:preTResult2}
\begin{aligned}
\|(\hat v_k,\hat W_k) \|_1    \leq \|(v_k,W_k)\|_1 +
\frac{1}{4}
 \| (v_k,W_k)-(\hat v_k,\hat W_k) \|_1.
\end{aligned}
\end{equation}

Let $\bar{W}$ denote an $(s_n,r_{1n})$-sparse approximation of $H^{-1}$.
Denote by
$\bar{W}_k$ the row of $\bar W$ corresponding to $k\in V_O$, and let
 $T_k$ be the set of indices of nonzero entries of $\bar{W}_{k\cdot}$. Note that by the definition of sparse approximation, we have $|T_k|\leq s_n$.
Moreover, there exists a sparse approximation where
$\bar{W}_{k T_k} = W_{kT_k}$
(henceforth, with some abuse of notation, for any matrix $A$ and a subset of its columns $S$, we denote by $A_{kS}$ the row vector, whose entries consist of
$\{A_{kj}\}_{j\in S})$. Let $\bar{W}$ be such an approximation,
and note that
$\| {W}_{kT_k^c} \|_1 \leq r_{1n} $.

Note that
$\|(\hat v_k,\hat W_k) \|_1  = \|(\hat v_k,\hat W_{kT_k}) \|_1 + \|\hat W_{kT_k^c} \|_1$,
and
$\|( v_k, W_k) \|_1  = \|( v_k, W_{kT_k}) \|_1 + \| W_{kT_k^c} \|_1$.
Moreover, by the triangle inequality, we have
$\|( v_k, W_{kT_k}) \|_1 - \|(\hat v_k,\hat W_{kT_k}) \|_1 \leq
\|(\hat v_k,\hat W_{kT_k})- ( v_k, W_{kT_k}) \|_1$.
Combining these observations with
\eqref{eq:preTResult2} we obtain
\begin{equation}\label{eq:preTResult3}
\begin{aligned}
 \|\hat W_{kT_k^c} \|_1
   \leq
\|(\hat v_k,\hat W_{kT_k})- ( v_k, W_{kT_k}) \|_1
 + \| W_{kT_k^c} \|_1
   + \frac{1}{4} \| (v_k,W_k)-(\hat v_k,\hat W_k) \|_1.
\end{aligned}
\end{equation}

To complete the proof, we consider two cases: (a) $\|(\hat v_k,\hat W_{kT_k})-(v_k,W_{kT_k}) \|_1 \leq 2r_{1n}$ and (b)
$\|(\hat v_k,\hat W_{kT_k})-(v_k,W_{kT_k}) \|_1 > 2r_{1n}$.
In the first case, \eqref{eq:preTResult3} implies that
\begin{equation}
\begin{aligned}
\|\hat W_{kT_k^c} \|_1
&\leq
3 r_{1n}
+ \frac{1}{4}
\| (v_k,W_k)-(\hat v_k,\hat W_k) \|_1
=
3 r_{1n}
+ \frac{1}{4}
\left(
\|(\hat v_k,\hat W_{kT_k})-(v_k,W_{kT_k}) \|_1
+
\|\hat W_{kT_k^c}-W_{kT_k^c} \|_1
\right)
\\
&\leq
3 r_{1n}
+ \frac{1}{4}
\left(2r_{1n}
+
\|\hat W_{kT_k^c} \|_1 +\| W_{kT_k^c} \|_1
\right) \leq
3 r_{1n}
+ \frac{1}{4}
\left(3r_{1n}
+
\|\hat W_{kT_k^c} \|_1
\right),
\end{aligned}
\end{equation}
where the first and third inequalities use $\| W_{kT_k^c} \|_1 \leq r_{1n}$, and
 the second one uses the
 triangle inequality.
Rearranging terms, we conclude
$\|\hat W_{kT^c_k} \|_1 \leq 5 r_{1n}$.
Together with our assumption in case (a)
(and using the notation in
\eqref{eq:deltaShortHand}), these observations yield
\begin{equation}\label{eq:lem_part_i}
\|\hat{\delta}^k\|_1=
\|(-\hat v_k,\hat W_{k})-(-v_k,W_{k}) \|_1 \leq \|(\hat v_k,\hat W_{kT_k})-(v_k,W_{kT_k}) \|_1 + \|\hat W_{kT^c_k} \|_1 + \|W_{kT_k^c}\|_1 \leq 8r_{1n},
\end{equation}
where we make use of the fact that negating an entry of a vector does not change its ($\ell_1$) norm.

Next consider case (b).
Note that in this case we have
$ 2\|W_{kT^c_k}\|_1 \leq 2r_{1n} <\|(\hat v_k,\hat W_{kT_k})-(v_k,W_{kT_k}) \|_1$.
Using this observation together with \eqref{eq:preTResult3} and the triangle inequality we have
\begin{equation}\label{eq:preTResult4}
\begin{aligned}
\|\hat W_{kT^c_k} - W_{kT^c_k} \|_1
&\leq
\|\hat W_{kT_k^c} \|_1
+
\|W_{kT^c_k}\|_1\\
&\leq
\|(\hat v_k,\hat W_{kT_k})- ( v_k, W_{kT_k}) \|_1
+ 2\| W_{kT_k^c} \|_1
+ \frac{1}{4} \| (v_k,W_k)-(\hat v_k,\hat W_k) \|_1\\
&\leq
(2+1/4)\|(\hat v_k,\hat W_{kT_k})- ( v_k, W_{kT_k}) \|_1
+ \frac{1}{4} \| W_{kT_k^c}-\hat{W}_{kT_k^c} \|_1.
\end{aligned}
\end{equation}
Thus, rearranging terms, we conclude that
\begin{equation}\label{eq:preTResult5}
\begin{aligned}
\|\hat W_{kT^c_k} - W_{kT^c_k} \|_1
&\leq
3\|(\hat v_k,\hat W_{kT_k})- ( v_k, W_{kT_k}) \|_1.
\end{aligned}
\end{equation}
The definition of the restricted eigenvalues (see \eqref{eq:kappaCond}) and  \eqref{eq:preTResult5} imply that
\begin{equation}\label{eq:K3}
\kappa^2_{3} \leq s_n \frac{(\hat\delta^k)^T \hat{\Sigma} (\hat\delta^k) }{\|\hat\delta^k \|_1^2}.
\end{equation}
We next use this inequality to obtain a bound on $\|\hat{\delta}_k\|_1$.

To this end, first observe
that
\begin{equation}
\begin{aligned}
\hat \Sigma \hat \delta^k
&= \hat \Sigma (-v_k, W_k)^T - \hat \Sigma (-\hat v_k, \hat W_k)^T \\
&=
\frac{1}{n}\sum_{t=1}^n ( -v_k +  W_kp_O^{(t)})(1;p_O^{(t)})
-
\frac{1}{n}\sum_{t=1}^n (-\hat v_k + \hat W_kp_O^{(t)})(1;p_O^{(t)}).
\end{aligned}
\end{equation}
Using this inequality, we conclude that
with probability
$1-o(1)$
the following holds:
\begin{equation}\label{eq:useRE}
\begin{array}{rl}
(\hat\delta^k)^T\hat \Sigma \hat\delta^k & \leq  \|\hat\delta^k\|_1 \|\hat\Sigma \hat\delta^k\|_\infty \\
&\leq   \|\hat\delta^k\|_1  \| \frac{1}{n}\sum_{t=1}^n (y_k^{(t)}-\hat v_k+\hat W_kp_O^{(t)})(1;p_O^{(t)})\|_\infty + \|\hat\delta^k\|_1\| \frac{1}{n}\sum_{t=1}^n \varepsilon_k^{(t)}(1;p_O^{(t)})\|_\infty\\
& \leq \|\hat\delta^k\|_1 \lambda \hat z_k + \|\hat\delta^k\|_1 \lambda z_k(v_k,W_k).\\
\end{array}
\end{equation}
Here the first inequality  follows from Holder's inequality, and the second one follows from the triangle inequality.
Recall that  $\{\hat v_k,\hat W_k,\hat z_k\}_k$ is optimal and $\{v_k,W_k,z_k(v_k,W_k)\}_k$ is feasible in
 (\ref{def:lasso})  with probability at least $1-o(1)$. These observations imply  that
 the third inequality holds with probability $1-o(1)$.

Using $\tau=1/(4M_n)$ to
rewrite the second inequality of
\eqref{eq:1NormTBound} we conclude that with probability $1-o(1)$ the following inequality holds:
\begin{equation}
\begin{aligned}
\hat{z}_k &\leq z_k(v_k,W_k) + 4M_n \| (v_k,W_k)\|_1 -4M_n  \| (\hat v_k,\hat W_k)\|_1 \\
&\leq
z_k(v_k,W_k) + 4M_n  \| (v_k,W_k)- (\hat v_k,\hat W_k)\|_1
=
z_k(v_k,W_k) + 4M_n \|\hat{\delta}^k\|_1.
\end{aligned}
\end{equation}
 Using this inequality to bound $\hat{z}_k$ in \eqref{eq:useRE}, we obtain
 \begin{equation}\label{eq:boundDeltaK}
 \begin{aligned}
 (\hat\delta^k)^T\hat \Sigma \hat\delta^k & \leq 2 \|\hat\delta^k\|_1 \lambda z_k(v_k,W_k) +
  4 \lambda M_n  \|\hat\delta^k\|_1^2.
 \end{aligned}
 \end{equation}

Lemma \ref{lem:Alg1:Step1} implies that  $z_k(v_k,W_k)$ is bounded with probability $1-o(1)$.
This observation together with
\eqref{eq:K3} and
\eqref{eq:boundDeltaK}
 yields
 \begin{equation}
 \begin{aligned}
\frac{\kappa^2_{3}\|\hat\delta^k \|_1^2}{s_n} & \leq  \lambda C'_1 \|\hat\delta^k\|_1 +
4 \lambda  M_n  \|\hat\delta^k\|_1^2,
 \end{aligned}
 \end{equation}
 for some constant $C'_1$.
Rearranging the terms this inequality can be written as follows:
 \begin{equation}
 \begin{aligned}
\left( \frac{\kappa^2_{3}}{s_n} - 4 \lambda  M_n \right)\|\hat\delta^k \|_1
 & \leq   \lambda C'_1
 .
 \end{aligned}
 \end{equation}
By Assumption \ref{Assumption:New},
 $ \frac{4 \lambda s_n  M_n }{\kappa^2_{3}} \leq 1/2$; hence the previous inequality  implies that
 \begin{equation}
 \begin{aligned}
 \frac{\kappa^2_{3}}{2s_n} \|\hat\delta^k \|_1
 & \leq   \lambda C'_1 .
 \end{aligned}
 \end{equation}
 Thus, for some constant $C'_2$ we obtain
\begin{equation}\label{eq:lem_part_ii}
\|\hat \delta^k\|_1 \leq C_2' \frac{\lambda s_n}{\kappa^2_{3}}.
\end{equation}
Combining \eqref{eq:lem_part_i} and \eqref{eq:lem_part_ii}
we conclude that for some constant $C_3'\geq0 $, we have
$\|\hat \delta^k\|_1 \leq C_3' r_{1n} + C_3' \lambda s_n/\kappa^2_{3}$, and the claim follows.
\hfill \halmos
\endproof

\proof{Proof of Theorem \ref{thm:OrthoHinv}.}

The identity $y_O^{(t)} = v_O - H^{-1}p_O^{(t)}+\varepsilon_O^{(t)}$  (Lemma \ref{lem:firstOrderO}),
together with Step 3 of Algorithm 1, implies that
\begin{equation}\label{eq:initialDiffThm1}
\begin{aligned}
(-\check v_k, \check W_{k,\cdot})^T & = (-\hat v_k, \hat W_{k,\cdot})^T -  \hat \Psi \En[\{y_k - (\hat v_k-\hat W_{k,\cdot} p_O)\}(1;p_O)]\\
& = (-\hat v_k, \hat W_{k,\cdot})^T -  \hat \Psi \En
\left[
\left(
\{y_k - ( v_k- \Hinvkdot p_O)\}
+ (v_k-\hat{v}_k,- \Hinvkdot + \hat{W}_{k,\cdot})(1;p_O)
\right)
(1;p_O) \right]
\\
& = (-\hat v_k, \hat W_{k,\cdot})^T +  \hat \Psi \En[(1;p_O)(1;p_O)^T \hat\delta^k] -  \hat \Psi \En[\varepsilon_k(1;p_O)]\\
& = (-v_k, H^{-1}_{k,\cdot})^T +  \{\hat \Psi \En[(1;p_O)(1;p_O)^T]-I\}\hat\delta^k - \hat \Psi\En[\varepsilon_k(1;p_O)]\\
& = (-v_k, H^{-1}_{k,\cdot})^T -  \hat \Psi\En[\varepsilon_k(1;p_O)] + (A_1),\\
\end{aligned}
\end{equation}
for all $k\in V_O$,
where $A_1:=  \{\hat \Psi \En[(1;p_O)(1;p_O)^T]-I\}\hat\delta^k$,
and $\hat\delta^k$ is as defined in \eqref{eq:deltaShortHand}.
Here, the second equality follows by adding/subtracting the same term, and the rest of the equalities are obtained via straightforward algebraic manipulation.

We bound $A_1$ using Holder's inequality as follows:
\begin{equation}\label{eq:boundA1}
\begin{aligned}
\|A_1\|_\infty & = \|\{\hat \Psi \En[(1;p_O)(1;p_O)^T]-I\}\hat\delta^k\|_\infty \\
& = \|\{\hat \Psi \hat \Sigma-I\}\hat\delta^k\|_{\infty}\\
& \leq \|\hat \Psi \hat \Sigma-I\|_{e,\infty} \|\hat\delta^k\|_1.\\
\end{aligned}
\end{equation}

By Lemma \ref{thm:DS:fast} with probability $1-o(1)$  we have
$$\max_{k\in V_O} \|\hat\delta^k\|_1
\leq \ C_1' r_{1n} + C_1' \lambda s_n/\kappa^2_{\bar c}
,$$
for some constant $C'_1 \geq 0$ and $\bar c=3$.
Using
Lemma \ref{lem:lambdaScale} and
the fact that $\kappa^2_{\bar c}$ is lower bounded by a constant
(Assumption \ref{Assumption:New}), this inequality implies that
$$\max_{k\in V_O} \|\hat\delta^k\|_1
\leq C'_2 s_n\sqrt{\log V_O/n} + C'_2r_{1n},$$
for some constant $C'_2 \geq 0$.

Moreover, by feasibility of $\{\hat \Psi_k, \hat z_k\}_k$ in \eqref{def:lasso:ortho}, Lemma \ref{lem:lambdaScale}, and Lemma \ref{lem:Alg1:DeBias}, with probability $1-o(1)$ we have that $$\|\hat \Psi \hat\Sigma-I\|_{e,\infty}\leq \lambda \|\hat z \|_\infty \leq C'_3\sqrt{\frac{\log |V_O|}{n}},$$
 for some constant $C_3'\geq 0$.
Using these observations with \eqref{eq:boundA1},
we conclude that
with probability $1-o(1)$ the following holds:
 $$ \|A_1\|_\infty \leq \frac{C_4's_n\log |V_O|}{n} + \frac{C_4' r_{1n}\sqrt{\log |V_O|}}{\sqrt{n}}, $$
for some constant $C_4'\geq 0$.

Thus, \eqref{eq:initialDiffThm1} implies that
with probability $1-o(1)$ we have
\begin{equation}
\label{eq:thm1FirstPart}
	\sqrt{n}\{ (-\check v_k, \check W_{k,\cdot})^T - (-v_k, H^{-1}_{k,\cdot})^T\} =  -\sqrt{n}\hat \Psi\En[\varepsilon_k(1;p_O)] +
	\sqrt{n} A_1,
	\end{equation}	
where
$\|\sqrt{n} A_1 \|_\infty= O(n^{-1/2}s_n\log |V_O| + r_{1n}\sqrt{\log |V_O|})$,
and the first part of the claim follows.

For the second part of the claim,
consider an $(s_n,r_{1n})$-sparse approximation of $H^{-1}$, denoted by $\bar{W}$.
For each row $k$, let $T_k$ denote the support of $\bar W_{k,\cdot}$.
Similarly, for each column $\ell$, let $\hat{T}_\ell$ denote the support of $\bar W_{\cdot,\ell}$.
Note that there always exists a sparse approximation
 $\bar{W}$
where
$\bar W_{kT_k}= H^{-1}_{kT_k}$
and
$\bar W_{\hat{T}_\ell\ell}= H^{-1}_{\hat{T}_\ell \ell}$
for all $k,\ell \in V_O$. Let $\bar{W}$ be such an approximation.

Note that we have
\begin{equation} \label{eq:Thm1LastBound}
\begin{aligned}
\| \check W_{k,\cdot}^\mu - H_{k,\cdot}^{-1} \|_1 & \leq \| \check W_{k,\cdot}^\mu -
\bWkdot
 \|_1 + \|\bWkdot - H_{k,\cdot}^{-1} \|_1\\
& \leq \| \check W_{k T_k}^\mu - H^{-1}_{kT_k} \|_1 + \| \check W_{k T_k^c}^\mu\|_1 + r_{1n},\\
\end{aligned}
\end{equation}
where
the first inequality follows from the
triangle inequality, and in the second inequality
we use the fact that $\|\bWkdot - H_{k,\cdot}^{-1} \|_1 \leq r_{1n}$ since
$\bar{W}$ is an
$(s_n,r_{1n})$-sparse approximation of $H^{-1}$.

Since $|T_k| \leq s_n$ it
can be readily seen that
\begin{equation} \label{eq:boundThresholdedDiff}
\| \check W_{k T_k}^\mu - H^{-1}_{kT_k} \|_1 \leq s_n \max_{j\in T_k} |\check W_{kj}^\mu-(H^{-1})_{kj}|.
\end{equation}
Also note that using the triangle inequality $\mu_{kj}\geq 2 |\check W_{kj} -(H^{-1})_{kj}|$ implies that $\mu_{kj}\geq  2|(H^{-1})_{kj}|-2 |\check W_{kj}|$.
Thus, if $|\check W_{kj}|\leq \mu_{kj}$, then $\check W_{kj}^\mu=0$, and $|(H^{-1})_{kj}| \leq \frac{3}{2} \mu_{kj}$. Hence, in this case, we obtain
$|\check W_{kj}^\mu-(H^{-1})_{kj}|\leq \frac{3}{2} \mu_{kj}$.
Conversely, if $|\check W_{kj}|> \mu_{kj}$, then $\check W_{kj}^\mu= \check W_{kj}$. Hence $|\check W_{kj}^\mu-(H^{-1})_{kj}|= |\check W_{kj}-(H^{-1})_{kj}|\leq \frac{1}{2} \mu_{kj}$.
Thus,
these cases (together
with the fact that
$\mu_{kj}\leq C \sqrt{\log|V_O|/n}$ for all $k,j$) imply that
\begin{equation}\label{eq:boundCheckWDiff}
|\check W_{kj}^\mu-(H^{-1})_{kj}| \leq \frac{3}{2} \mu_{kj}\leq C'_5 \sqrt{\frac{\log|V_O|}{n}},
\end{equation}
for some constant $C'_5 \geq 0$.
Hence, using
\eqref{eq:boundThresholdedDiff}, we obtain
\begin{equation}\label{eq:Thm1LastBound2}
\| \check W_{k T_k}^\mu - H^{-1}_{kT_k} \|_1 \leq C'_5 s_n  \sqrt{\frac{\log |V_O|}{n}}.
\end{equation}%

Finally, using the triangle inequality,
$\mu_{kj}\geq 2 |\check W_{kj} -(H^{-1})_{kj}|$ also implies that $\mu_{kj}\geq 2 |\check W_{kj}|- 2|(H^{-1})_{kj}|$.
Hence, if  $|(H^{-1})_{kj}|\leq  \mu_{kj}/2$, then $\mu_{kj} \geq |\check W_{kj}|$.
Using this observation it follows that
\begin{equation}\label{eq:Thm1LastBound3}
\begin{aligned}
\| \check W_{k T_k^c}^\mu\|_1 &=\sum_{j \in T_k^c} |\check W_{k j}^\mu| =  \sum_{j \in T_k^c} |\check W_{k j}| \mathbf{1}\{ \check W_{k j}>\mu_{kj} \} \\
& \leq \sum_{j \in T_k^c} \{|\check W_{kj}-H^{-1}_{kj}| + |H^{-1}_{kj}|\} \mathbf{1}\{|(H^{-1})_{kj}|> \mu_{kj}/2\} \\
& \leq \|H^{-1}_{kT_k^c}\|_1 +  \sum_{j \in T_k^c} \frac{\mu_{kj}}{2} \mathbf{1}\{|(H^{-1})_{kj}|> \mu_{kj}/2\} \\
& \leq r_{1n} +  \sum_{j \in T_k^c} (\mu_{kj}/2) \frac{|(H^{-1})_{kj}|}{(\mu_{kj}/2)} \leq 2r_{1n}.
\end{aligned}
\end{equation}
Here the last line makes use of the fact that $\|H^{-1}_{kT_k^c}\|_1 \leq r_{1n}$ (approximate sparsity).

	Using \eqref{eq:Thm1LastBound2} and \eqref{eq:Thm1LastBound3}, we obtain from
	\eqref{eq:Thm1LastBound}
	that
	$$
\| \check W_{k,\cdot}^\mu - H_{k,\cdot}^{-1} \|_1  \leq  C'_5 s_n \sqrt{\frac{\log |V_O|}{n}} + 3 r_{1n}.
	$$
	Thus, we obtain $
	\| \check W^\mu - H^{-1} \|_\infty =
	\max_{k\in V_O} \| \check W_{k,\cdot}^\mu - H_{k,\cdot}^{-1} \|_1 \leq C'_5 s_n \sqrt{\frac{\log |V_O|}{n}} + 3 r_{1n}$, as claimed.

Following a similar approach
to \eqref{eq:Thm1LastBound}
for columns of $\check W^\mu$ and $H^{-1}$
we obtain
\begin{equation} \label{eq:Thm1LastBoundC}
\begin{aligned}
\| \check W_{\cdot,\ell}^\mu - H_{\cdot,\ell}^{-1} \|_1
& \leq \| \check W_{ \hat{T}_\ell \ell }^\mu - H^{-1}_{\hat{T}_\ell \ell} \|_1 + \| \check W_{\hat{ T}_\ell^c \ell}^\mu\|_1 + r_{1n}.\\
\end{aligned}
\end{equation}
In addition,
since $|\hat{T}_\ell | \leq s_n$, by
repeating the approach in \eqref{eq:boundThresholdedDiff} and using
\eqref{eq:boundCheckWDiff}
 it can also be shown that
\begin{equation} \label{eq:boundThresholdedDiff2}
 \| \check W_{ \hat{T}_\ell \ell }^\mu - H^{-1}_{\hat{T}_\ell \ell} \|_1
\leq s_n \max_{j\in \hat{T}_\ell } |\check W_{j\ell}^\mu-(H^{-1})_{j\ell}| \leq
C'_5 s_n \sqrt{\frac{\log |V_O|}{n}}.
\end{equation}
Finally, repeating the steps of \eqref{eq:Thm1LastBound3} to bound
$\| \check W_{ \hat{T}_\ell^c\ell}^\mu\|_1$, we also obtain $\| \check W_{ \hat{T}_\ell^c\ell}^\mu\|_1 \leq 2 r_{1n}$.
Together with \eqref{eq:Thm1LastBoundC} and \eqref{eq:boundThresholdedDiff2}, this inequality implies that
	$$
\|\check W^\mu -H^{-1}\|_1 = \max_{\ell \in V_O}	\| \check W_{\cdot,\ell}^\mu - H_{\cdot,\ell}^{-1} \|_1  \leq  C'_5 s_n \sqrt{\frac{\log |V_O|}{n}} + 3 r_{1n}.
	$$
Hence, the claim follows.
\hfill\halmos

\endproof

\subsection{Proof of Theorem \ref{cor:Thm1}}

\proof{Proof of Theorem \ref{cor:Thm1}.}
By Theorem \ref{thm:OrthoHinv}(i) and the triangle inequality, with probability $1-o(1)$ we have
$$\max_{k\in V_O} |\check v_k - v_k | \vee  \max_{k,j\in V_O} | \check W_{kj} - H^{-1}_{kj} | \leq \max_{k\in \bar{V}_O,j\in V_O}\left|
\hPkdot \frac{1}{n} \sum_{t=1}^n (1;p_O^{(t)}) \varepsilon_j^{(t)}\right| + n^{-1/2}\max_{k \in V_O} \|R_n^k\|_\infty ,  $$
where
$\max_{k \in V_O}\|R^k_n\|_\infty = O(n^{-1/2}s_n\log |V_O| + r_{1n}\sqrt{\log |V_O|})$.
Note that
 $n^{-1/2}s_n\sqrt{\log |V_O|} =o(1) $
(by Assumption~\ref{assumption:basicRandom}).
{Moreover,
	by Lemma \ref{lem:MInv0} $H^{-1}$ is a submatrix of $M^{-1}$.
	Since the latter matrix  has bounded matrix $1$ and $\infty$-norms, so does $H^{-1}$.
Thus, it follows that
it is always possible to find a sparse approximation with
$r_{1n} \leq C_1'$ for some constant $C_1' \geq 0$ .}
Using these observations, we have that $\max_{k \in V_O} \|R^k_n\|_\infty \leq 2C_1'\sqrt{\log |V_O|}$, which in turn implies that
\begin{equation}\label{eq:Corr1_Bound1}
\max_{k\in V_O} |\check v_k - v_k | \vee\max_{k,j\in V_O} | \check W_{kj} - H^{-1}_{kj} | \leq \max_{k\in \bar{V}_O,j\in V_O}\left|
\hPkdot
 \frac{1}{n} \sum_{t=1}^n (1;p_O^{(t)}) \varepsilon_j^{(t)}\right| +2 C_1' \sqrt{\frac{\log V_O}{n}},
\end{equation}
with probability $1-o(1)$.
Next, we bound the first term on the right-hand side of
\eqref{eq:Corr1_Bound1}.

Observe that 	
by Lemma \ref{lem:Alg1:DeBias}(v), with probability $1-o(1)$ we have
\begin{equation} \label{eq:4thMomentBounded}
 \frac{1}{n}\sum_{t=1}^n|\hat \Psi_{k\cdot} (1;p_O^{(t)} )|^4 \leq C_2'
\end{equation}
for all $k\in \bar{V}_O$ and some constant $C_2'>0$.
Moreover, with probability $1-o(1)$ we also have
\begin{equation}
\frac{1}{n}\sum_{t=1}^n|\hat \Psi_{k\cdot} (1;p_O^{(t)} )|^2
=
 \Phk \hat\Sigma  \Phk^T \geq (1 - \lambda \hat z_k)^2/\hat\Sigma_{kk}
 \geq c_1'
\end{equation}
	for all $k$ and some constant $c_1'>0$.
	Here, the equality follows from the definition of $\hat{ \Sigma}$, and
	the first inequality follows from Lemma \ref{lem:Alg1:DeBias}(iv).
	The second inequality follows from the fact that covariates
	and hence $\hat{\Sigma}_{kk}$
	are bounded by a constant,
	$\hat z_k$ is bounded from above by a constant (Lemma \ref{lem:Alg1:DeBias}(ii)),
	and $\lambda = o(1)$ by Lemma \ref{lem:lambdaScale} and Assumption \ref{assumption:basicRandom}.
	Thus, we conclude that the  event
$\mathcal{E}=\{  \frac{1}{n}\sum_{t=1}^n|\hat \Psi_{k\cdot} (1;p_O^{(t)} )|^4 \leq C'_2, \frac{1}{n}\sum_{t=1}^n|\hat \Psi_{k\cdot} (1;p_O^{(t)} )|^2 \geq c'_1, \ \mbox{for all} \ k\in \bar V_O \}$, occurs with probability $1-o(1)$.
In what follows,
with some abuse,
 we use the notation $\{\tilde{p}_O^{(t)}\} \in \mathcal{E}$ to express that the  sequence  of covariates $\{\tilde{p}_O^{(t)}\}$ satisfies the conditions of this event.

Let $Z^{(t)}_{ij}=
\hat \Psi_{i\cdot} (1;p_O^{(t)} ) \varepsilon_j^{(t)}$ for $t\in[n]$, and $i\in \bar{V}_O, j\in V_O$.
Denote by
$X^{(t)}$  a  vector of length $|\bar V_O| \times |{V}_O|$ whose entries consist of
$\{Z^{(t)}_{ij}\}_{ i\in \bar{V}_O,j \in {V}_O }$.
Note that Step 2 of Algorithm 1
and $\hat \Psi$
rely only on $\{p_O^{(t)}\}$.
Hence, the definition of $Z_{ij}^{(t)}$
and Assumption \ref{assumption:basicRandom}
imply that
$E[Z^{(t)}_{ij} \mid  \{p_O^{(t)} \}]=
\hat \Psi_{i\cdot} (1;p_O^{(t)} )
E[\varepsilon_j^{(t)}\mid p_O^{(t)}]
=0$.
Moreover, Assumption \ref{assumption:basicRandom} yields
$\min_{j\in V_O}E[|\varepsilon_j^{(t)}|^2 \mid p_O^{(t)}] \geq c$ and  $\max_{j\in V_O}E[|\varepsilon_j^{(t)}|^4 \mid p_O^{(t)}] \leq C$.
Thus, for $\{p_O^{(t)}\} \in \mathcal{E}$
we have
$$\frac{1}{n}\sum_{t=1}^nE[|Z^{(t)}_{ij}|^2  \mid  \{p_O^{(t)} \} ] = \frac{1}{n}\sum_{t=1}^n |\hat \Psi_{i\cdot} (1;p_O^{(t)} )|^2E[|\varepsilon^{(t)}_{j}|^2\mid p_O^{(t)}] \geq c'_2,  $$
$$\frac{1}{n}\sum_{t=1}^nE[|Z^{(t)}_{ij}|^4  \mid  \{p_O^{(t)} \} ] = \frac{1}{n}\sum_{t=1}^n |\hat \Psi_{i\cdot} (1;p_O^{(t)} )|^4E[|\varepsilon^{(t)}_{j}|^4\mid p_O^{(t)}] \leq C'_3 , $$
for some constants $c'_2,C'_3>0$.

Observe that   $\{X^{(t)}\}$ are independent vectors conditionally on $\{p_O^{(t)} , t\in[n]\}$.
Furthermore,
for $\{p_O^{(t)}\} \in \mathcal{E}$,
we have
$$E\left[\max_{t \in [n]} \|X^{(t)}\|_\infty^4 \mid \{p_O^{(t)} \} \right]
\leq
\left(\max_{t \in [n],{i \in \bar{V}_O}}
|\hat \Psi_{i\cdot} (1;p_O^{(t)} )|^4
\right)\left(
E\left[\max_{t \in [n], j\in V_O} |\varepsilon_j^{(t)}|^4 \mid p_O^{(t)}\right]
\right)
\leq C'_2 M_\varepsilon,
$$
where we use
 Assumption
\ref{assumption:basicRandom}.
Finally,
by Assumption \ref{assumption:basicRandom} we also have
$M_\varepsilon \log |V_O| = o(n)$,
and $\log |V_O|/n = o(1)$,  $C'_5 \log |V_O| \geq \log n$  for some constant $C_5'$.
These observations collectively imply that  $\{Z_{ij}^{(t)} \}$ conditional on $\{p_O^{(t)} \}\in {\mathcal E}$ satisfy the conditions of Lemma \ref{lem:ozanVersion} (with $L=4$).
Hence, together with the fact that $\mathcal{E}$ occurs with  probability $1-o(1)$, we obtain
\begin{equation} \label{eq:Thm2DetailStepPre}
\begin{aligned}
\lambda \geq
\max_{i\in \bar{V}_O, j\in V_O}
\frac{   \left|   \frac{1}{n} \sum_{t=1}^n \hat\Psi_{i\cdot}(1;p_O^{(t)} ) \varepsilon_j^{(t)}\right| }{
	\sqrt{
		\frac{1}{n} \sum_{t=1}^n
		\left|
		\hat\Psi_{i\cdot}(1;p_O^{(t)} ) \varepsilon_j^{(t)}\right|^2
	}
} .
\end{aligned}
\end{equation}
This in turn yields
\begin{equation} \label{eq:Thm2DetailStep}
\begin{aligned}
\max_{i\in \bar{V}_O, j\in V_O} \left|   \frac{1}{n} \sum_{t=1}^n \hat\Psi_{i\cdot}(1;p_O^{(t)} ) \varepsilon_j^{(t)}\right|
&\leq \lambda
\max_{i\in \bar{V}_O, j\in V_O}
\sqrt{
	\frac{1}{n} \sum_{t=1}^n
	\left|
	\hat\Psi_{i\cdot}(1;p_O^{(t)} ) \varepsilon_j^{(t)}\right|^2
}\\
&\leq
{\lambda }
\max_{i\in \bar{V}_O, j\in V_O}
\sqrt[4]{\left(\frac{1}{n}\sum_{t=1}^n
	\left|
	\hat\Psi_{i\cdot}(1;p_O^{(t)} ) \right|^4\right)
	\left(\frac{1}{n}\sum_{t=1}^n
	\left|
	\varepsilon_j^{(t)} \right|^4\right)
},
\end{aligned}
\end{equation}
where, the second inequality follows from the Cauchy--Schwarz inequality.

Note that
under Assumption \ref{assumption:basicRandom}, $\{\varepsilon_O^{(t)}\}$ vectors also satisfy the conditions of Lemma~\ref{lem:ozanVersion} (with $L=4$),
which implies that with probability $1-o(1)$, we have
$\frac{1}{n}\sum_{t=1}^n
\left|
\varepsilon_j^{(t)} \right|^4 \leq C_6'$ for some constant $C_6'>0$.
This observation, together with
\eqref{eq:Thm2DetailStep},
implies that
$\max_{i\in \bar{V}_O, j\in V_O} \left|   \frac{1}{n} \sum_{t=1}^n \hat\Psi_{i\cdot}(1;{p}_O^{(t)} ) \varepsilon_j^{(t)}\right| \leq C_7' \lambda$ for some constant $C_7' > 0$.
Hence, Lemma \ref{lem:lambdaScale}
implies that
\begin{equation} \label{eq:MDSN_bound3_o3}
\begin{aligned}
\max_{i\in \bar{V}_O, j\in V_O} \left|   \frac{1}{n} \sum_{t=1}^n \hat\Psi_{i\cdot}(1;{p}_O^{(t)} ) \varepsilon_j^{(t)}\right|
\leq C_7' \lambda \leq C_8' \sqrt{\frac{\log |V_O|}{n}},
\end{aligned}
\end{equation}
for some constant $C_8' > 0$, with probability $1-o(1)$.
Substituting the bound (\ref{eq:MDSN_bound3_o3}) in \eqref{eq:Corr1_Bound1}, we conclude that with probability $1-o(1)$ we have
$\max_{k\in V_O} |\check v_k - v_k | \vee\max_{k,j\in V_O} | \check W_{kj} - H^{-1}_{kj} |  \leq C_9' \sqrt{\frac{\log |V_O|}{n}},$ for some constant $C'_9 > 0$.

Next we prove the second claim.
Let $\mu_{kj}$ be defined as in \eqref{eq:thresholds},
and set $\bar \sigma_{kj}^2 := \frac{1}{n} \sum_{t=1}^n |\hat\Psi_{k\cdot}(1;p_O^{(t)}) \varepsilon_j^{(t)}|^2$.
Note that Theorem \ref{thm:OrthoHinv}(i) implies that
$$
|(\check W - H^{-1})_{kj}| 	 \leq  \left|\frac{1}{n} \sum_{t=1}^n \hat\Psi_{k\cdot}(1;p_O^{(t)}) \varepsilon_j^{(t)}\right| + n^{-1/2}|(R^k_{n})_j|,
$$
for all $k,j\in V_O$ with probability $1-o(1)$.
Using this observation together with  \eqref{eq:Thm2DetailStepPre} implies that
with probability $1-o(1)$,  uniformly over $k,j \in   V_O$, we have
\begin{equation}\label{eq:AddSubSigmakj}
\begin{array}{rl}
|(\check W - H^{-1})_{kj}| &
 \leq  \lambda \left( \frac{1}{n} \sum_{t=1}^n |\hat\Psi_{k\cdot}(1;p_O^{(t)}) \varepsilon_j^{(t)}|^2\right)^{1/2} +n^{-1/2}|(R^k_{n})_j|\\
 & \leq  \lambda \bar\sigma_{kj} +  n^{-1/2}|(R^k_{n})_j| \\
 & \leq  \lambda \hat\sigma_{kj} + \lambda|\bar\sigma_{kj}-\hat\sigma_{kj}| + n^{-1/2}|(R^k_{n})_j| \\
 & \leq   \lambda \hat\sigma_{kj}
 +\frac{1}{\log n} \lambda \hat\sigma_{kj}
 -\frac{1}{\log n} \lambda \hat\sigma_{kj}
 +  \lambda|\bar\sigma_{kj}-\hat\sigma_{kj}| + n^{-1/2}|(R^k_{n})_j| \\
 & \leq  \frac{1}{2}\mu_{kj} - \frac{1}{\log n} \lambda \hat\sigma_{kj}+
\lambda|\bar\sigma_{kj}-\hat\sigma_{kj}| + n^{-1/2}|(R^k_{n})_j|, \\
\end{array}
\end{equation}
where the second inequality uses the definition of $\bar{ \sigma}_{kj}$, the third one is obtained by the triangle inequality, and
the fourth one is obtained by adding/subtracting the same term. Finally,
the
last inequality follows from
the definition of
$\mu_{kj}$ in \eqref{eq:thresholds}.

Note that
\begin{equation} \label{eq:Thm2Aux0}
\begin{aligned}
|\bar\sigma_{kj}-\hat\sigma_{kj}| &=
\sqrt{\frac{1}{n} \sum_{t=1}^n |\hat\Psi_{k\cdot}(1;p_O^{(t)}) \varepsilon_j^{(t)}|^2}
-
\sqrt{
\frac{1}{n} \sum_{t=1}^n |\hat\Psi_{k\cdot}(1;p_O^{(t)})
(y_j^{(t)}-(\hat v_j-\hat W_{j,\cdot}p_O^{(t)}))
|^2
}\\
&\leq
\sqrt{
	\frac{1}{n} \sum_{t=1}^n |\hat\Psi_{k\cdot}(1;p_O^{(t)}) (\varepsilon_j^{(t)}-y_j^{(t)}+(\hat v_j-\hat W_{j,\cdot}p_O^{(t)}))|^2
}\\
&\leq
\left(\max_{t\in [n]}\left|(v_j-\hat v_j+
\hat W_{j,\cdot}p_O^{(t)}-
H^{-1}_{j,\cdot}p_O^{(t)})\right|
\right)
\sqrt{
	\frac{1}{n} \sum_{t=1}^n |\hat\Psi_{k\cdot}(1;p_O^{(t)}) |^2
},
\end{aligned}
\end{equation}
where the first inequality follows by observing that for any two vectors $B_1$, $B_2$, we have
$\|B_1\|_2 - \|B_2\|_2 \leq
\|B_1 -B_2\|_2$, and the second inequality follows from Lemma \ref{lem:firstOrderO}.
On the other hand,
\begin{equation} \label{eq:Thm2Aux1}
\left|(v_j-\hat v_j+
\hat W_{j,\cdot}p_O^{(t)}-
H^{-1}_{j,\cdot}p_O^{(t)})\right| \leq \|(\hat v_j,\hat{W}_{j,\cdot}) - (v_j, H^{-1}_{j,\cdot})\|_1  \bar{p},
\end{equation}
using Holder's inequality and the fact that $\|(1;p_O^{(t)}) \|_\infty \leq \bar{p}$.
 Also note that for all $k\in V_O$ with probability $1-o(1)$, we have
\begin{equation}\label{eq:Thm2Aux2}
\begin{aligned}
	\left(\sum_{t=1}^n \frac{1}{n} |\hat\Psi_{k\cdot}(1;p_O^{(t)}) |^2\right)^2
	\leq {n}\sum_{t=1}^n
	\frac{1}{n^2}
	|\hat\Psi_{k\cdot}(1;p_O^{(t)}) |^4
	\leq C_2',
\end{aligned}	
\end{equation}
where the first inequality follows from the Cauchy--Schwarz inequality, and the second one follows from
\eqref{eq:4thMomentBounded}.
Using \eqref{eq:Thm2Aux0}, \eqref{eq:Thm2Aux1},  \eqref{eq:Thm2Aux2}, and Lemma \ref{thm:DS:fast}, we conclude that
\begin{equation}
|\bar\sigma_{kj}-\hat\sigma_{kj}| \leq C'_{10}  \|(\hat v_j,\hat{W}_{j,\cdot}) - (v_j, H^{-1}_{j,\cdot})\|_1 \leq
 C_{11}' r_{1n} + C_{11}' \lambda s_n/\kappa^2_{3}
\end{equation}
for
all $k,j \in V_O$ and
some constants $C_{10}',C_{11}'>0$ with probability $1-o(1)$.
Since Assumption \ref{assumption:basicRandom} implies Assumption~\ref{Assumption:New}
(see Lemma \ref{lem:assBasicImpliesNew}), and  Assumption \ref{Assumption:New} implies that $\kappa^2_{3} $ is lower bounded by a constant, we equivalently  obtain
\begin{equation} \label{eq:boundSigmaDiff}
|\bar\sigma_{kj}-\hat\sigma_{kj}| \leq
C_{11}' r_{1n} + C_{12}'  \lambda s_n
\end{equation}
for some constant $C_{12}'>0$.
Using this together with \eqref{eq:AddSubSigmakj}, we conclude that
\begin{equation}\label{eq:AddSubSigmakj2}
\begin{array}{rl}
|(\check W - H^{-1})_{kj}| \leq  \frac{1}{2}\mu_{kj} - \frac{1}{\log n}\lambda \hat\sigma_{kj}+
 \lambda
\left( C_{11}' r_{1n} + C_{12}'\lambda s_n \right)
+ n^{-1/2}|(R^k_{n})_j|, \\
\end{array}
\end{equation}
for all $k,j\in V_O$ with probability $1-o(1)$.
Note that if
\begin{equation}\label{aux:cond:SNmu2}
\lambda
\left( C_{11}' r_{1n} + C_{12}'\lambda s_n\right)
+ n^{-1/2} \max_{k,j\in V_O}
|(R^k_{n})_j| \leq  \frac{1}{\log n}\lambda \min_{k,j\in V_O} \hat\sigma_{kj},
\end{equation}
then \eqref{eq:AddSubSigmakj2} implies that
\begin{equation} \label{eq:muTargetCond}
 \mu_{kj} \geq 2|(\check W - H^{-1})_{kj}| \ \  \mbox{simultaneously over} \ k,j\in V_O.
\end{equation}

Suppose that $\min_{k,j\in V_O} \hat\sigma_{kj}$ is bounded away from zero. We next show that in this case \eqref{aux:cond:SNmu2} and hence \eqref{eq:muTargetCond} hold.
To see this, note that Theorem \ref{thm:OrthoHinv} implies that
$\max_{k,j\in V_O} |(R^k_{n})_j|= O(n^{-1/2}s_n\log |V_O| + r_{1n}\sqrt{\log |V_O|})$.
Moreover, by Lemma \ref{lem:lambdaScale}, we have
 $C_{13}'
  \sqrt{\frac{\log |V_O|}{n}} \geq
 \lambda \geq c_2' \sqrt{\frac{\log{(n|V_O|)}}{n}}$ for some constants $c_2',C_{13}'>0$. Thus, \eqref{aux:cond:SNmu2} holds if
 \begin{equation} \label{eq:asymptoticCond}
 r_{1n} +  \sqrt{\frac{\log |V_O|}{n}}  s_n + \frac{{n^{-1/2}s_n\log |V_O| + r_{1n}\sqrt{\log |V_O|}} }{{\sqrt{\log(n|V_O|)}} } = o(1/\log n).
 \end{equation}
Note that  $\log(n|V_O|) = \log n + \log(|V_O|) \geq \log(|V_O|)$.
In addition,
for Theorem \ref{cor:Thm1}
it is assumed that
$r_{1n} = o(1/\log n)$, and by Assumption \ref{assumption:basicRandom}
we have
 $ \frac{ s_n^2(\log |V_O|) }{n}  = o(1/\log n^3)	$.
 These observations
imply that \eqref{eq:asymptoticCond} holds, which  in turn implies
\eqref{aux:cond:SNmu2} and \eqref{eq:muTargetCond}.

We proceed by establishing that $\min_{k, j\in V_O} \hat\sigma_{kj}$ is bounded away from zero with probability $1-o(1)$. Observe that
since
$r_{1n}=o(1/\log n)$, and  $\lambda s_n \leq C_{13}' \sqrt{\frac{\log |V_O|}{n}} s_n=o(1/\log n)$ by Lemma \ref{lem:lambdaScale} and Assumption~\ref{assumption:basicRandom},
\eqref{eq:boundSigmaDiff} also implies that
$
|\bar\sigma_{kj}-\hat\sigma_{kj}|  = o(1/\log n)$
for all $k,j\in V_O$ with probability $1-o(1)$.
Using this observation, it follows that
\begin{equation}\label{eq:boundSigmaHat}
 \min_{k\in \bar V_O, j\in V_O} \hat\sigma_{kj} \geq \min_{k\in \bar V_O, j\in V_O} \bar \sigma_{kj} - C_{14}'  (1/\log n),
\end{equation}
with probability $1-o(1)$
for some constant $C_{14}'>0$.
We next
show that $\min_{k\in \bar V_O, j\in V_O} \bar \sigma_{kj}$ is bounded away from zero with probability $1-o(1)$   using a truncation argument.

 Let $A_{kj} = \frac{1}{n}\sum_{t=1}^n E[|\hat\Psi_{k\cdot}(1;p_O^{(t)}) \varepsilon_j^{(t)}|^2 \mathbf{1}\{|\varepsilon_j^{(t)}|\leq \sqrt{\log n}\}\mid  \{p_O^{(t)} \}]$.
Recalling that $\{p_O^{(t)}\} \in \mathcal{E}$ with probability $1-o(1)$,
and using the union bound, we obtain
\begin{equation}\label{eq:preHoefding}
\begin{aligned}
P(\exists k,j : \bar\sigma_{kj}^2 < A_{kj} - \bar{\gamma}) & \leq P(\exists k,j : \bar\sigma_{kj}^2 < A_{kj} - \bar{\gamma} \mid \mathcal{E}) + o(1) \\
& \leq |V_O|^2\max_{k,j}
P\left(\frac{1}{n}\sum_{t=1}^n |\hat\Psi_{k\cdot}(1;p_O^{(t)}) \varepsilon_j^{(t)}|^2 < A_{kj}-\bar{\gamma} \mid \mathcal{E} \right)+o(1)   \\
& \leq |V_O|^2\max_{k,j} P\left(\frac{1}{n}\sum_{t=1}^n |\hat\Psi_{k\cdot}(1;p_O^{(t)}) \varepsilon_j^{(t)}|^2
\mathbf{1}\{|\varepsilon_j^{(t)}|\leq \sqrt{\log n}\}
< A_{kj}-\bar{\gamma} \mid \mathcal{E} \right)+o(1)  ,
\end{aligned}
\end{equation}
where $\bar{\gamma}$
is a parameter to be specified later.

Note that
conditional on $\{p_O^{(t)}\}$,
$\{|\hat\Psi_{k\cdot}(1;p_O^{(t)}) \varepsilon_j^{(t)}|^2
\mathbf{1}\{|\varepsilon_j^{(t)}|\leq \sqrt{\log n}\}\}_t$ are independent random variables, and
the expectation of
$\frac{1}{n}\sum_{t=1}^n |\hat\Psi_{k\cdot}(1;p_O^{(t)}) \varepsilon_j^{(t)}|^2
\mathbf{1}\{|\varepsilon_j^{(t)}|\leq \sqrt{\log n}\}$ is given by $A_{kj}$.
Furthermore,
$|\hat\Psi_{k\cdot}(1;p_O^{(t)}) \varepsilon_j^{(t)}|^2
\mathbf{1}\{|\varepsilon_j^{(t)}|\leq \sqrt{\log n}\}$ is bounded by
$|\hat\Psi_{k\cdot}(1;p_O^{(t)}) \sqrt{\log n}|^2$.
Thus,
using Hoeffding's inequality conditional on $\{p_O^{(t)}\}$,
we obtain
\begin{equation*}
P \left(\frac{1}{n}\sum_{t=1}^n |\hat\Psi_{k\cdot}(1;p_O^{(t)}) \varepsilon_j^{(t)}|^2
\mathbf{1}\{|\varepsilon_j^{(t)}|\leq \sqrt{\log n}\}
< A_{kj}-\bar{\gamma} \mid \{p_O^{(t)}\} \right) \leq
\exp \left(\frac{-2 \bar{\gamma}^2 n^2}{ \sum_{t=1}^n
|\hat\Psi_{k\cdot}(1;p_O^{(t)})|^4 ({\log n})^2}
\right).
\end{equation*}
Taking the expectation over $\{p_O^{(t)}\}\in \mathcal{E}$, and using this inequality together with
\eqref{eq:preHoefding} yields
\begin{equation}\label{eq:Hoefding}
\begin{aligned}
P(\exists k,j : \bar\sigma_{kj}^2 < A_{kj} - \bar{\gamma})
\leq
|V_O|^2\max_{k,j}E \left[\exp\left({-2\bar{\gamma}^2n} \Big/\{\frac{1}{n}\sum_{t=1}^n|\hat\Psi_{k\cdot}(1;p_O^{(t)})|^4\log^2 n\}\right)\mid \mathcal{E} \right] + o(1).
\end{aligned}
\end{equation}
Finally, using \eqref{eq:4thMomentBounded} together with this inequality, we conclude that
\begin{equation}\label{eq:Hoefding2}
\begin{aligned}
P(\exists k,j : \bar\sigma_{kj}^2 < A_{kj} - \bar{\gamma})
\leq
|V_O|^2
\exp\left(\frac{-C_{15}'\bar{\gamma}^2n}{\log^2 n}\right) + o(1),
\end{aligned}
\end{equation}
for some constant $C_{15}'>0$.

Note that the definition of $A_{kj}$ implies that
$$
 \begin{array}{rl}
 \frac{1}{n}\sum_{t=1}^n E[|\hat\Psi_{k\cdot}(1;p_O^{(t)}) \varepsilon_j^{(t)}|^2\mid \{p_O^{(t)}\}] - A_{kj} & \leq  \frac{1}{n}\sum_{t=1}^n E[|\hat\Psi_{k\cdot}(1;p_O^{(t)}) \varepsilon_j^{(t)}|^2 \mathbf{1}\{|\varepsilon_j^{(t)}|> \sqrt{\log n}\}\mid \{p_O^{(t)}\}] \\
  & \leq  \frac{1}{n}\sum_{t=1}^n |\hat\Psi_{k\cdot}(1;p_O^{(t)})|^2 E[
  | \varepsilon_j^{(t)}|^2 \mathbf{1}\{|\varepsilon_j^{(t)}|> \sqrt{\log n}\}
  \mid \{p_O^{(t)}\}]\\
 & \leq  \frac{1}{n}\sum_{t=1}^n |\hat\Psi_{k\cdot}(1;p_O^{(t)})|^2 E[|\varepsilon_j^{(t)}|^4/\log n\mid \{p_O^{(t)}\}],
 \end{array}$$
 where in the third line we make use of the fact that when
 $|\varepsilon_j^{(t)}|> \sqrt{\log n}$, we have
 $|\varepsilon_j^{(t)}|^2/{\log n}>1$.
 Note that Assumption \ref{assumption:basicRandom} implies that
 $E[|\varepsilon_j^{(t)}|^4 \mid \{p_O^{(t)}\}]$ is bounded by a constant for all $j\in V_O$.
 Similarly Lemma~\ref{lem:Alg1:DeBias} implies that
with probability
 $1-o(1)$
 we have that
 $\frac{1}{n}\sum_{t=1}^n |\hat\Psi_{k\cdot}(1;p_O^{(t)})|^2$ is also bounded by a constant  for all $k\in V_O$. Thus, with probability $1-o(1)$, uniformly over $k,j\in V_O$, we obtain
\begin{equation} \label{eq:AkjBound1}
  \frac{1}{n}\sum_{t=1}^n E[|\hat\Psi_{k\cdot}(1;p_O^{(t)}) \varepsilon_j^{(t)}|^2\mid \{p_O^{(t)}\}] - A_{kj} \leq C_{16}'/\log n
\end{equation}
for some constant $C_{16}'>0$.
On the other hand, with probability $1-o(1)$
 we have
\begin{equation} \label{eq:AkjBound2}
\begin{aligned}
\frac{1}{n}\sum_{t=1}^n E[|\hat\Psi_{k\cdot}(1;p_O^{(t)}) \varepsilon_j^{(t)}|^2\mid \{p_O^{(t)}\}] &=
\frac{1}{n}\sum_{t=1}^n
|\hat\Psi_{k\cdot}(1;p_O^{(t)})|^2
E[| \varepsilon_j^{(t)}|^2\mid p_O^{(t)}] \\
&\geq
c\frac{1}{n}\sum_{t=1}^n
|\hat\Psi_{k\cdot}(1;p_O^{(t)})|^2\\
& \geq
{c}
(1-\lambda \hat{z}_k)^2/ \hat{ \Sigma}_{kk}\\
&\geq c_3'>0,
\end{aligned}
\end{equation}
for some constant $c_3'$.
 Here, the first inequality follows from Assumption \ref{assumption:basicRandom}, and the second one follows from Lemma~\ref{lem:Alg1:DeBias}. The third inequality uses Lemma \ref{lem:lambdaScale} as well as
 the fact that
 $\hat{z}_k$
and
$\hat{ \Sigma}_{kk}$
are bounded by  constants (respectively due to Lemma \ref{lem:Alg1:DeBias} and the fact that the covariates are bounded by~$\bar{p}$).

  Observe that by letting
$\bar{\gamma}  :=
  C_{17}'(\log n)(\log(|V_O|^2n))^{1/2}/n^{1/2}
  $ for some constant $C_{17}'>0$ we obtain from \eqref{eq:Hoefding2} that
  with probability $1-o(1)$, for all $k,j\in V_O$, we have
$ \bar\sigma_{kj}^2 > A_{kj} - \bar{\gamma}$.
Hence, it follows that
$\min_{k,j\in V_O} \bar\sigma_{kj}^2 > \min_{k,j\in V_O} A_{kj} - \bar{\gamma}$
with probability $1-o(1)$.
On the other hand,
  \eqref{eq:AkjBound1} and \eqref{eq:AkjBound2} imply that
  with probability
  $1-o(1)$, we also have
  $\min_{k,j\in V_O} A_{kj} >c_4'$ for some constant $c_4'>0$.
  Combining these observations, and also noting that
  $\bar{\gamma}=o(1)$, we obtain
  $ \bar\sigma_{kj}^2 >c_5'$ with probability $1-o(1)$ for some constant $c_5'>0$ and all $k,j\in V_O$.
  By \eqref{eq:boundSigmaHat}, we conclude that
  $\min_{k\in \bar V_O, j\in V_O} \hat\sigma_{kj}$ is bounded away from zero with probability $1-o(1)$.
  Hence,
  \eqref{aux:cond:SNmu2} and \eqref{eq:muTargetCond} hold with probability $1-o(1)$, as argued before.

  Finally, observe that following an identical approach to \eqref{eq:preHoefding}, and using Hoeffding's inequality (with the same $\bar{\gamma}$), we also obtain
  with probability $1-o(1)$, uniformly over all $k,j$, that
  \begin{equation}\label{eq:preHoefding3}
 \bar\sigma_{kj}^2 < A_{kj} + \bar{\gamma} .
  \end{equation}
  Also note that with probability $1-o(1)$ for all $k,j\in V_O$
  and some constant $C_{18}'>0$
   we have
   \begin{equation} \label{eq:AkjBound3}
   \begin{aligned}
   A_{kj} \leq  \frac{1}{n}\sum_{t=1}^n E[|\hat\Psi_{k\cdot}(1;p_O^{(t)}) \varepsilon_j^{(t)}|^2 \mid  \{p_O^{(t)} \} ]
   &\leq
   \frac{1}{n}\sum_{t=1}^n
   |\hat\Psi_{k\cdot}(1;p_O^{(t)})|^2
   E[|\varepsilon_j^{(t)}|^2 \mid  \{p_O^{(t)} \} ] \\
   &\leq
   \frac{\sqrt{C}}{n}\sum_{t=1}^n
   |\hat\Psi_{k\cdot}(1;p_O^{(t)})|^2
   \leq C_{18}'.
   \end{aligned}
   \end{equation}
   Here, the last line follows from Assumption \ref{assumption:basicRandom}
   (which by Jensen's inequality implies that
   $E[|\varepsilon_j^{(t)}|^2 \mid  p_O^{(t)}  ] \leq C^{1/2}$)
   and \eqref{eq:Thm2Aux2}.
Observe that
\eqref{eq:preHoefding3}, \eqref{eq:AkjBound3},  and the fact that $\bar{\gamma}=o(1)$
imply
that
for all $k,j\in V_O$,
$\bar\sigma_{kj}$ is bounded by a constant with probability $1-o(1)$.
Since
$
|\bar\sigma_{kj}-\hat\sigma_{kj}|  = o(1/\log n)$
(for all $k,j\in V_O$ with probability $1-o(1)$), we conclude
that
for all $k,j\in V_O$,
$\hat\sigma_{kj}$ is bounded by a constant with probability $1-o(1)$.
Thus, by the definition of $\mu_{kj}$ (see \eqref{eq:thresholds}) we obtain
  $\mu_{kj} \leq 3 \lambda \hat{\sigma}_{kj} \leq C_{19}'  \lambda$,
  for some constant $C_{19}'>0$.
  By Lemma \ref{lem:lambdaScale}, we conclude that
  with probability $1-o(1)$,   $\mu_{kj}\leq C_{20}' \sqrt{\log|V_O|/n}$ for some constant $C_{20}'>0$. Hence, the constructed $\{\mu_{kj}\}$ satisfy
  \eqref{eq:thresholdConstraints} with probability $1-o(1)$.
  Note that this readily implies the claim in \eqref{eq:Thm2Rates0}.
Moreover, Theorem \ref{thm:OrthoHinv} implies the rates claimed in \eqref{eq:Thm2Rates}, completing the proof.~\hfill \halmos	
\endproof

\section{Proofs of Section \ref{se:applications}}

\begin{proof}{Proof of Theorem \ref{thm:optimalAdv}}
	By definition we have
	${{p}}_O^\star=
	-\frac{1}{\chi} \tilde H^{-T} e
	$ and
	$
	\hat {{p}}_O =
	\left(-\frac{1}{\chi} (\check{W}^\mu)^T e  \wedge \bar p \cdot e_O\right)_+$.
	Thus,
	it follows from Theorem \ref{thm:OrthoHinv} that with probability at least $1-o(1)$ we have
	\begin{equation}\label{eq:price1NormBoundadv}
		\begin{aligned}
			\|   {{p}}_O^\star- \hat{{p}}_O \| _2  & =
			\frac{1}{\chi}
			\| - \tilde H^{-T} e -\left(-(\check{W}^\mu)^T  e \wedge \bar p \cdot e_O \right)_+\|_2
			\leq
			\| \tilde H^{-T} e -(\check{W}^\mu)^T  e \|_2  \\
			&\leq 		\| \tilde H^{-T}  -(\check{W}^\mu)^T   \|_2 \|e\|_2
			\leq
			\sqrt{|V_O|} \sqrt{\|\tilde H^{-1}-\check{W}^\mu \|_1  \|\tilde H^{-1}-\check{W}^\mu \|_\infty}\\
			&		\leq
			s_nC'_1 \sqrt{|V_O|\log(|V_O|)/n} +	\sqrt{|V_O|}  3 r_{1n},
		\end{aligned}
	\end{equation}
	for some constant $C_1'$.
	
	Since the payoff $\Pi$ is quadratic in $p_O$, we have
	\begin{equation}
		\begin{aligned}
			\Pi({p}_O) &=  \Pi({p}_O^\star) + ({p}_O - {p}_O^\star)^T \nabla \Pi({p}_O^\star) +
			({p}_O - {p}_O^\star)^T \frac{\nabla^2 \Pi({p}_O^\star)}{2}
			({p}_O - {p}_O^\star) \\
			&= \Pi({p}_O^\star) -  \frac{\chi }{2}
			\|{p}_O - {p}_O^\star\|_2^2
			,
		\end{aligned}
	\end{equation}
	where we use the optimality condition $\nabla \Pi({p}_O^\star)=0$ and the fact that $-\nabla^2 \Pi({p}_O^\star)  = \chi I$.
	
	This observation
	together with \eqref{eq:price1NormBoundadv} implies that
	\begin{equation}\label{eq:finalProfitBoundadv}
		\begin{aligned}
			\Pi({{p}}^\star_O) -\Pi(\hat{{p}}_O) &\leq
			\frac{\chi}{2}\left(			s_nC'_1 \sqrt{|V_O|\log(|V_O|)/n} +	\sqrt{|V_O|}  3 r_{1n} \right)^2.
		\end{aligned}
	\end{equation}

	On the other hand,
	since $z_i \geq \underline{z} >0$,
setting $p_O=c e_O$ for small enough $c>0$
	guarantees that $y_i \geq   c z_i >c^2 /\chi>0$  for all $i$ (even when network effects are ignored). Thus,
	we have
	$\Pi({p}_O^\star)=\Omega(|V_O|)$.
	This observation together with
	\eqref{eq:finalProfitBoundadv} implies that
	for some constant $C_2'>0$, we have
	\[R(\hat{{p}}_O)\leq
	C_2'\left(
	s_n^2  \frac{\log(|V_O|)}{n} +  r_{1n}^2 \right)
	.\]
	Hence, the claim follows.
	\hfill \halmos
\end{proof}

   \noindent\proof{Proof of Theorem \ref{thm:optimalprices}.}
   Let $\hat A := ( \check{W}^\mu+ (\check{W}^\mu)^T)$, and  $A := (H^{-1}+H^{-T})$.
   By Lemma \ref{lem:optPrices}, the   optimal prices are given by
   ${{p}}_O^\star=A^{-1}  {v}_O$.
   Using a similar notation, the price vector $\hat {{p}}_O$ can alternatively be given by
   $\hat {{p}}_O=([\hat{A}^{-1} \check{{v}}_O] \wedge {\bar{p}} {e}_O)_+$.

   Observe that by the assumption that $0 < {{p}}_O^\star< \bar{p}{e}_O$, we obtain
   \begin{equation}\label{eq:price1NormBound}
   \begin{aligned}
   \|   {{p}}_O^\star- \hat{{p}}_O \| _2  & =
   \|A^{-1} {v}_O - ([\hat{A}^{-1} \check{{v}}_O] \wedge {\bar{p}} {e}_O)_+\|_2 \leq
   \|A^{-1} {v}_O -  \hat{A}^{-1} \check{{v}}_O\|_2 \\
   & \leq \| ( A^{-1} - \hat A^{-1}) (\check{{{v}}}_O -{{v}}_O) \|_2 + \| (A^{-1} - \hat A^{-1}) {{v}}_O \|_2 + \|A^{-1} ({{v}}_O-\check{{{v}}}_O) \|_2 \\
   & \leq \| \hat A^{-1} - A^{-1}\|_2  \|  \check{{{v}}}_O -{{v}}_O  \|_2 + \| \hat A^{-1} - A^{-1} \|_2 \|{{v}}_O\|_2+ \|A^{-1}\|_2  \| \check{{{v}}}_O -{{v}}_O \|_2. \\
   \end{aligned}
   \end{equation}
   Here, the second line follows from the triangle inequality, and the third one follows from the definition of matrix $p$-norms.

   {By Lemma \ref{lem:Ainv}} of Appendix \ref{subse:MatrixLemma}, we have
   $ \|A^{-1}\|_2 \leq \bar{\lambda}^4/2\zeta^3$.	
   Note that it follows from Theorem \ref{thm:OrthoHinv} that with probability at least $1-o(1)$ we have
   \begin{equation}\label{eq:AhatADiff}
   \begin{aligned}
   \|A-\hat{A}\|_2 &=\|H^{-1}-\check{W}^\mu +H^{-T}-(\check{W}^\mu)^T\|_2 \leq
   2\|H^{-1}-\check{W}^\mu \|_2  \\
   &\leq  2  \sqrt{\|H^{-1}-\check{W}^\mu \|_1  \|H^{-1}-\check{W}^\mu \|_\infty}
   \leq
   2s_nC'_1 \sqrt{\log(|V_O|)/n} + 6 r_{1n},
   \end{aligned}
   \end{equation}
   for some constant $C_1'>0$.
   In addition,
   using the results of \citet{dwyer1953errors} (see equation (4.3)), it follows that
   \begin{equation}\label{eq:boundAinvDiff}
   \|\hat{A}^{-1} - A^{-1}\|_2 \leq \frac{\|\hat{A} -A\|_2 \|{A}^{-1} \|_2^2}{1- \|\hat{A} -A\|_2 \|A^{-1}\|_2},
   \end{equation}
   provided that $\|\hat{A} -A\|_2 \|A^{-1}\|_2 <1$.
   The latter condition holds, since
   $ \|A^{-1}\|_2 \leq \bar{\lambda}^4/2\zeta^3$ and
   by the assumptions of the theorem we have
   $2s_nC_1' \sqrt{\log(|V_O|)/n} + 6 r_{1n}=o(1)$
   which together with  \eqref{eq:AhatADiff} implies the inequality.
   Expressing \eqref{eq:boundAinvDiff} more explicitly, once again  using the fact that
   $ \|A^{-1}\|_2 \leq \bar{\lambda}^4/2\zeta^3$ together with
   \eqref{eq:AhatADiff}, we obtain
   \begin{equation}\label{eq:finBound_1nNew}
   \begin{aligned}
   \|\hat{A}^{-1} - A^{-1}\|_2 &\leq
   \left(\frac{\bar{\lambda}^4}{2\zeta^3}\right)^2
   \frac{      2s_nC_1' \sqrt{\log(|V_O|)/n} + 6 r_{1n}}{1- \frac{\bar{\lambda}^4}{2\zeta^3}(2s_nC_1' \sqrt{\log(|V_O|)/n} + 6 r_{1n})}\\
   &\leq
   4\left(\frac{\bar{\lambda}^4}{2\zeta^3}\right)^2
   \left(     s_nC_1' \sqrt{\log(|V_O|)/n} + 3 r_{1n}\right).
   \end{aligned}
   \end{equation}
   Similarly, by   Theorem
   \ref{thm:OrthoHinv},
    with probability $1-o(1)$
   we have
   $\|\check{{v}}_O-{{v}}_O \|_2 \leq C_2'  \sqrt{|V_O|}\sqrt{\frac{\log |V_O|}{n}},$
   for some constant $C_2'>0$.
   Combining these inequalities
   with \eqref{eq:price1NormBound},
   \eqref{eq:finBound_1nNew}, and the fact that
   $ \|A^{-1}\|_2 \leq \bar{\lambda}^4/2\zeta^3$
   we obtain
   \begin{equation}\label{eq:price1NormBound2}
   \begin{aligned}
   \|   {{p}}_O^\star- \hat{{p}}_O \| _2  & \leq
   \|  \check{{{v}}}_O -{{v}}_O  \|_2 ( \| \hat A^{-1} - A^{-1}\|_2 + \|A^{-1}\|_2  )
   +  \| \hat A^{-1} - A^{-1} \|_2  \|{{v}}_O\|_2 \\
   &\leq C_2'
   \sqrt{|V_O|}\sqrt{\frac{\log |V_O|}{n}} \left(
   4\left(\frac{\bar{\lambda}^4}{2\zeta^3}\right)^2
   \left(     s_n C_1' \sqrt{\log(|V_O|)/n} + 3 r_{1n}\right) +\frac{\bar{\lambda}^4}{2\zeta^3}\right)\\
   & \qquad	+ 	  4\left(\frac{\bar{\lambda}^4}{2\zeta^3}\right)^2
   \left(     s_nC_1' \sqrt{\log(|V_O|)/n} + 3 r_{1n}\right)  \|{{v}}_O\|_2 \\
   &\leq C_3' \sqrt{|V_O|}  \sqrt{\frac{{\log(|V_O|)}}{n}} +
    C_4' \|{v}_O\|_2  \left(     s_n \sqrt{\log(|V_O|)/n} +  r_{1n}  \right),
   \end{aligned}
   \end{equation}
for some constants $C_3', C_4'$,
 where we use
  $ s_n  \sqrt{\log(|V_O|)/n} +  r_{1n}=o(1)$.

To complete the proof, we make use of the following lemma, whose proof is given at the end of this proof.
   \begin{lemma} \label{lem:profitDiff}
   	Suppose that $0<{p}_O^\star< \bar{p} \cdot {e}_O$. Then
   	for any other feasible price vector  $p_O$  we have
   	\begin{equation}
   	\Pi({p}_O^\star)-\Pi({p}_O) \leq \frac{1}{\zeta}  \| {p}_O^\star - {p}_O\|_2^2.
   	\end{equation}
   \end{lemma}

   Lemma \ref{lem:profitDiff}
   together with \eqref{eq:price1NormBound2} implies that
   \begin{equation}\label{eq:finalProfitBound}
   \begin{aligned}
   \Pi({{p}}^\star_O) -\Pi(\hat{{p}}_O) &\leq
   \frac{1}{\zeta} \left(C_3' \sqrt{|V_O|}  \sqrt{\frac{{\log(|V_O|)}}{n}} + C_4'  \|{v}_O\|_2 \left(     s_n \sqrt{\log(|V_O|)/n} +  r_{1n}  \right) \right)^2 \\
   &\leq
   \frac{1}{\zeta} \left( 2(C_3')^2 {|V_O|} \frac{{\log(|V_O|)}}{n}
   + 2  (C_4')^2 \|{v}_O\|_2^2 \left(     s_n \sqrt{\log(|V_O|)/n} +  r_{1n}  \right)^2
   \right)\\
   &\leq
   \frac{1}{\zeta} \left( 2(C_3')^2 {|V_O|} \frac{{\log(|V_O|)}}{n}
   + 4  (C_4')^2 \|{v}_O\|_2^2 \left(     s_n^2 {\log(|V_O|)/n} +  r_{1n}^2  \right)
   \right),
   \end{aligned}
   \end{equation}
   where we use $(x+y)^2 \leq 2 x^2 + 2 y^2$.

   On the other hand, we have
   $\Pi({p}_O^\star)=\Omega(|V_O|)$.
   This readily  follows
   since setting prices ${p}_O=\frac{\bar{p}}{2}$ guarantees
   $\Pi({p}_O) = \Omega(|V_O|)$ (and by Assumptions  \ref{ass:DiagDominance} and \ref{ass:OutsideOpt}
   we have $\bar{a} \geq  a_i>\bar{p}>0$ for all $i$).
   Moreover, it follows from Lemma \ref{lem:boundVo} that $\|{v}_O \|_2^2 \leq C'_5 |V_O|$ for some constant $C'_5>0$.
   These observations together with
   \eqref{eq:finalProfitBound} imply that
   for some constant $C_6'>0$, we have
   \[R(\hat{{p}}_O)\leq
  C_6'\left(
   s_n^2  \frac{\log(|V_O|)}{n} +  r_{1n}^2 \right)
   .\]
   Hence, the claim follows.
   \hfill
   \halmos
   \endproof	

   \noindent\proof{Proof of Lemma \ref{lem:profitDiff}.}
   By Lemma \ref{lem:interiorCons}, the consumption decisions are positive for $p_O \leq  \bar p \cdot e_O$.
Thus,
Lemma \ref{lem:firstOrderO}, implies that the expected revenues for a given price vector ${p}_O \leq \bar p \cdot e_O$ can be expressed as
   \[
   \Pi({p}_O)= \mathbb{E} [ \langle {p}_O, {y}_O({p}_O)\rangle]=
   {p}_O^T {v}_O- \frac{1}{2} {p}_O^T(H^{-1}+H^{-T}) {p}_O,
   \]
and \eqref{eq:sellerOpt} can be rewritten as follows:
   \begin{equation}\label{eq:sellerOpt2}
   	\begin{aligned}
   		\max_{ 0\leq  {p}_O \leq \bar{p} \cdot {e}_O} \qquad & {p}_O^T {v}_O- \frac{1}{2} {p}_O^T(H^{-1}+H^{-T}) {p}_O\\ %
   	\end{aligned}
   \end{equation}
By the assumption of the lemma, the constraints are not binding. Since $(H^{-1}+H^{-T}) $
is positive definite (Lemma \ref{lem:MInv})
the objective is concave, which means that $p_O^\star$ solves the optimization problem obtained after relaxing the constraints, i.e.,
   \begin{equation}\label{eq:sellerOpt3}
	\begin{aligned}
		\max_{   {p}_O } \qquad & {p}_O^T {v}_O- \frac{1}{2} {p}_O^T(H^{-1}+H^{-T}) {p}_O.\\ %
	\end{aligned}
\end{equation}
   Since revenues are quadratic in prices, second-order Taylor expansion (around ${p}^\star_O$) is exact. Thus, we have
   \begin{equation}
   \begin{aligned}
   \Pi({p}_O) &=  \Pi({p}^\star_O) + ({p}_O - {p}^\star_O)^T \nabla \Pi({p}^\star_O) +
   ({p}_O - {p}^\star_O)^T \frac{\nabla^2 \Pi({p}^\star_O)}{2}
   ({p}_O - {p}^\star_O) \\
   &= \Pi({p}^\star_O) -
   ({p}_O - {p}^\star_O)^T \frac{(H^{-1}+H^{-T}) }{2}
   ({p}_O - {p}^\star_O),
   \end{aligned}
   \end{equation}
   where we use the optimality condition $\nabla \Pi({p}^\star_O)=0$ (which holds since ${p}^\star_O$ is optimal in the unconstrained optimization problem).
   Rearranging terms, this implies that
   \[
   |\Pi({p}_O^\star) - \Pi({p}_O)|\leq
   \left| ({p}_O - {p}^\star_O)^T \frac{(H^{-1}+H^{-T}) }{2}
   ({p}_O - {p}^\star_O) \right|
   \leq \frac{1}{2}  \| H^{-1}+H^{-T} \|_2 \|{p}_O - {p}^\star_O\|_2^2.
   \]

   By Lemma   \ref{lem:matBounds}
   $||(H^{-1}+H^{-T})||_2\leq
   2||H^{-1}||_2 \leq \frac{2}{\zeta }$, which together with the previous inequality immediately implies the desired result.
   \hfill
   \halmos
   \endproof

\section{Proofs of Section \ref{se:estimationExamples}} \label{ref:SecEstEx}
\noindent\proof{Proof of Lemma \ref{lem:m-banded}.}
Let $M^\ell$ be a matrix such that $(M^{\ell})_{\ell(i),\ell(j)}=M_{ij}$ for all $i,j\in V$. Observe that by definition $M^\ell$ is $m$-banded.
Moreover, it can be seen that
$M^\ell= \tilde{P}^T M \tilde{P}$ for some permutation matrix $\tilde{P}$.
This implies that $(M^\ell)^{-1}=  \tilde{P}^T M^{-1} \tilde{P}$, and hence $(M^\ell)^{-1}_{\ell(i),\ell(j)}=(M^{-1})_{ij}$ for all $i,j \in V$.

Let $\sigma_{\max}(A)$ and $\sigma_{\min}(A)$ denote respectively the largest and smallest singular values of some matrix $A$.
Using Assumption \ref{ass:DiagDominance}, it is possible to bound singular values of $A=M M^T$. In particular, we have
\[
\sigma_{\max}(M M^T) \leq \| M M^T \|_2 \leq \| M\|_2^2 \leq (2 \bar{\lambda}  - \zeta)^2,
\]
where the last inequality follows from Lemma \ref{lem:matBounds} of Appendix \ref{subse:MatrixLemma}.
In addition, observing
$\sigma_{\min} (M M^T) = 1/\sigma_{\max}( (M M^T)^{-1}  ) = 1/ \sigma_{\max}( M^{-T} M^{-1})$, we have
\[
\sigma_{\min} (M M^T) = 1/ \sigma_{\max}( M^{-T} M^{-1}) =
1/ \| M^{-T} M^{-1} \|_2 \geq 1/   \| M^{-1} \|_2^2 \geq
\zeta^2,
\]
where once again the last inequality follows from Lemma \ref{lem:matBounds}.
Let $u=(2 \bar{\lambda}  - \zeta)^2$ and $l=\zeta^2$ denote the upper and lower bounds on  	$\sigma_{\max}(M M^T) $ and $	\sigma_{\min}(M M^T) $ respectively, and let
\begin{equation} \label{eq:ratio}
r=u/l =  \left( \frac{2 \bar{\lambda}  - \zeta}{\zeta} \right)^2
\end{equation}
denote their ratio.
Observe that $ (M^\ell) (M^\ell)^T =\tilde{P}^T M M^T \tilde{P}$. Since
$\tilde{P}$ is a unitary matrix
($\tilde{P} \tilde{P}^T=I$), it follows that
$(M^\ell) (M^\ell)^T$ and $M M^T$ share the same singular values, and hence the aforementioned singular value bounds   also apply to $(M^\ell) (M^\ell)^T$.

Proposition 2.3 in \cite{demko1984decay} implies that
when $M^\ell$ is $m$-banded and satisfies the aforementioned singular value bounds, the entries of
$(M^\ell)^{-1}$  exhibit exponential decay as we get away from the diagonal. That is, we have
\begin{equation}\label{eq:m-banded}
|(M^{\ell})^{-1}_{ij}|
\leq C' \lambda_1^{|i-j|}.
\end{equation}
Here
\[
\lambda_1 =\left( \frac{\sqrt{r}-1}{\sqrt{r}+1}\right)^{1/m} = \left(\frac{\frac{2 \bar{\lambda}  - \zeta}{\zeta} -1}{ \frac{2 \bar{\lambda}  - \zeta}{\zeta}+1}\right)^{1/m} = \left(\frac{ \bar{\lambda}-\zeta}{ \bar{\lambda}}
\right)^{1/m},
\]
where the last quantity is strictly positive by Assumption \ref{ass:DiagDominance} as long as the set of edges in the network is nonempty.
In addition,
\begin{equation} \label{eq:upBC0}
\begin{aligned}
C'&= (m+1) \|M^\ell\|_2 \lambda_1^{-m} \max \left\{\frac{1}{l}, \frac{(1+\sqrt{r})^2}{2lr}\right\} \\
&\leq
(m+1) \|M\|_2
\left(\frac{ \bar{\lambda}}{ \bar{\lambda}-\zeta}
\right)
\max \left\{\frac{1}{l}, \frac{(1+\frac{2 \bar{\lambda}  - \zeta}{\zeta})^2}{2u} \right\} \\
&\leq
(m+1) (2 \bar{\lambda}  - \zeta)
\left(\frac{ \bar{\lambda}}{ \bar{\lambda}-\zeta}
\right) \max\left\{\frac{1}{\zeta^2}, \frac{(\frac{2 \bar{\lambda}}{\zeta})^2}{2(2 \bar{\lambda}  - \zeta)^2} \right\}\\
&		
\leq
(m+1) (2 \bar{\lambda}  - \zeta)
\left(\frac{ \bar{\lambda}}{\bar{\lambda}-\zeta}
\right) \max \left\{\frac{1}{\zeta^2}, \frac{2}{ \zeta^2} \right\}
=
2(m+1) \frac{ \bar{\lambda}(2 \bar{\lambda}  - \zeta)}{\zeta^2 ( \bar{\lambda}-\zeta)} =C_1	,
\end{aligned}
\end{equation}
where in the
first inequality we use the fact that $\|M^\ell \|_2 = \|\tilde{P}^T M \tilde{P}\|_2 \leq \|\tilde{P}\|_2^2 \|M\|_2 \leq \|M\|_2$,
in the second inequality we use Lemma \ref{lem:matBounds}, and in the last inequality we use the fact that $2\bar{\lambda}-\zeta \geq  \bar{\lambda}$, which follows from Assumption \ref{ass:DiagDominance}.

Note that \eqref{eq:m-banded} implies that
\[
|M^{-1}_{ij}| = |(M^{\ell})^{-1}_{\ell(i),\ell(j)}|  \leq C' \lambda_1^{|\ell(i)-\ell(j)|}.
\]
Substituting the upper bound of $C'$ from \eqref{eq:upBC0} and the expression for $\lambda_1$ in
this inequality, we establish the first part of   the claim.
Note that by
Lemma \ref{lem:MInv0} of Appendix \ref{subse:MatrixLemma},
$H^{-1} = [M^{-1}]_{OO}$.
Thus, the bound on $H^{-1}_{ij}$ also follows from the bound on $M^{-1}_{ij}$.
\hfill	\Halmos
\endproof

\noindent\proof{Proof of Lemma \ref{lem:decayMbanded}.}	
Observe that  Lemma \ref{lem:m-banded} implies that
\begin{equation} \label{eq:boundInv_mBanded}
\begin{aligned}
\| \bar{W} - H^{-1} \|_\infty &= \max_{k\in V_O}
\sum_{j : |\ell(k)-\ell(j)|> (s_n-1) /2}|H^{-1}_{kj}|
\leq 2\sum_{j= \lfloor s_n/2 \rfloor }^{\infty} \tilde{C}_1 \lambda_1^{j} \leq \frac{2 \tilde{C}_1 \lambda_1^{\lfloor s_n/2 \rfloor}}{1-\lambda_1} \leq \frac{2 \tilde{C}_1 \lambda_1^{s_n/2}}{\lambda_1(1-\lambda_1)} \\
\| \bar{W} - H^{-1} \|_1 &= \max_{k\in V_O}
\sum_{j : |\ell(k)-\ell(j)|> (s_n-1)/2}|H^{-1}_{jk}| \leq 2\sum_{j= \lfloor s_n/2 \rfloor }^{\infty} \tilde{C}_1 \lambda_1^{j} \leq \frac{2 \tilde{C}_1 \lambda_1^{\lfloor s_n/2 \rfloor}}{1-\lambda_1} \leq \frac{2 \tilde{C}_1 \lambda_1^{s_n/2}}{\lambda_1(1-\lambda_1)} \\
\end{aligned}
\end{equation}
for all $k \in V_O$.
Here, we use the fact that summation over $j$ such that $|\ell(k)-\ell(j)| > (s_n-1)/2$
is smaller than summation over $j$ such that $|\ell(k)-\ell(j)| \geq (s_n-1)/2$.
The latter is equal to summation over $j$ such that
$|\ell(k)-\ell(j)| \geq \lceil (s_n-1)/2 \rceil $, and
$\lceil (s_n-1)/2 \rceil\geq \lfloor s_n/2  \rfloor$.

Let
$\tilde{C}_2>0$ be a constant such that
$$s_n=\tilde{C}_2 \max\left\{m,  \log {n}  \right\}
\geq  \frac{2}{\log \lambda_1^{-1}} \log \left( \frac{2\tilde C_1 n}{\lambda_1(1-\lambda_1)} \right)  . $$
For such  $\tilde{C}_2$, we obtain
\[
\frac{2 \lambda_1^{s_n/2}}{\lambda_1(1-\lambda_1)}
\leq
\frac{2}{\lambda_1(1-\lambda_1)}  \frac{\lambda_1 (1-\lambda_1)}{2 \tilde{C}_1 n} \leq \frac{1}{\tilde{C}_1 n}.%
\]
These inequalities, together with \eqref{eq:boundInv_mBanded},
imply that
$\bar{W}$ is an $(s_n,r_{1n})$-sparse approximation
of  $H^{-1}$
for
$r_{1n}=1/n$.
\hfill
\halmos
\endproof

\noindent\proof{Proof of Lemma \ref{lem:jaffardDecay}.}
Recall that $M=\Lambda-G$, $G_{ij}\geq 0$, and $\Lambda_{ii}=\lambda_i \leq \bar{\lambda}$.
Thus,
using this observation with \eqref{eq:decayCond}, and
choosing $C_0'=\max\{\tilde{C}, \bar \lambda\}$, we conclude that
\begin{equation}\label{eq:decayPolM_result}
|M|_{ij}\leq  \frac{ C_0'}{(1+ |\ell(i)-\ell(j)|)^\theta}.
\end{equation}

As before,
let $M^\ell$ be a matrix such that $(M^{\ell})_{\ell(i),\ell(j)}=M_{ij}$ for all $i,j\in V$.
Observe that
\eqref{eq:decayPolM_result} implies that
\begin{equation}\label{eq:decayPolM_result2}
|M^\ell_{ij}|\leq  \frac{ C_0'}{(1+ |i-j|)^\theta},
\end{equation}
for any $i,j$.
Also recall that
$M^\ell= \tilde{P}^T M \tilde{P}$ for some permutation matrix $\tilde{P}$, and
$(M^\ell)^{-1}=  \tilde{P}^T M^{-1} \tilde{P}$.
This in turn implies that   $(M^\ell)^{-1}_{\ell(i),\ell(j)}=M^{-1}_{ij}$ for all $i,j \in V$.

By Lemma \ref{lem:jaffardDecayOrig} this implies that
\begin{equation}\label{eq:boundMInvDecay}
|(M^\ell)^{-1}_{ij}|\leq  C_1'/(1+ |i-j|)^\theta,
\end{equation}
where
$C_1'$ depends on $C_0'$, $\theta$, and $\|(M^{\ell})^{-1}\|_2$.
On the other hand, by Lemma \ref{lem:matBounds},
we have
\[
\|(M^{\ell})^{-1}\|_2 =
\|\tilde{P}^T M^{-1} \tilde{P}\|_2
\leq \|M^{-1}\|_2  \leq 1/\zeta,
\]
where we use the fact that $\tilde{P}$ is a permutation matrix and hence $\|\tilde{P}\|_2=1$.
Thus, recalling that $C_0'=\max\{\tilde{C}, \bar \lambda \}$, we conclude that
$C_1'$ depends only on $\tilde{C},\bar{\lambda}, \theta, \zeta$.

Thus, using \eqref{eq:boundMInvDecay}, we immediately obtain
that for $i,j\in V_O$,
\begin{equation}\label{eq:tempMvsH}
|M^{-1}_{ij} |= |(M^\ell)^{-1}_{\ell(i),\ell(j)} |\leq C_1'/(1+ |\ell(i)-\ell(j)|)^\theta.
\end{equation}
Moreover,
by Lemma \ref{lem:MInv0}, $H^{-1}$ is a diagonal subblock of $M^{-1}$.
Hence,  for $i,j\in V_O$ we also have $|H^{-1}_{ij}| =|M^{-1}_{ij} |$.
This observation, together with \eqref{eq:tempMvsH}, implies that
the matrix $H^{-1}$ itself exhibits the polynomial decay property (with identical parameters to $M^{-1}$).
\hfill \halmos

\noindent\proof{Proof of Lemma \ref{lem:polyDecay}.}
Using
Lemma \ref{lem:jaffardDecay}, we obtain
\begin{equation*}\label{eq:decayBoundInv}
\begin{aligned}
\| \bar{W} - H^{-1} \|_\infty &= \max_{k\in V_O}
\sum_{j : |\ell(k)-\ell(j)|> (s_n-1)/2}|H^{-1}_{kj}|
\leq \max_{k\in V_O}
\sum_{j : |\ell(k)-\ell(j)|\geq \lceil (s_n-1)/2 \rceil}|H^{-1}_{kj}|
\leq 2\sum_{j= \lceil (s_n-1)/2 \rceil }^{\infty} \frac{C_1}{(1+j)^{\theta}}
\\
&= 2{\tilde{C}_1} \sum_{j= \lceil (s_n-1)/2 \rceil+1 }^{\infty} {j^{-\theta}}	\leq 	2\tilde{C}_1 \int_{  s_n/2 }^\infty x^{-\theta}dx	
\\
&\leq
2\frac{\tilde{C}_1}{\theta-1}  \left(  \frac{s_n}{2}\right)^{-\theta+1}
=
\frac{\tilde{C}_1 2^\theta }{\theta-1}   {s_n}^{-\theta+1}
.
\end{aligned}
\end{equation*}
Here, the second line uses the fact that
$\left(\lceil (s_n-1)/2 \rceil+1 \right) - \left( s_n/2  \right) \geq 1/2$, and
$\ell^{-\theta}\leq   \int_{\ell-1/2}^{\ell+1/2} x^{-\theta} dx$  for any $\ell\geq 1$ since $x^{-\theta}$ is convex in $x \geq 0$.

Since the column and row entries
of $H^{-1}$
decay at the same rate under the assumptions of the lemma, we also obtain
\begin{equation}\label{eq:decayBoundInv2}
\begin{aligned}
\| \bar{W} - H^{-1} \|_1 & \leq
\frac{\tilde{C}_1 2^\theta }{\theta-1}   {s_n}^{-\theta+1}.
\end{aligned}
\end{equation}

From
\eqref{eq:decayBoundInv2}
it follows that
$\bar W$ is an
$(s_n,r_{1n})$-sparse approximation
of  $H^{-1}$ where
\begin{equation} \label{eq:rBoundsDecay}
\begin{aligned}
r_{1n} &\leq \frac{\tilde{C}_1 2^\theta }{\theta-1}   {s_n}^{-\theta+1} .
\end{aligned}
\end{equation}
Let $s_n=   \{n/\log(V_O)\}^{1/2\theta}  $ as assumed in the statement of the lemma.
Then \eqref{eq:rBoundsDecay} yields
\begin{equation} \label{eq:rBoundsDecay2}
\begin{aligned}
r_{1n} &\leq \frac{\tilde{C}_1 2^\theta }{\theta-1}
\left(\frac{\log |V_O| }{n}\right)^{(\theta-1)/2\theta},
\end{aligned}
\end{equation}
and the claim follows. \hfill \halmos
\endproof

\noindent\proof{Proof of Lemma \ref{lem:appxDiagonalizable}.}

(i)
Let
$
k = \lceil \log_{d_e/q}(\sqrt{n|V_O|}/\log(|V_O|)) \rceil,
$
and recall that
$\bar W$ is the $|V_O| \times |V_O|$ submatrix
of $g^\star_k(M)$.
Observe that any matrix polynomial of degree $k$ is such that its $(i,j)$th entry is nonzero only if $\rho(i,j)\leq k$.
By the exponential growth assumption, the number of nonzero entries in each row/column
of $\bar W$ is bounded by
\begin{equation*}
\begin{aligned}
C_e d_e^{ \lceil \log_{d_e/q}(\sqrt{n|V_O|} /\log(|V_O|)) \rceil}
&\leq
C_e d_e d_e^{ \log_{d_e/q}( \sqrt{n|V_O|} /\log(|V_O|))  }
=
C_e d_e
\left(\frac{ \sqrt{n|V_O|}}{\log(|V_O|)} \right)^{\log_{d_e/q}(d_e)}.
\end{aligned}
\end{equation*}

Using the shorthand notation
$\nu=\log_{d_e/q}(d_e)$,
these observations collectively imply that
$\bar W$ is
$s_n= C_e d_e
\left(\frac{ \sqrt{n|V_O|}}{\log(|V_O|)} \right)^{\nu}
$-sparse.

Observe that
using the aforementioned choice of $k$, we have
\begin{equation*}
\begin{aligned}
q^k \leq
q^{\log_{d_e/q}(\sqrt{ n|V_O|}/\log(|V_O|)) }
&=
\left( \frac{ \sqrt{n|V_O|}}{\log(|V_O|)} \right)^{\log_{d_e/q}(q)}
=
\left( \frac{\log(|V_O|)}{ \sqrt{n|V_O|}} \right)^{\log_{d_e/q}(1/q)}\\
& =
\left( \frac{\log(|V_O|)}{ \sqrt{n|V_O|}} \right)^{1-\log_{d_e/q}(d_e)}
=
\left( \frac{\log(|V_O|)}{ \sqrt{n|V_O|}} \right)^{1-\nu}.
\end{aligned}
\end{equation*}
By \eqref{eq:polyApproxHinv}
we have
\begin{equation}\label{eq:bound2NormHinvFin}
\| H^{-1}_{\ell ,\rowdot}-\bar{W}_{\ell,\rowdot} \|_2,
\| H^{-1}_{\rowdot,\ell}-\bar{W}_{\rowdot,\ell} \|_2
\leq  \|
H^{-1} - \bar{W}\|_2
\leq
\tilde{C}
q^k
\leq
\tilde{C}
\left( \frac{\log(|V_O|)}{ \sqrt{n|V_O|}} \right)^{1-\nu}.
\end{equation}
Note that by \eqref{eq:bound2NormHinvFin}, we obtain
\begin{equation}\label{eq:boundWHinv}
\begin{aligned}
\| \bar{W}  - H^{-1}\|_\infty
&= \max_\ell \| \bar{W}_{\ell ,\rowdot}  - H_{\ell ,\rowdot}^{-1}\|_1
\leq  {|V_O|}^{1/2}\max_\ell \| \bar{W}_{\ell ,\rowdot}  - H_{\ell ,\rowdot}^{-1}\|_2
\leq  {|V_O|}^{1/2} \| \bar{W}  - H^{-1}\|_2\\
&\leq \tilde{C} {|V_O|}^{1/2}
\left( \frac{\log(|V_O|)}{ \sqrt{n|V_O|}} \right)^{1-\nu}
=
\tilde{C}
{|V_O|}^{\nu/2}
\left( \frac{\log(|V_O|)}{\sqrt{n} } \right)^{1-\nu}
.
\end{aligned}
\end{equation}
Using the fact that $\|A\|_1=\|A^T\|_\infty$ for any matrix $A$, and
proceeding similarly, it also follows that
$\| \bar{W}  - H^{-1}\|_1 \leq  \tilde{C}
{|V_O|}^{\nu/2}
\left( \frac{\log(|V_O|)}{\sqrt{n} } \right)^{1-\nu}$.
These  inequalities  imply that
$$r_{1n}=
\tilde{C}{|V_O|}^{\nu/2}
\left( \frac{\log(|V_O|)}{\sqrt{n} } \right)^{1-\nu}
.$$
Hence, the claim follows.

(ii)
Let
\[
k =
\left\lfloor
\sqrt[d_\star]{ {(n/\log(|V_O|))} }
\right\rfloor,
\]
for some $d_\star \geq 0$, which we specify later.
Once again recalling that
$\bar W$ is the $|V_O| \times |V_O|$ submatrix
of $g^\star_k(M)$, and noting that  any matrix polynomial of degree $k$ is such that its $(i,j)$th entry is nonzero only if $\rho(i,j)\leq k$, we can bound the  number of nonzero entries in each row/column
of $\bar W$  by
\[
C_p \left(\frac{n}{\log(|V_O|)} \right)^{d_p/d_\star}
.
\]

Observe that
using the aforementioned choice of $k$, we have
\[
q^k \leq
q^{\sqrt[d_\star]{ {(n/\log(|V_O|))} }-1}.
\]

From \eqref{eq:polyApproxHinv}
we have
\begin{equation}\label{eq:bound2NormHinvFinNew2}
\| H^{-1}_{\ell ,\rowdot}-\bar{W}_{\ell,\rowdot} \|_2,
\| H^{-1}_{\rowdot,\ell}-\bar{W}_{\rowdot,\ell} \|_2
\leq  \|
H^{-1} - \bar{W}\|_2
\leq
\tilde{C}
q^k
\leq
\tilde{C}
q^{\sqrt[d_\star]{ {(n/\log(|V_O|))} }-1}.
\end{equation}

Finally,
as before, we have
$\| \bar{W}  - H^{-1}\|_\infty=\max_\ell \| \bar{W}_{\ell ,\rowdot}  - H_{\ell ,\rowdot}^{-1}\|_1  \leq  \sqrt{V_O} \| \bar{W}  - H^{-1}\|_2 \leq \tilde{C} \sqrt{|V_O|}
q^{\sqrt[d_\star]{ {(n/\log(|V_O|))} }-1}
$, where we use \eqref{eq:bound2NormHinvFinNew2}.
Proceeding similarly, it also follows that
$\| \bar{W}  - H^{-1}\|_1 \leq
\tilde{C} \sqrt{|V_O|}
q^{\sqrt[d_\star]{ {(n/\log(|V_O|))} }-1}$.
These  inequalities  imply that
$r_{1n}=
\tilde{C}
{\sqrt{|V_O|}}
q^{\sqrt[d_\star]{ {(n/\log(|V_O|))} }-1}
$.
The claim follows by choosing $d_\star=4 d_p$.
\hfill \halmos \endproof

\section{Simultaneous Confidence Intervals via Bootstrap and Setting Threshold Parameters}\label{Sec:SimultaneousBootstrap}

In this section, we discuss the construction of simultaneous confidence intervals for all $|V_O|^2$ coefficients of the $H^{-1}$ matrix. This is of interest in itself, as it allows for  controlling the noise in  estimation,
	which sheds light on the  impact
	that a change in the covariate of
	one agent has on the outcome of another agent.
	 Moreover, the same construction will allow us to provide data-driven choices for the threshold values $\mu_{kj}, k,j \in V_O$ used in Algorithm \ref{alg:Alg1}.

	The construction of such simultaneous confidence intervals with asymptotic exact coverage is nonstandard as (i) there are model selection mistakes in the regularized estimators used in Algorithm \ref{alg:Alg1}, and (ii) the number of coefficients is potentially much larger than the sample size. In particular, the latter implies that as the sample size grows so does the dimension of the estimation errors and hence there is no limiting distribution. Instead, we construct a sequence of Gaussian processes that provide a suitable approximation of the maximum estimation error via a multiplier bootstrap approximation. Although the maximum error is a scalar, it is not clear that there is a limiting distribution. However, our approximations hold for each $n$ with errors vanishing sufficiently fast so that the distortions on the coverage of the confidence regions vanish too. This result relies on the recent central limit theorems for the maximum of high-dimensional vectors developed in \cite{chernozhukov2012gaussian,chernozhukov2012comparison,chernozhukov2013gaussian,chernozhukov2014honestbands}.

		We start by stating an additional assumption on which our analysis in this section relies:
	
	\begin{assumption}\label{assumption:HL:Psi}
		With probability $1-o(1)$ we have that
		$ \max_{k \in \bar{V}_O } \|
		\hPkdot  -
		\bPkdot\|_1 \leq
		d_n $, for some (fixed) sequence $d_n \to 0$.
	\end{assumption}

		Assumption \ref{assumption:HL:Psi} is implied by standard assumptions in the high-dimensional statistics literature. Indeed, since for each $k\in V_O$, $\hPkdot$ is estimated via an $\ell_1$-regularized procedure, it suffices to have  $\bPkdot$  approximately sparse and the expected value of the design matrix associated with the covariate vector to have eigenvalues  bounded away from zero and from above; see, e.g., \cite{BickelRitovTsybakov2009,BelloniChernozhukovHansen2011,c.h.zhang:s.zhang,GRD2014,javanmard2014confidence}. It is plausible that we can relax this condition as   done in the analysis of Theorem \ref{cor:Thm1}.

	Our starting point is the (approximate) linear representation established in Theorem \ref{thm:OrthoHinv}, namely,
	\begin{equation}\label{eq:MAM} \sqrt{n}\{ \check W^T - H^{-T}\} =  -
	\frac{1}{\sqrt{n}}\sum_{t=1}^n \hat \Psi_{V_O,\cdot} (1;{p}_O^{(t)})(\varepsilon^{(t)})^T  +
	\tilde R_n,  \end{equation}
	where $\tilde R_n$ collects all the approximation errors except in the first row of $[R_n^1, \cdots, R_n^{|V_O|}]$,
	and
	$\hat \Psi_{V_O,\cdot}$ is the submatrix of
	$\hat \Psi$ obtained by restricting attention to rows $\{\hat \Psi_{k,\cdot}\}_{k\in V_O}$
	. We will approximate ${\rm cv}_{(1-\alpha)}$, i.e., the $(1-\alpha)$-quantile of the random variable
	$$ \max_{k,j\in V_O} \left|  \frac{1}{\sqrt{n}}\sum_{t=1}^n \frac{\hat \Psi_{k,\cdot}(1;{p}^{(t)})\varepsilon_j^{(t)}}{\sigma_{kj}} \right|,$$
	where $\sigma_{kj}^2 = \frac{1}{n}\sum_{t=1}^n E[\{\bar \Psi_{k,\cdot}(1;{p}_O^{(t)})\varepsilon_j^{(t)}\}^2]$. The confidence intervals  will then ensure that with provability converging to $1-\alpha$, simultaneously over $k,j\in V_O$, we have \begin{equation}\label{def:SCI} \check W_{kj} - {\rm cv}_{(1-\alpha)} \frac{\sigma_{kj}}{\sqrt{n}} \leq H^{-1}_{kj} \leq   \check W_{kj} + {\rm cv}_{(1-\alpha)} \frac{\sigma_{kj}}{\sqrt{n}}, \end{equation}
	provided that the approximation errors $\tilde R_n$ do not impact the coverage.\footnote{It follows that a sufficient condition for that is $\max_{k,j\in V_O}|(\tilde R_{n})_{kj}| = o(\log^{-1/2} |V_O|)$, which is implied by our conditions and Theorem \ref{thm:OrthoHinv}.}
	
	To approximate the quantile ${\rm cv}_{(1-\alpha)}$ and obtain the correct coverage one needs to account for the correlation structure. This is done through the use of a multiplier bootstrap procedure conditional on the data as follows.
For each $k,j\in V_O$, define
	$$ T_{kj} := \frac{1}{\sqrt{n}}\sum_{i=1}^n \hat\xi^{(t)} \frac{\hat \Psi_{k,\cdot}(1;p_O^{(t)})\{ y_j^{(t)} - (\hat v_j - \hat W_{j,\cdot} p_O^{(t)})\}}{\hat\sigma_{kj}} $$
	where $\hat\sigma_{kj}^2 = \frac{1}{n}\sum_{t=1}^n \{\hat \Psi_{k,\cdot}(1;{p}_O^{(t)})[ y_j^{(t)} - (\hat v_j - \hat W_{j,\cdot} p_O^{(t)})]\}^2$, and $\hat\xi^{(t)}$ are i.i.d. standard normal random variables independent of the data.  The associated critical value we are interested in is
	$$ {\rm cv}^*_{(1-\alpha)} = \mbox{conditional} \  (1-\alpha)\mbox{-quantile of} \ \max_{k,j\in V_O} |T_{kj} | \ \ \mbox{given the data}.$$

	The next result shows that despite the high dimensionality, the Gaussian multiplier bootstrap can be used to approximate the quantiles ${\rm cv}_{(1-\alpha)}$ of the distribution and obtain simultaneous confidence intervals of the form of (\ref{def:SCI}). To state the result concisely we let $B_n \geq 1$ be a parameter such that
	\begin{equation}
	B_n^4 \geq E[\max_{k,j\in [V_O]} |\bar \Psi_{k\cdot}(1;p_O^{(t)})\epsilon_j^{(t)}|^4 ].
	\end{equation}
In the proof of our result, we
 	leverage new central limit theorems
 	of \cite{chernozhukov2013gaussian,chernozhukov2013testing,chernozhukov2014honestbands,chernozhukov2012gaussian,chernozhukov2012comparison}
 	 for the maximum of the entries of the average of high-dimensional vectors, and the linear
 	  representation derived in Theorem \ref{thm:OrthoHinv}.
This allows us to construct  confidence intervals that are simultaneously valid even if the number of components exceeds the sample size.
 	  In what follows,
 $\{\delta_n \}$ denotes a fixed sequence satisfying
 $\delta_n\to 0$.

	\begin{theorem}\label{thm:inferenceAlg1}
		In addition to Assumptions \ref{assumption:basicRandom} and \ref{assumption:HL:Psi}, suppose that $d_n\log(V_On) + n^{-1/2}s_n\log^{3/2}(V_O)+r_{1n}\log(nV_O)\leq \delta_n$ and  $\{n^{-1}B_n^4\log^7(n|V_O|)\}^{1/6}\leq \delta_n$.
		Then, the critical value ${\rm cv}^*_{(1-\alpha)}$ computed via the multiplier bootstrap procedure satisfies
		$$ \lim_{n\to \infty} \left|P \left( \check W_{kj} - \frac{{\rm cv}^*_{(1-\alpha)} \hat\sigma_{kj}}{\sqrt{n}}\leq H^{-1}_{kj} \leq \check W_{kj} + \frac{{\rm cv}^*_{(1-\alpha)} \hat \sigma_{kj}}{\sqrt{n}}, \ \ \mbox{for all} \  k,j\in V_O\right) - (1-\alpha)\right| = 0.$$
		Therefore, setting $\mu_{kj} = 2{\rm cv}^*_{(1-\alpha)}\hat\sigma_{kj}/\sqrt{n}$, for $k,j\in V_O$, with probability converging to $1-\alpha$ we have
		$$ \mu_{kj} \geq 2|(\check W - H^{-1})_{kj}| \ \  \mbox{simultaneously over} \ k,j\in V_O,$$
		as required in Theorem \ref{thm:OrthoHinv}(ii).
	\end{theorem}

	\proof{Proof of Theorem \ref{thm:inferenceAlg1}.}
	By Theorem \ref{thm:OrthoHinv} we have that (\ref{eq:MAM}) holds. This corresponds to the many approximate means setting considered in \cite{belloni2018high} with $p=|V_O|^2$. It will  be convenient for us to associate each $j'\in[p]$ with $(k,j)\in V_O\times V_O$, where $j'=(k-1)V_O+j$. We will verify Conditions M, E, and A in \cite{belloni2018high} (which collectively imply Condition W in \cite{belloni2018high} using the choice of weights $w_{j'}=1/\sigma_{kj}$ and $\hat w_{j'}=1/\hat\sigma_{kj}$) and invoke Theorem 2.6 in \cite{belloni2018high} to establish the claims.
	
	Define $Z_{tj'} = \bar\Psi_{k\cdot}(1;p_O^{(t)})\varepsilon_j^{(t)}$ and $\hat Z_{tj'} = \hat\Psi_{k\cdot}(1;p_O^{(t)})\{y_j-(\hat v_j-\hat W_{j\cdot}p_O^{(t)})\}$
	for $j'\in [p]$.
	Note that
	Condition E of \cite{belloni2018high} immediately holds by our assumptions on $B_n$.
	
	To verify Condition M of \cite{belloni2018high}, note that
	$$ \begin{array}{rl}
	\frac{1}{n}\sum_{t=1}^n E[Z_{tj'}^2] & =\sigma_{kj}^2 =  E[ \{\bar\Psi_{k\cdot}(1;p_O^{(t)})\}^2E[ \varepsilon_j^2 \mid p_O^{(t)}]] \\
		& \geq c E[ \{\bar\Psi_{k\cdot}(1;p_O^{(t)})\}^2]\\
	&=c
	\bar\Psi_{k,\cdot} \bar{ \Psi} ^{-1} \bar\Psi_{k,\cdot}^T =
	\bar \Psi_{k,k}\\

	& \geq c/\bar p^2,
	\end{array}$$
	where we used Assumption \ref{assumption:basicRandom}(i) in the second line, and
	the fact that $\bar\Psi=E[(1;p_O^{(t)})(1;p_O^{(t)})^T]^{-1}$in the third and fourth lines.
Thus $\sigma_{kj}^2=\frac{1}{n}\sum_{t=1}^n E[Z_{tj'}^2]$ are bounded away from zero, satisfying
(after appropriate normalization)
the first requirement of Condition M.
	
	The second part of Condition M requires bounds on the third and fourth moments of $Z_{tj'}$. For the third moment we have for all $j'\in[p]$ that $$ \begin{array}{rl}
	\frac{1}{n}\sum_{t=1}^n E[|Z_{tj'}|^3] & =  E[ |\bar\Psi_{k\cdot}(1;p_O^{(t)})|^3E[ |\varepsilon_j|^3 \mid p_O^{(t)}]] \leq C_1',\\
	\end{array}$$
	for some constant $C_1'>0$
	under Assumptions \ref{assumption:basicRandom}(i)--(iii).
	Similarly, for the fourth moments we have $$ \begin{array}{rl}
	\frac{1}{n}\sum_{t=1}^n E[|Z_{tj'}|^4] & \leq  B_n^4,\\
	\end{array}$$
by definition of $B_n$.
These observations imply that the second part of condition M also follows.
	
	Next we verify Condition A. We rewrite (\ref{eq:MAM}) to obtain
	\begin{equation}\label{eq:MAMprime} \sqrt{n}\{ \check W^T - H^{-T}\} = - \frac{1}{\sqrt{n}}\sum_{t=1}^n \bar \Psi_{V_O,\cdot} (1;{p}_O^{(t)})(\varepsilon^{(t)})^T  + \tilde R_n',  \end{equation}
	where $\tilde R_n' = \tilde{R}_n + (\bar \Psi_{V_O,\cdot}-\hat \Psi_{V_O,\cdot}) \frac{1}{\sqrt{n}}\sum_{t=1}^n (1;{p}_O^{(t)})(\varepsilon^{(t)})^T $. Note that with probability $1-o(1)$ we have
	$$
	\begin{array}{rl}
	\max_{k,j\in V_O}| (\tilde R_{n}')_{{kj}}| & \leq \max_{k,j\in V_O}|(\tilde R_{n})_{kj}| + \max_{k\in V_O}\|\bar \Psi_{k,\cdot}-\hat\Psi_{k,\cdot}\|_1 \max_{j\in V_O}\left\| \frac{1}{\sqrt{n}}\sum_{t=1}^n (1;{p}_O^{(t)})\varepsilon_j^{(t)} \right\|_\infty\\
	& \leq C_2'\{n^{-1/2}s_n\log |V_O| + r_{1n}\sqrt{\log |V_O|}\}  + C_2'd_n \sqrt{\log (V_O)},\\
	\end{array}
	$$
	for some constant $C_2'>0$.
 Here we apply Theorem \ref{thm:OrthoHinv} to control $\tilde R_n$.
To control the second term we apply Holder's inequality, and use Lemma \ref{lemma:auxBoundsLinfty} and Assumption \ref{assumption:HL:Psi} to bound each term.
 Using the
 assumptions of the theorem on $s_n$, $r_{1n}$, $d_n$, and $\delta_n$ we have that with probability $1-o(1)$
	$$
	\begin{array}{rl}
	\max_{k,j\in V_O}|(\tilde R_{n}')_{kj}| & \leq C_3' \delta_n  \log^{-1/2}(V_On),\\
	\end{array}
	$$
	for some constant $C_3'>0$,
	which establishes the first part of Condition A.
	
	To show the second part of Condition A, note that
	$$
	\begin{array}{l}
	{\displaystyle\max_{j'\in[p]}}\frac{1}{n}\sum_{t=1}^n(\hat Z_{tj'}-Z_{tj'})^2 \leq 2\max_{k,j\in V_O} \frac{1}{n}\sum_{t=1}^n (\{\hat\Psi_{k,\cdot}-\bar\Psi_{k,\cdot}\}(1;p_O^{(t)})\varepsilon_j^{(t)})^2\\
	+ 2\max_{k,j\in V_O} \frac{1}{n}\sum_{t=1}^n \{\hat \Psi_{k,\cdot}(1;p_O^{(t)})\{v_j-\hat v_j +(\hat W_{j,\cdot}-H^{-1}_{j\cdot})p_O^{(t)}\}\}^2 \\
	\leq 2\max_{k\in V_O} \|\hat\Psi_{k,\cdot}-\bar\Psi_{k,\cdot}\|_1^2\max_{t\in[n]}\|(1;p_O^{(t)})\|_\infty^2\max_{j\in V_O}\frac{1}{n}\sum_{t=1}^n (\varepsilon_j^{(t)})^2\\
	+ 2\max_{j\in V_O}\|(\hat v_j,\hat W_{j\cdot})-(v_j,H^{-1}_{j\cdot})\|_1^2\max_{t\in[n]}\|(1;p_O^{(t)})\|_\infty^2\max_{k\in V_O} \frac{1}{n}\sum_{t=1}^n \{\hat\Psi_{k\cdot}(1;p_O^{(t)})\}^2\\
	\leq 2 d_n^2 \bar p^2 C_4' + C_4'\{ n^{-1/2}\sqrt{s_n^2\log(V_On)}+r_{1n}\}^2\bar p^2 C_4'
	\end{array}$$
	for some constant $C_4'>0$,
	where the last inequality holds with probability $1-o(1)$.
	Here, we make use of the fact that
	 with probability $1-o(1)$  we have
 $\max_{k\in V_O} \|\hat\Psi_{k\cdot}-\bar\Psi_{k\cdot}\|_1\leq d_n$ by Assumption \ref{assumption:HL:Psi};  $\max_{j\in V_O}\|(\hat v_j,\hat W_{j\cdot})-(v_j,H^{-1}_{j\cdot})\|_1\leq C_5'\{n^{-1/2}\sqrt{s_n^2\log(V_On)}+r_{1n}\}$
	for some constant $C_5'>0$  by Lemma \ref{thm:DS:fast};  $\max_{j\in V_O}\frac{1}{n}\sum_{t=1}^n (\varepsilon_j^{(t)})^2\leq C_6'$
	for some constant $C_6'>0$
	by Lemma \ref{lemma:auxBoundsLinfty}; $\max_{k\in V_O} \frac{1}{n}\sum_{t=1}^n \{\hat\Psi_{k\cdot}(1;p_O^{(t)})\}^2 \leq C_7'$
	for some constant $C_7'>0$
	 by Lemma \ref{lem:ozanVersion};
as well as  the fact that
	  $|p_j^{(t)}|\leq \bar p$.
	
Since $\bar p$ is a bounded constant,  and $d_n\log(V_On) + n^{-1/2}s_n\log^{3/2}(V_O)+r_{1n}\log(nV_O)\leq \delta_n$ by the assumptions of the theorem,
	with probability $1-o(1)$ we have$$
	\begin{array}{l}
	{\displaystyle\max_{j'\in[p]}}\frac{1}{n}\sum_{t=1}^n(\hat Z_{tj'}-Z_{tj'})^2 \leq C_8'\delta_n^2/\log^2(V_On),
	\end{array}
	$$
	for some constant $C_8'>0$.
	This completes the verification of Condition A part (ii).
	Thus, Theorem 2.6 of \cite{belloni2018high}
	applies, and readily implies the first claim of the theorem. The second claim then follows by definition of the (simultaneous) confidence regions.
	\hfill
	\Halmos
	\endproof

\section{Consumption Equilibria and Structure of Optimal Prices} \label{app:NetworkInsight}

In this section, we focus on the problem introduced in Section
\ref{subse:estimationPrices}, and derive some properties of the consumption equilibria.
In addition, we shed light on the structure of optimal prices in this problem.

Our first result establishes that under Assumption \ref{ass:OutsideOpt}, at the consumption equilibrium all agents have nonzero consumption.

\begin{lemma} \label{lem:interiorCons}
	Suppose that Assumption \ref{ass:OutsideOpt} holds and that ${p} \leq \bar{p} \cdot {e}$.
	Then, in the corresponding consumption equilibrium ${y}$,
	we have $y_i>0$ for all $i\in V$.
\end{lemma}
\noindent\proof{Proof of Lemma \ref{lem:interiorCons}.}
Assume for contradiction that there exists an agent $i$ whose equilibrium consumption level is $y_i=0$. Note that the marginal payoff of this agent is given by
\[\frac{\partial u_i^{(t)}(y_i,y_{-i})}{\partial y_i}= a_i+\xi_i^{(t)}- 2b_i y_i + \sum_j G_{ij} y_j -p_i=a_i+\xi_i^{(t)}+ \sum_j G_{ij}y_j -p_i \geq a_i+\xi_i^{(t)}-p_i.\]
Here the second equality follows since $y_i=0$, and the inequality follows since $G_{ij},y_j \geq 0$.
By Assumption \ref{ass:OutsideOpt},
for any agent $i\in V$ we have  $a_i+\xi_i^{(t)}>\bar p \geq p_i$. Thus, we conclude that the agent can improve her payoff by increasing $y_i$.
Hence, we obtain a contradiction to the assumption that ${y}$ is a consumption equilibrium, and the claim follows. \hfill \Halmos	
\endproof

We next provide a bound on agents' equilibrium consumption.
\begin{lemma} \label{lem:boundedConsumption}
	Suppose that Assumption \ref{ass:OutsideOpt} holds.
	Let ${y}^{(t)}$ denote the consumption equilibrium under price vector ${p}^{(t)}$.
	We have $\max_i |y^{(t)}_i|\leq \frac{1}{\zeta} \max_i |a_i+\xi_i^{(t)}-p^{(t)}_i|$.
\end{lemma}	
\proof{Proof of Lemma \ref{lem:boundedConsumption}.}
To simplify exposition, throughout the proof we suppress the time index $t\in \mathbb{Z}_{++}$.
By Lemma \ref{lem:interiorCons},
we have that equilibrium consumption levels satisfy ${y}>0$. Hence, the consumption levels are given by \eqref{eq:firstOrder}, and we obtain ${y} = M^{-1} ({a}+{\xi}-{p})$.
Note that $\|{y}\|_\infty \leq \| M^{-1}\|_\infty \| {a}+{\xi} -{p}\|_\infty$.
By Lemma \ref{lem:matBounds}, we have $\| M^{-1}\|_\infty \leq \frac{1}{\zeta}$. In addition, by the definition of the infinity norm we have
$\|{y}\|_\infty=\max_i |y_i|$  and $ \| {a}+{\xi}-{p}\|_\infty = \max_i |a_i+{\xi}_i-p_i|$. Hence it follows that
$\max_i |y_i|\leq \frac{1}{\zeta} \max_i |a_i+\xi_i-p_i|$.
\hfill \halmos
\endproof

We next  derive the properties of the optimal price vector for the problem introduced in Section \ref{subse:estimationPrices},
and shed light on its dependence on the  network structure.
We start with a simple observation: when
the constraint $0\leq {p}_O\leq \bar p \cdot {e}_O$ are not binding, an optimal solution to  problem \eqref{eq:sellerOpt} can be obtained in closed form.
\begin{lemma} \label{lem:optPrices}
	{	Suppose that Assumption \ref{ass:OutsideOpt} holds and $0<{p}_O^\star < \bar p \cdot {e}_O$; then}
	\begin{equation*}
		{p}_O^\star = (H^{-1}+H^{-T})^{-1}
		H^{-1} ({a}_O-S_{OL}({a}_L-{p}_L) )=
		H^{T}(H+H^{T})^{-1}
		({a}_O-S_{OL}({a}_L-{p}_L)) .
	\end{equation*}
\end{lemma}
\proof{Proof of Lemma \ref{lem:optPrices}.}
Using Lemma \ref{lem:firstOrderO} to express agents' equilibrium consumption levels, the optimization problem in \eqref{eq:sellerOpt} can be rewritten as follows:
\[
\max_{ 0\leq {p}_O \leq \bar{p} \cdot {e}_O} \qquad  \mathbb{E}_{\xi}[\langle {p}_O,
H^{-1} ({a}_O+{\xi}_O  -{p}_O) -H^{-1} S_{OL} ({a}_L+{\xi}_L  -{p}_L)
\rangle]
\]
Recalling that $ \mathbb{E}_{\xi} [{\xi}]=0$, this problem can alternatively be written as follows:
\begin{equation}\label{eq:sellerOptExp}
	\begin{aligned}
		\max_{ 0\leq{p}_O \leq \bar{p} \cdot {e}_O} \qquad &  \langle {p}_O,
		H^{-1} ({a}_O -{p}_O - S_{OL} ({a}_L -{p}_L) )
		\rangle .
	\end{aligned}
\end{equation}
Since, the optimal solution is assumed to satisfy
$0<{p}_O^\star < \bar{p} \cdot {e}_O$, first-order conditions yield
\[
H^{-1} ( {a}_O - {p}^\star_O - S_{OL} ( {a}_L - {p}_L)) -H^{-T}  {p}^\star_O=0.
\]
Rearranging terms, we obtain
\begin{equation} \label{eq:auxFOC}
	H^{-1} ( {a}_O  - S_{OL} ( {a}_L - {p}_L))=
	(H^{-1} + H^{-T})  {p}^\star_O.
\end{equation}
By Lemma \ref{lem:MInv}, $(H^{-1} + H^{-T})$ is positive definite, and invertible.
Thus, the previous equality implies that
\[
{p}^\star_O=       (H^{-1} + H^{-T})^{-1}H^{-1}(  {a}_O  - S_{OL} ( {a}_L - {p}_L))
.
\]
Since $(H^{-1} + H^{-T})^{-1}H^{-1}= H^{T}(H+H^T)^{-1}$,
this can alternatively be expressed as
\[{p}^\star_O=      H^{T}(H+H^T)^{-1}( {a}_O  - S_{OL} ({a}_L -{p}_L)),\] as claimed.
\hfill      \Halmos
\endproof

If the constraints
$0\leq p_O\leq \bar p \cdot  {e}_O$
are binding
for   some set $S\subset V_O$ of agents, then
for all   $i\in S$   we have  $p_i=\bar p$
or $p_i=0$, and for $i\in V_O \setminus S$, a similar characterization to the one in Lemma~\ref{lem:optPrices}   can be obtained.

We next present two corollaries of this lemma.
First, assuming that the influence structure is symmetric,
agents share identical $\{a_i,b_i\}$ parameters,
and $\bar p $ is not small, we
show that the optimal solution is always interior, and prices  admit a  simpler characterization:
\begin{corollary}\label{lem:optPriceSym}
	Suppose that   $G=G^T$,
	$a_i=\tilde{a}$, $b_i=\tilde{b}$ for all $i\in V$. Let
	\begin{equation} \label{eq:candPrice}
		{q}_O :=
		\frac{\tilde{a}}{2}{e}_O-\frac{1}{2} S_{OL} {e}_L(\tilde{a}-\bar{p}).
	\end{equation}
	If $0<{p}_O^\star < \bar p \cdot{e}_O$,
	then  the optimal prices are given by ${p}_O^\star= {q}_O$. Moreover,  for sufficiently large
	$\bar p<\tilde{a}$, this is always the case.
\end{corollary}
\proof{Proof of Corollary \ref{lem:optPriceSym}.}
Observe that when $G=G^T$, we have $M=M^T$, $M_{OO}=M_{OO}^T$, $M_{LL}=M_{LL}^T$, $M_{OL}=M_{LO}^T$, and $H=H^T$.
In this case, we have the following identity $(H^{-1}+H^{-T})^{-1} H^{-1}=I/2 $.
Moreover, when $0<{p}_O^\star<\bar p \cdot {e}_O$, ${p}_O^\star$ is given as in Lemma \ref{lem:optPrices}.
Using the characterization in this lemma, together with the aforementioned identity, yields
\[
{p}_O^\star = \frac{1}{2}
({a}_O-S_{OL}({a}_L-{p}_L) )={q}_O,
\]
where the last equality follows
since ${a}_O=\tilde a \cdot {e}_O$,
${a}_L=\tilde{a} \cdot {e}_L$, and ${p}_L=\bar{p} \cdot {e}_L$ under the assumptions of the corollary.

Thus, to complete the proof, it suffices to show that $0<{p}_O^\star < \bar p \cdot {e}_O$ for sufficiently large $\bar p$.
Observe that  as $\bar p \rightarrow \tilde{a}$, we have ${q}_O \rightarrow \frac{\tilde{a}}{2} {e}_O$.
Hence,  for sufficiently large $\bar p< \tilde{a}$, we have ${q}_O \approx   \frac{\tilde{a}}{2} {e}_O < \bar p \cdot {e}_O$.

Recall that the platform's optimization problem is equivalently given by \eqref{eq:sellerOptExp}. On the other hand, the constructed ${q}_O$ is feasible in this problem, and it can be readily checked that it satisfies the first-order optimality conditions (given in \eqref{eq:auxFOC}). Hence, $0<{p}_O^\star={q}_O<\bar p \cdot {e}_O$ is an optimal solution for sufficiently large $\bar p< \tilde a$.
\hfill \Halmos
\endproof

It was established in \citet{candogan2012optimal} that
when all agents are observable, the optimal prices set by the platform are independent of the network structure, whenever the underlying influence structure is symmetric (and $a_i=\tilde a, b_i=\tilde b$ for all $i\in V$). Interestingly,
Corollary \ref{lem:optPriceSym} shows that this is no longer the case when there are latent nodes.
In this case, the network structure  manifests itself
through the $S_{OL}=M_{OL}M^{-1}_{LL}$ term, and impacts the optimal prices. In order to gain further insight into the impact of the network structure on prices, we first introduce the notion of \emph{Bonacich centrality} of agents
\citep[see][]{ballester2006s,candogan2012optimal}).

\begin{definition}[Bonacich Centrality]
	For a network with (weighted) adjacency matrix G and scalar $\alpha$, the Bonacich centrality vector of parameter $\alpha$ is given by
	$\mathcal{K}(\alpha,G)=(I-\alpha G)^{-1} \mathbf{1}$,
	provided that $(I-\alpha G)^{-1}$ is well defined and nonnegative.
\end{definition}
We use the shorthand notations $\mathcal{K}_L(\alpha):= \mathcal{K}(\alpha, G_{LL})$ and $\mathcal{K}_O(\alpha):=\mathcal{K}(\alpha, G_{OO})$ to  denote the centrality of agents after restricting attention to the latent and observable components of the networks, respectively.
The next result shows that the optimal prices in the symmetric case can be expressed in terms of  the Bonacich centrality of the latent component of the network.
\begin{corollary}\label{prop:symCase}
	Suppose that   $G=G^T$,
	$a_i=\tilde a$, $b_i=\tilde b$ for all $i\in V$.
	Then, for sufficiently large
	$\bar p < \tilde a$, the optimal prices are given as follows:
	\begin{equation*}
		{p}_O^\star=
		\frac{\tilde a}{2}{e}_O+
		\frac{\tilde a-\bar{p}}{4\tilde b}
		G_{OL} \mathcal{K}_L\left(\frac{1}{2\tilde b}\right)
		.
	\end{equation*}
\end{corollary}

\proof{Proof of Corollary \ref{prop:symCase}.}
Recall that for sufficiently large $\bar p< \tilde a$, the optimal price vector is as given in Corollary \ref{lem:optPriceSym}.
Also recall that
$S_{OL}=M_{OL} M_{LL}^{-1}$, and
$M_{LL}^{-1}{e}_L= (2\tilde{b}I-G_{LL})^{-1} {e}_L=\frac{1}{2b}  \mathcal{K}_L(\frac{1}{2\tilde{b}})$.
Substituting this expression in the optimal price vector provided in Corollary \ref{lem:optPriceSym}, it follows that
$	{p}_O^\star=
\frac{\tilde a}{2}{e}_O-
\frac{\tilde a-\bar p}{4\tilde{b}}
M_{OL}   \mathcal{K}_L(\frac{1}{2\tilde{b}})$. The result follows by noting that $M=\Lambda-G=2\tilde{b}I-G$; thus, $M_{OL}$ (which consists of the off-diagonal entries of  $M$) is equal to $-G_{OL}$. \hfill \Halmos
\endproof    	

Thus, in the symmetric case, the optimal prices of the platform have a simple and intuitive structure. In particular, the platform
first offers a nominal price of $\tilde a/2$ to all observable nodes (captured by the $\frac{\tilde a}{2} {e}_O$ term).
Then, she
considers the centralities of latent agents (captured by $\mathcal{K}_L(\frac{1}{2\tilde b})$), and increases the prices offered to observable agents, proportional to how much they are \emph{influenced by} the ``central'' latent agents  (captured by the term $G_{OL}
\mathcal{K}_L\left(\frac{1}{2\tilde b}\right)$). Note that this markup term suggests that not all observable agents should receive the same price. In particular, if an observable agent is strongly influenced by a latent agent (and hence the relevant entry of $G_{OL}$ is large), and if this latent agent is central (and hence the corresponding entry of $\mathcal{K}_L(\frac{1}{2\tilde b})$ is large), then the platform should consider a significant markup for this agent.
Intuitively, this is the case since
such observable agents have a strong incentive to consume the product (due to the positive  influence of the latent agents on them), and the platform can
improve her profits by charging higher prices to those agents.

\section{Auxiliary Results}  \label{app:auxResults}

In this section, we provide some auxiliary results and technical lemmas
that are used in our analysis throughout the paper.
In particular, in Section \ref{subse:MatrixLemma}
we derive various properties of
 $M$ and $H$ matrices and their inverses.
 In Section \ref{subse:probalisticResults}, we
 derive some technical lemmas on which the analysis of Algorithm~1 builds.
 In Section \ref{subse:Assumptions}, we focus on Assumptions	
  \ref{assumption:basicRandom} and  \ref{Assumption:New}, and establish that the former implies the latter.
 Finally, in Section \ref{subse:litThms}, for completeness, we state some known
 concentration bounds  and other useful results from the literature, which
we leverage when deriving our results.
 	
 	\subsection{Matrix Identities and Preliminary Results} \label{subse:MatrixLemma}

Our first two results,
Lemma \ref{lem:MInv0} and
Lemma \ref{lem:MInv},
characterize  properties of $M$ and $H$ as well as their inverses.
Some of the characterizations assume strong connectivity of the underlying network. We say that a directed network is strongly connected if it is possible to reach  any node from every other node by traversing directed edges in the underlying network. 	
 	
 	\begin{lemma}\label{lem:MInv0}
 	  Let $H= (M_{OO}-M_{OL}M_{LL}^{-1}M_{LO})$, $S_{OL}=M_{OL}M^{-1}_{LL}$ and $S_{LO}=M_{LL}^{-1}M_{LO}$,
 	  and suppose that $M$ and $M_{LL}$ are invertible. We have:
 	  \begin{equation}
 	  	\hspace{-.5in}
 	  	M^{-1}=
 	  	\begin{bmatrix}
 	  		H^{-1}, & \quad  -H^{-1}S_{OL} \\
 	  		-S_{LO} 	H^{-1}, &\quad
 	  		M_{LL}^{-1}+ S_{LO} 	H^{-1} S_{OL}
 	  	\end{bmatrix}.
 	  \end{equation}
 	\end{lemma}
 	\proof{Proof of Lemma \ref{lem:MInv0}.}
 	Since $M_{LL}$ is invertible,
 	it can be verified that  $M$ can be expressed  in terms of the following matrix multiplication:
 	\begin{equation}\label{eq:M_UDL}
 		M= \begin{bmatrix}
 			I_O, & M_{OL} M_{LL}^{-1}  \\
 			0,  & I_L
 		\end{bmatrix}
 		\begin{bmatrix}
 			M_{OO}-M_{OL}M_{LL}^{-1}M_{LO}, & 0\\
 			0,  & M_{LL}
 		\end{bmatrix}
 		\begin{bmatrix}
 			I_{O}, &  0\\
 			M_{LL}^{-1}M_{LO}, & I_{L}
 		\end{bmatrix}.
 	\end{equation}
 	Since $M$ is invertible, it follows that the middle term is also invertible (as otherwise $M$ would necessarily be rank-deficient). It can be seen that its inverse is given   by
 	$$
 	\begin{bmatrix}
 		M_{OO}-M_{OL}M_{LL}^{-1}M_{LO}, & 0\\
 		0,  & M_{LL}
 	\end{bmatrix}^{-1}
 	=
 	\begin{bmatrix}
 		H, &  0\\
 		0, & M_{LL}
 	\end{bmatrix}^{-1}
 	=
 	\begin{bmatrix}
 		H^{-1}, &  0\\
 		0, & M_{LL}^{-1}
 	\end{bmatrix}.
 	$$
 	On the other hand,   the inverses of the matrices on left/right-hand side of \eqref{eq:M_UDL} can be written as
 	$$
 	\begin{bmatrix}
 		I_{O}, &  0\\
 		M_{LL}^{-1}M_{LO}, & I_{L}
 	\end{bmatrix}^{-1}
 	=
 	\begin{bmatrix}
 		I_{O}, &  0\\
 		-M_{LL}^{-1}M_{LO}, & I_{L}
 	\end{bmatrix}
 	=
 	\begin{bmatrix}
 		I_{O}, &  0\\
 		-S_{LO}, & I_{L}
 	\end{bmatrix}
 	$$ and
 	$$
 	\begin{bmatrix}
 		I_O, & M_{OL} M_{LL}^{-1}  \\
 		0 ,  & I_L
 	\end{bmatrix}^{-1}
 	=
 	\begin{bmatrix}
 		I_O, & -M_{OL} M_{LL}^{-1}  \\
 		0  , & I_L
 	\end{bmatrix}
 	=
 	\begin{bmatrix}
 		I_O, & -S_{OL}  \\
 		0  , & I_L
 	\end{bmatrix}.
 	$$
 	Using these observations together with \eqref{eq:M_UDL}, we get
 	\begin{equation}
 		M^{-1}= \begin{bmatrix}
 			I_{O}, &  0\\
 			-S_{LO}, & I_{L}
 		\end{bmatrix}
 		\begin{bmatrix}
 			H^{-1}, &  0\\
 			0, & M_{LL}^{-1}
 		\end{bmatrix}
 		\begin{bmatrix}
 			I_O, & -S_{OL}  \\
 			0  , & I_L
 		\end{bmatrix}.
 	\end{equation}
 	The claim follows by  multiplying these matrices.
 	\hfill \halmos
 	
\begin{lemma}\label{lem:MInv}
	Under Assumption \ref{ass:DiagDominance}, we have the following:
	\begin{itemize}
		\item[(i)] 	$M+M^T$ is positive definite.
		
		\item[(ii)] 	$M_{LL}$, $M_{OO}$, and $M$ are invertible.

		\item[(iii)] $M^{-1}+M^{-T}$ and $H^{-1}+H^{-T}$ are positive definite.
		
		\item[(iv)] $M^{-1} = \left(
		\sum_{k=0}^\infty (\Lambda^{-1 }G)^k \right) \Lambda^{-1}$. Moreover, if the underlying network is strongly connected and the edge weights are positive, then all entries of
		$M^{-1}$ and $H^{-1}$
		are positive.
		
		\item[(v)] Suppose that
		the edge weights are positive, and
		the induced subnetwork of latent agents is strongly connected.
		Let  $i,j\in V_O$, $i\neq j$ be such that $(i,j)\notin E$.
		If there is a directed edge from $i$ to a latent agent and another edge from a latent agent to $j$, then
		$H_{ij}<0$.
	\end{itemize}
	
\end{lemma}
\proof{Proof of Lemma \ref{lem:MInv}.}

(i) Under Assumption \ref{ass:DiagDominance}, the matrix
$M+M^T= (\Lambda - G )+(\Lambda -G^T) $ is (strictly) diagonally dominant. The
Gershgorin circle theorem (see, e.g., \cite{horn2012matrix}) implies that
all of the matrix's eigenvalues are positive, and hence positive definiteness readily follows.

(ii)  Assumption \ref{ass:DiagDominance} also implies that   $M_{LL}+M_{LL}^T$ is  diagonally dominant and hence positive definite.
Note that we have
$$
{y}_L^T M_{LL} {y}_L= \frac{1}{2} {y}_L^T (M_{LL}+M_{LL}^T) {y}_L>0
$$
for ${y}_L\neq 0$. Thus, we conclude  that $M_{LL}$ is nonsingular and invertible. Repeating same argument for $M_{OO}$ and $M$, we obtain that these matrices are also nonsingular and invertible.

(iii) Fix any ${y}\in \mathbb{R}^{|V|}$ such that
${y}\neq 0$.
Define $\hat{{y}}=M^{-1}{y}$, and observe that since $M^{-1}$ is invertible and full rank, we have $\hat{{y}}\neq 0$.
We have
\[{y}^T (M^{-1}+M^{-T}) {y} =
{y}^T M^{-T} (M+M^T) M^{-1} {y}=
\hat{{y}}^{T}
(M+M^T) \hat{{y}}>0,\]
where the inequality follows from part (i).
Since ${y}\neq 0$ is arbitrary, it follows that $(M^{-1}+M^{-T}) $ is positive definite.
By part Lemma
\ref{lem:MInv0}
 $H^{-1}+H^{-T}$ is a submatrix of $(M^{-1}+M^{-T}) $, and hence it follows that this matrix is also positive definite.

(iv)
By  Assumption \ref{ass:DiagDominance}, $\Lambda^{-1}G$ is a matrix with nonnegative entries and row sums strictly bounded by one. Hence,
the Perron--Frobenious theorem (see, e.g., \citet{horn2012matrix})  implies that the spectral radius of $\Lambda^{-1}G$ is bounded by one. Thus, the inverse of $M^{-1}$ admits the following power series representation, which by the bound on the spectral radius of $\Lambda^{-1}G$ is convergent:
\begin{equation}\label{eq:powerSeries}
M^{-1}=(\Lambda-G)^{-1}=(I-\Lambda^{-1} G)^{-1} \Lambda^{-1}=  \left(\sum_{k=0}^{\infty} (\Lambda^{-1}G)^k \right) \Lambda^{-1}.
\end{equation}

Since $\Lambda$ is a diagonal matrix with strictly positive diagonal entries, and the edge weights are positive, \eqref{eq:powerSeries} implies that if the underlying network is strongly connected, then for large enough $k$,  all entries of $(\Lambda^{-1} G)^k$ are strictly positive.
Hence,  all entries of $M^{-1}$ are also strictly positive.
The result for $H^{-1}$ trivially follows since $H^{-1}$ is a submatrix of $M^{-1}$ (by Lemma \ref{lem:MInv0}).

(v)
Recall that
$H=(M_{OO}-M_{OL}M_{LL}^{-1} M_{LO})$.
Following the same approach as before, we conclude that $M_{LL}=\Lambda_{LL}-G_{LL}$ is diagonally dominant, and
since the induced subnetwork of latent agents is strongly connected
we conclude that all entries of  $M_{LL}^{-1}$ are positive.
Thus, if $i,j\in V_O$
are
such that $i\neq j$ and
there is a directed edge from $i$ to latent agents and another edge from a latent agent to $j$, then
$T_{ij}>0$, where $T=M_{OL}M_{LL}^{-1} M_{LO}$.
If, in addition, there is no edge between $i$, $j$, then $(M_{OO})_{ij}=0$.
Since $H=M_{OO}-T$, it follows that $H_{ij}=-T_{ij}<0$.
\hfill \Halmos
\endproof

 We proceed by characterizing norms of matrices   $M$ and $H$ as well as some related matrices.
 It is possible to provide bounds on
 various norms of
 a matrix
 $A$
 (and its inverse)
 when the matrix satisfies diagonal dominance assumptions (similar to Assumption \ref{ass:DiagDominance}). We next summarize such a result due to \citet{VARAH19753}.

 \begin{theorem}[\citet{VARAH19753}] \label{theo:implicationsOfDiagDom}
 	Assume that for some $\zeta >0$, matrix $A$ is such that
 	either (a)
 	$A_{ii} \geq \sum_{j| j\neq i} |A_{ij}|  + \zeta $, or
 	(b) 	$A_{ii} \geq \sum_{j| j\neq i} |A_{ji}| + \zeta$.
 	We have the following bounds on the norm of $A$ and~$A^{-1}$:
 	\begin{itemize}
 		\item[(i)] If (a) holds, then
 		$||A||_\infty \geq \zeta$ and
 		$||A^{-1} || _\infty \leq \frac{1}{\zeta}$.
 		\item[(ii)] If  (b) holds, then
 		$||A||_1 \geq \zeta$ and
 		$||A^{-1}||_1 \leq \frac{1}{\zeta} $.
 		\item[(iii)] If  both (a) and (b) hold, then
 		$|| A^{-1}||_2^{-1}= \sigma_{min}(A) \geq \zeta  $, where $\sigma_{min}(A)$ denotes the smallest singular value of $A$.
 	\end{itemize}
 \end{theorem}

 Using this theorem, we next provide a result on norms of matrices  $M$ and $H$.
 \begin{lemma} \label{lem:matBounds}
 	Under Assumption \ref{ass:DiagDominance}, we have the following:
 	\begin{itemize}
 		\item[(i)] $\| M \|_p \leq 2 \bar{\lambda}-\zeta $ and $\| M^{-1} \|_p \leq \frac{1}{\zeta}$ for $p\in\{1,\infty\}$.
 		\item[(ii)] $\| H \|_p \leq \frac{\bar \lambda^2}{\zeta}$ and $\| H^{-1} \|_p \leq  \frac{1}{\zeta}$ for $p\in\{1,\infty\}$.
 		\item[(iii)] $\|M\|_2\leq 2 \bar{\lambda}-\zeta$,  $\|H\|_2\leq \frac{\bar \lambda^2}{\zeta}$, $\|M^{-1}\|_2  \leq \frac{1}{\zeta}$, and $\|H^{-1}\|_2\leq \frac{1}{\zeta}$.
 		\item[(iv)] Singular values of $M$ belong to the interval $[\zeta, 2\bar{\lambda}-\zeta]$.
 	\end{itemize}
 \end{lemma}
 \proof{Proof of Lemma \ref{lem:matBounds}.}
 (i) By definition,
 $||M||_\infty = \max_i \sum_{j} |M_{ij}|\leq 2\bar \lambda -\zeta$, where the inequality follows since $\sum_{j} |M_{ij}|=\lambda_i + \sum_{j| j\neq i} |M_{ij}| \leq 2\lambda _i-\zeta$ and $\bar{\lambda}\geq \lambda_i$ for all $i$.
 The same argument also yields $||M||_1 = \max_j \sum_{i} |M_{ij}|\leq 2 \bar{\lambda}-\zeta$.

 Using Assumption \ref{ass:DiagDominance},
 Theorem \ref{theo:implicationsOfDiagDom} readily implies that $\| M^{-1} \|_p \leq \frac{1}{\zeta}$ for $p\in\{1,\infty\}$.

 (ii) Recall that
 $H=M_{OO}-M_{OL}M_{LL}^{-1}M_{LO}$.
 Using the fact that matrix norms are subadditive and submultiplicative, we obtain
 $\|H\|_p \leq \|M_{OO}\|_p + \|M_{OL}\|_p \|M_{LL}^{-1}\|_p  \|M_{LO}\|_p$.
 Observe that by Assumption \ref{ass:DiagDominance}, we have that absolute row/column sums of $M_{OL}$ and $M_{LO}$ are bounded by $2\bar \lambda-\zeta$. Thus, for $p\in \{1,\infty \}$, we have $\|M_{OL}\|_p,\|M_{LO}\|_p\leq 2\bar \lambda-\zeta$.
 Since $M_{OO}$ and $M_{LL}$ are diagonal blocks of $M$, they also satisfy Assumption \ref{ass:DiagDominance}. Thus, using the first part of the lemma, we obtain
 $\| M_{OO} \|_p \leq 2 \bar{\lambda}-\zeta$, and $\| M_{LL}^{-1} \|_p \leq \frac{1}{\zeta}$.
 Combining these observations, for $p\in\{1,\infty \}$, we obtain
 \[
 \|H\|_p \leq \|M_{OO}\|_p + \|M_{OL}\|_p \|M_{LL}^{-1}\|_p  \|M_{LO}\|_p
 \leq 2 \bar{\lambda}-\zeta  +   \frac{(\bar \lambda -\zeta)^2}{\zeta}=  \frac{\bar \lambda^2}{\zeta} .
 \]

 Observing that $H^{-1}$ is a submatrix of $M^{-1}$ (see Lemma \ref{lem:MInv0}), we have that
 $\|H^{-1} \|_p \leq \| M^{-1} \|_p $ for any $p$-norm. By the first part of the lemma, we obtain $\|H^{-1} \|_p \leq \frac{1}{\zeta}$.

 (iii) Using Holder's inequality, we have $\|A\|_2 \leq \sqrt{\|A\|_1 \|A\|_\infty }$ for any matrix $A$. The result then immediately follows from parts (i) and (ii) of the lemma.

 (iv) Since the largest singular value of $M$ is given by $\|M\|_2$, part (i) immediately implies that
 it is upper bounded by $2\bar{\lambda}-\zeta$.
 On the other hand, employing the version of the
 classic Gershgorin theorem for singular values, we have that
 the smallest singular value is bounded from below by
 $\min_i \left(|M_{ii} - \frac{1}{2} \left(\sum_{j\neq i} |M_{ij}| + \sum_{j\neq i} |M_{ji}| \right) \right)$
 (see \citet{johnson1989gersgorin}).
 Once again, using
 Assumption
 \ref{ass:DiagDominance}
 we conclude that the smallest singular value of $M$ is bounded from below by  $\min_i 2\lambda_i-(2\lambda_i-\zeta)=\zeta$. Hence, singular values of $M$ are contained in the interval  $[\zeta, 2\bar{\lambda}-\zeta]$.
 \hfill \halmos
 \endproof	
 	
 	Leveraging the characterization in this result, we next provide a useful characterization of the norms of the inverse of the $H^{-1} + H^{-T}$ matrix, which plays a key role in our analysis of the seller's revenues.
 	
\begin{lemma}\label{lem:Ainv}
	Let $A	=H^{-1} + H^{-T}$.
	 	Under Assumption \ref{ass:DiagDominance}
	we have
	$\|A^{-1}\|_2\leq \frac{ \bar{\lambda}^4}{2\zeta^3}$ and $\|A^{-1}\|_1\leq  \frac{ \bar{\lambda}^4}{2\zeta^3}$.
\end{lemma}
\noindent\proof{Proof of Lemma \ref{lem:Ainv}.}
It follows from the definition of $A$ that
\begin{equation} \label{eq:AinvBound}
\|A^{-1}\|_2=
\|(H^{-1}+H^{-T})^{-1}\|_2=
\|H^T(H^{T}+H)^{-1}H\|_2
\leq \|H\|_2^2  \|(H^{T}+H)^{-1}\|_2.
\end{equation}

Let ${I_O}$ be the
$|V_O|\times |V_O|$
identity matrix, and let $D$  be a $|V|\times|V|$ matrix
$D=[I_O, 0; 0, 0]$, where $0$'s denote blocks of appropriate size whose entries are all equal to zero.
Consider some $\zeta'\in (0,\zeta)$.
Observe that $M_{\zeta'}=M- {\zeta'} D $ is strictly (row and column) diagonally dominant by Assumption \ref{ass:DiagDominance}.

Consider the matrix $H_{\zeta'}=M_{OO}-{\zeta'}I_O-M_{OL} M_{LL}^{-1} M_{LO}=H-{\zeta'}I_O$.
Observe that $H_{\zeta'}$ is the Schur complement of $M_{LL}$ in $M_{\zeta'}$, i.e., $H_{\zeta'}= M_{\zeta'}/M_{LL}$, where `$/$' denotes the Schur complement.
It follows from \cite{lei2003schur} that when $M_{\zeta'}$ is strictly (row/column) diagonally dominant, so is
$H_{\zeta'}$.
Since this is true
for any ${\zeta'}\in(0,\zeta)$, and
$H_{\zeta'}=H-{\zeta'}I_O$,
it follows that $H$ is also (row and column) diagonally dominant, with a diagonal dominance gap of at least $\zeta$.
This implies that
$H^{T}+H$ is diagonally dominant with a diagonal dominance gap of $2\zeta$. Thus, from Gershgorin's theorem, it follows that all singular values are greater than $2\zeta$. Thus
$\|(H^{T}+H)^{-1}\|_2\leq \frac{1}{2\zeta}$.

By Lemma \ref{lem:matBounds} we  have
$\|H\|_2 \leq \frac{\bar{\lambda}^2}{\zeta}$.	
Now, using \eqref{eq:AinvBound}, we obtain
\[\|A^{-1}\|_2
\leq \|H\|_2^2  \|(H^{T}+H)^{-1}\|_2
\leq
\left(\frac{\bar{\lambda}^2}{\zeta}\right)^2 \frac{1}{2\zeta}=
\frac{\bar{\lambda}^4}{2\zeta^3}.
\]
Hence the first claim follows.

Following a similar approach, note that
$\|A^{-1}\|_1 \leq \|H\|^2_1  \|(H^{T}+H)^{-1}\|_1
$.
As before, observe that $H^{T}+H$ is diagonally dominant, with a gap of $2\zeta$ between the diagonal entries and the sum of the corresponding off-diagonal entries. Thus, by Theorem \ref{theo:implicationsOfDiagDom}, we have
$||(H^{T}+H)^{-1}||_1\leq 1/2\zeta$.
Lemma \ref{lem:matBounds} yields
\[
\|H\|_1 \leq \frac{\bar{\lambda}^2}{\zeta}.
\]
Using this to bound $||A^{-1}||_1$, we obtain
\[
\|A^{-1}\|_1 \leq \frac{\bar{\lambda}^4}{2\zeta^3} .
\]
Hence, the second part of the claim also follows.
\hfill \halmos
\endproof

For our findings in Section \ref{se:estimation} we  assumed that $M^{-1}$ has bounded absolute row/column sums, i.e., $\| M^{-1} \|_{1}, \|M^{-1}\|_\infty$ are bounded.
We also assumed that parameters  such as $a_i$ are bounded.
We next establish that these imply that $\|v_O\|_\infty$ is bounded.
As a side note
we emphasize that
the property that $M^{-1}$ has bounded absolute row/column sums
is weaker than Assumption \ref{ass:DiagDominance}, as can be seen from
Lemma \ref{lem:matBounds}. This implies that  the latter assumption is sufficient for the next lemma.
\begin{lemma}\label{lem:boundVo}
	$\|{v}_O\|_\infty \leq  \|M^{-1}\|_\infty		(2\bar a + \bar p)  $, and hence $\|{v}_O\|_\infty$ is bounded.
\end{lemma}
	\proof{Proof of Lemma \ref{lem:boundVo}.}
	Using \eqref{eq:defVo} it readily follows that
	$\|{v}_O\|_\infty \leq \|H^{-1}\|_\infty   \|{a}_O\|_\infty + \|H^{-1} S_{OL}\|_\infty \|({a}_L -{p}_L) \|_\infty$.
	On the other hand, Lemma \ref{lem:MInv0}
	implies that
	$\max\{\|H^{-1}\|_\infty ,  \|H^{-1} S_{OL}\|_\infty\} \leq \|M^{-1}\|_\infty$.
	Since $|a_i|\leq \bar a$ and $|p_i| \leq \bar p$ for all $i$, these observations imply that
		$\|{v}_O\|_\infty \leq
\|M^{-1}\|_\infty		(2\bar a + \bar p)$, and the claim follows. \hfill \halmos
	\endproof

 	\subsection{Technical Lemmas} \label{subse:probalisticResults}

 	\begin{lemma} \label{lem:lambdaScale}
 		Suppose that $C' \log |V_O| \geq \log n$. Then,
 		$\lambda \leq  \sqrt{  C_1 \frac{\log |V_O|}{n}}$ for some constant $C_1 >0$.
 		{Furthermore,
 			$\lambda \geq c_1 \sqrt{\frac{ \log(n|V_O|)}{n}}$ for some constant $c_1>0$.}
 	\end{lemma}
 	\proof{Proof of Lemma \ref{lem:lambdaScale}.}
 	The first desired inequality can be stated as
 	$\Phi^{-1}(1-n^{-1}/3|V_O|^2) \leq \sqrt{C_1  \log{|V_O|}}$.
 	Since $\Phi$ is a monotone function, this is equivalent to
 	$1-\Phi( \sqrt{C_1 \log{|V_O|}})\leq \frac{n^{-1}}{3|V_O|^2}$.
 	By applying the Chernoff bound  to the standard normal
 	we obtain
 	\begin{equation}\label{eq:lem9Step1}
 	1-\Phi(\sqrt{C_1\log{|V_O|}}) \leq e^{-(C_1\log|V_O|)/2} =
 	|V_O|^{-C_1/2}.
 	\end{equation}
 	Since
 	$C' \log |V_O| \geq \log n$, we have $|V_O|^{-C'} \leq n^{-1}$. Combining this with \eqref{eq:lem9Step1} for sufficiently large $C_1$  yields
 	$$1-\Phi(\sqrt{C_1 \log{|V_O|}}) \leq
 	|V_O|^{-C_1/2} \leq
 	\frac{n^{-1}}{3|V_O|^2}
 	. $$
 	Hence, the first inequality follows.
 	
 	Using the definition of $\lambda$, the second desired inequality can be stated as
 	$\Phi^{-1}(1-n^{-1}/3|V_O|^2) \geq c_1 \sqrt{ \log{(n|V_O|)}}$. This can equivalently be written as
 	$1- \Phi (c_1 \sqrt{ \log{(n|V_O|)} }) \geq n^{-1}/3|V_O|^2$.
 	Using bounds on the complementary CDF of the standard normal (see, e.g., \citet{chang2011chernoff})
 	we have $1- \Phi (c_1 \sqrt{ \log{(n|V_O|)}}) \geq \alpha_1 \exp(-\alpha_2 c_1^2  \log{(n|V_O|)} )$
 	for some constants $\alpha_1,\alpha_2>0$.
 	Choosing $c_1$ sufficiently small, this implies that
 	$1- \Phi (c_1 \sqrt{ \log{(n|V_O|)}}) \geq 1/3 n |V_O|$. The claim follows since
 	$ 1/3 n |V_O| \geq 1/3n|V_O|^2 $.
 	\hfill \halmos
 	\endproof
 	
 	\begin{lemma}\label{lem:hBound}
 		Let $\rho_1^{(t)},\rho_2^{(t)}\in \mathbb{R}$, and $\gamma^{(t)}\in \mathbb{R}^k$ be given real-valued scalars/vectors for $t=1,\dots,n$.
 		Suppose that $h(\beta):=\sqrt{\frac{1}{n}\sum_{t=1}^n  (\rho_1^{(t)})^2(\rho_2^{(t)} -\beta^T \gamma^{(t)})^2} $ for $\beta\in \mathbb{R}^k$.
 		Then, $|h(\beta_1)-h(\beta_2)| \leq
 		M_\gamma \cdot
 		\|\beta_1-\beta_2\|_1 $,
 		where
 		$M_\gamma:= \max_i{\sqrt{\frac{1}{n}\left(\sum_{t=1}^n (\rho_1^{(t)})^2 (\gamma_i^{(t)})^2 \right) } }$.
 	\end{lemma}
 	\proof{Proof of Lemma \ref{lem:hBound}.}
 	Observe that
 	\begin{equation}\label{eq:hNewBound1}
 	\begin{aligned}
 	\Bigg| \sqrt{n} \frac{\partial h(\beta)}{\partial \beta_i}
 	\Bigg|
 	&\leq
 	\frac{|\sum_{t=1}^n(\rho_1^{(t)})^2\gamma_i^{(t)}(\rho_2^{(t)} -\beta^T \gamma^{(t)})|}{\sqrt{\sum_{t=1}^n  (\rho_1^{(t)})^2(\rho_2^{(t)} -\beta^T \gamma^{(t)})^2}}\\
 	&\leq
 	\frac{\sqrt{\left(\sum_{t=1}^n (\rho_1^{(t)})^2 (\gamma_i^{(t)})^2 \right)
 			\left( \sum_{t=1}^n
 			(\rho_1^{(t)})^2	(\rho_2^{(t)} -\beta^T \gamma^{(t)})^2
 			\right)	} }{\sqrt{ \sum_{t=1}^n  (\rho_1^{(t)})^2(\rho_2^{(t)} -\beta^T \gamma^{(t)})^2}}\\
 	& \leq
 	{\sqrt{\left(\sum_{t=1}^n (\rho_1^{(t)})^2 (\gamma_i^{(t)})^2 \right) } },
 	\end{aligned}	
 	\end{equation}
 	where the second inequality follows from the Cauchy--Schwarz inequality.
 	Using \eqref{eq:hNewBound1}, it readily follows that
 	\begin{equation}
 	\| \nabla h(\beta)\|_\infty\leq
 	\max_i{\sqrt{\frac{1}{n}\left(\sum_{t=1}^n (\rho_1^{(t)})^2 (\gamma_i^{(t)})^2 \right) } }=M_\gamma.
 	\end{equation}
 	Thus, $h$ is Lipschitz continuous, and using Lipschitz continuity we obtain
 	\begin{equation}
 	|h(\beta_1)-h(\beta_2)| \leq M_\gamma \|\beta_1-\beta_2\|_1,
 	\end{equation}
 	as claimed. \hfill \halmos
 	\endproof
 	
 	\begin{lemma} \label{lem:ozanVersion}
 		{Let $\{Z^{(t)}_{ij}\}_{t\in [n], i\in S_1,j \in S_2 }$,
 			where $|S_\ell|\leq | \bar{V}_O|$
 			for $\ell \in \{1,2\}$,
 			be a collection of random variables,
 			and $c',C'$ be some constants such that $C'>c'>0$.
 			Fix some $L \geq 2$.
 		}
 		For each $t\in[n]$ denote by
 		$X^{(t)}$  a  vector of length $|S_1| \times |S_2|$ whose entries consist of
 		$\{Z^{(t)}_{ij}\}_{ i\in S_1,j \in S_2 }$.
 		Suppose that the following assumptions hold:
 		\begin{enumerate}[label=A\arabic*.]
 			\item $E[Z^{(t)}_{ij}]=0$,
 			$\frac{1}{n}\sum_{t=1}^nE[|Z^{(t)}_{ij}|^2]\geq c'$, and
 			$\frac{1}{n}\sum_{t=1}^nE[|Z^{(t)}_{ij}|^3]\leq C'$
 			for all $ i\in S_1,j \in S_2 $, $t\in [n]$.
 			
 			\item $\{X^{(t)}\}_{t\in[n]}$ are independent random vectors.
 			\item
 			$\frac{1}{n}\sum_{t=1}^nE[|Z^{(t)}_{ij}|^L]\leq C'$
 			for all $ i\in S_1,j \in S_2 $. Moreover,
 			$\bar{M}_L:=E[\max_{t\in[n] } \|X^{(t)}\|_\infty^L ]$,
 			with
 			$ \bar{M}_L \frac{\log |V_O|}{n} = o(1)$.
 			
 			\item $ C' \log |V_O| \geq  \log n$ and $\frac{\log |V_O|}{n}=o(1)$.
 			
 		\end{enumerate}
 		Then,
 		with probability at least $1-o(1)$
 		the following hold:
 		
 		\begin{enumerate}[label=\roman*.]
 			
 			\item {$\lambda  \geq  \max_{i\in S_1, j\in S_2}
 				\left(
 				{   \left|\frac{1}{n} \sum_{t=1}^n  Z^{(t)}_{ij}  \right| }\Big /{   \sqrt{\frac{1}{n}
 						\sum_{t=1}^n \{Z^{(t)}_{ij} \}^2}}
 				\right),
 				$
 				where
 				$\lambda = \frac{1}{\sqrt{n}}  \Phi^{-1}\left(1-\frac{1}{3n|V_O|^2}\right)$, and $\Phi$
 				denotes the cumulative density function of the standard normal distribution.}
 			
 			\item
 			$\max_{i\in S_1,j\in S_2}
 			\frac{1}{n}\sum_{t=1}^n|Z^{(t)}_{ij}|^L
 			\leq C''$  for some constant $C''>0$.
 			
 		\end{enumerate}

 	\end{lemma}
 	
 	\proof{Proof of Lemma \ref{lem:ozanVersion}.}
 	Observe that
 	using the bounds
 	on
 		$\frac{1}{n}\sum_{t=1}^nE[|Z^{(t)}_{ij}|^2]$ and
 	$\frac{1}{n}\sum_{t=1}^nE[|Z^{(t)}_{ij}|^3]$,
 	A1 readily implies that
 	$$0<C_1':=\frac{\sqrt{c'}}{(C')^{1/3}} \leq \left \{\frac{1}{n} \sum_{t=1}^n E [ |Z^{(t)}_{ij}|^2 ] \right \}^{1/2} \Big / \left \{\frac{1}{n} \sum_{t=1}^n E[|Z^{(t)}_{ij}|^{3}] \right\}^{1/3}.$$
 	Since A1 also implies that $E[Z^{(t)}_{ij}]=0$,
 	using Lemma  \ref{Lemma: MDSN} we obtain
 	\begin{equation} \label{eq:MDSN_bound_o}
 	P\left(  \frac{\left|\sum_{t=1}^n  Z^{(t)}_{ij}  \right|}{\sqrt{ \sum_{t=1}^n \{Z^{(t)}_{ij} \}^2}} \geq x \right) \leq
 	2(1-\Phi(x))\left( 1+\frac{A}{\ell_n^3} \right),
 	\end{equation}
 	where $A$ is
 	a universal
 	constant (i.e., a constant independent of the model primitives), $0<\ell _n\leq C_1' n^{1/6}$, and $0\leq x \leq  C_1' \frac{n^{1/6}}{\ell_n}-1$.
 	We let
 	$$\ell_n=\frac{C_1' n^{1/6}}{1+ \sqrt{n} \lambda}=\frac{C_1' n^{1/6}}{1+\Phi^{-1}(1-n^{-1}/(3|V_O|^2)) } \mbox{\quad and \quad } x=\sqrt{n}  \lambda=   \Phi^{-1}(1-n^{-1}/(3|V_O|^2)) .$$
 	Observe that this choice of $\ell_n$ and $x$ implies that $x= C_1' \frac{n^{1/6}}{\ell_n}-1$.
 	Moreover, since $1+ \sqrt{n} \lambda \geq 1$, we also have $\ell_n\leq C_1' n^{1/6}$.
 	Thus,  the bound in \eqref{eq:MDSN_bound_o}
 	applies with this choice of $\ell_n$ and $x$.
 	Moreover, for this choice of $\ell_n$ and $x$,
 	it follows  from
 	Lemma \ref{lem:lambdaScale} %
 	and
 	\eqref{eq:MDSN_bound_o} that
 	\begin{equation} \label{eq:MDSN_bound2_o}
 	\begin{aligned}
 	P\left(
 	\frac{\left|\sum_{t=1}^n  Z^{(t)}_{ij}  \right|}{\sqrt{ \sum_{t=1}^n \{Z^{(t)}_{ij} \}^2}}  \geq
 	\sqrt{n} \lambda
 	\right)
 	&\leq
 	\frac{2 n^{-1}}{(3|V_O|^2)}
 	\left( 1+\frac{C'_2 (1+\sqrt{n} \lambda)^3}{n^{1/2}}\right) \\
 	&\leq
 	\frac{2 n^{-1}}{(3|V_O|^2)}
 	\left( 1+\frac{C'_3 (\log|V_O|)^{3/2}}{n^{1/2}}\right)
 	\end{aligned}
 	\end{equation}
 	for some constants $C'_2,C'_3>0$.
 	Using the
 	union bound, we obtain
 	\begin{equation} \label{eq:MDSN_bound3_0}
 	\begin{aligned}
 	n^{-1}
 	\left( 1+\frac{C'_3 (\log|V_O|)^{3/2}}{n^{1/2}}\right)
 	&\geq
 	P\left(\max_{ i\in S_1,j \in S_2 }
 	\frac{\left|\sum_{t=1}^n  Z^{(t)}_{ij}  \right|}{\sqrt{ \sum_{t=1}^n \{Z^{(t)}_{ij} \}^2}}
 	\geq  \sqrt{n} \lambda
 	\right) \\
 	&={
 		P\left(\max_{ i\in S_1,j \in S_2 }
 		\frac{\left|\frac{1}{n}\sum_{t=1}^n  Z^{(t)}_{ij}  \right|}{\sqrt{ \frac{1}{n}\sum_{t=1}^n \{Z^{(t)}_{ij} \}^2}}
 		\geq \lambda
 		\right)
 	}\\
 	\end{aligned}
 	\end{equation}
 	Since
 	$\frac{\log |V_O|}{n}=o(1)$ by A4, the first claim follows.

 	By A3,  we have
 	$\frac{1}{n}\sum_{t=1}^nE[|Z^{(t)}_{ij}|^L]\leq C'$  for
 	all
 	$ i\in S_1,j \in S_2 $.
 	Thus, it follows that
 	\begin{equation} \label{eq:mkBound}
 	\bar{m}_L:=\max_{ i\in S_1,j \in S_2 } \frac{1}{n} \sum_{t=1}^n E[|Z^{(t)}_{ij}|^L]
 	\leq
 	C'_4,
 	\end{equation}
 	for some constant $C_4'>0$.
 	Moreover, by A2, $\{X^{(t)}\}_{t\in[n]}$ are independent vectors.
 	Using these observations and applying
 	Lemma \ref{lem:m2bound}, we obtain
 	\begin{equation}\label{eq:applyLemmaAppx_o}
 	{E}\left[\max_{ i\in S_1,j \in S_2 } \frac{1}{n}\left|\sum_{t=1}^n|Z^{(t)}_{ij}|^L-{E}[|Z^{(t)}_{ij}|^L]\right|\right] \leq
 	C'_5 \frac{\log  |\bar{V}_O| }{n}
 	\bar{M}_{L}+C'_5\sqrt{\bar{M}_{L} \frac{\log  |\bar{V}_O|}{n}},
 	\end{equation}
 	where $C'_5 > 0$ is a constant that depends on $C'_4$, and
 	$\bar{M}_L=E[\max_{t\in[n] } \|X^{(t)}\|_\infty^L ]$.

 	Using Markov's inequality with 	\eqref{eq:applyLemmaAppx_o}, we obtain
 	\begin{equation}\label{eq:alternativeMarkov_o}
 	\begin{aligned}
 	P\left(
 	\max_{ i\in S_1,j \in S_2 }  \frac{1}{n}\left(\sum_{t=1}^n|Z^{(t)}_{ij}|^L-{E}[|Z^{(t)}_{ij}|^L]\right) >c'_1
 	\right)
 	&\leq
 	P\left(
 	\max_{ i\in S_1,j \in S_2 }  \frac{1}{n}\left| \sum_{t=1}^n|Z^{(t)}_{ij}|^L-{E}[|Z^{(t)}_{ij}|^L]\right| >c'_1
 	\right)
 	\\
 	&\leq \frac{1}{c'_1}\left(
 	C_5' \frac{\log  |\bar{V}_O| }{n} \bar M_{L}+C_5'\sqrt{ \bar M_{L} \frac{\log  |\bar{V}_O|}{n}}\right).
 	\end{aligned}
 	\end{equation}
 	Note that
 	\eqref{eq:mkBound}
 	implies that $\frac{1}{n}\sum_{t=1}^n{E}[|Z^{(t)}_{ij}|^L] \leq C'_4$
 	for all
 	$ i\in S_1,j \in S_2 $.
 	It follows from this observation and
 	\eqref{eq:alternativeMarkov_o} that for some constants $C'_6,C'_7>0$, we have
 	\begin{equation}\label{eq:alternativeMarkov2_o}
 	P\left(
 	\max_{ i\in S_1,j \in S_2 }
 	\frac{1}{n}\sum_{t=1}^n|Z^{(t)}_{ij}|^L >C_6'
 	\right) \leq
 	C_7' \frac{\log  |\bar{V}_O| }{n} \bar M_{L}+C_7'\sqrt{\bar M_{L} \frac{\log  |\bar{V}_O|}{n}} .
 	\end{equation}
 	By A3 we also have $\bar M_{L} \frac{\log  |{V}_O|}{n} = o(1)$.
 	Hence, we conclude from \eqref{eq:alternativeMarkov2_o} that
 	with probability $1-o(1)$ the following inequality holds:
 	\begin{equation}\label{eq:boundedXtrSq_o}
 	\max_{ i\in S_1,j \in S_2 }
 	\frac{1}{n}\sum_{t=1}^n|Z^{(t)}_{ij}|^L
 	\leq C'_6.
 	\end{equation}
 	Thus, the second claim also follows.
 	\hfill\halmos
 	\endproof

 	\begin{lemma}\label{lemma:auxBoundsLinfty}
 		Under Assumption
 		\ref{assumption:basicRandom}  with probability $1-o(1)$ we have
 		\begin{equation}\label{eq:lemmaBounds}\| \En[ \varepsilon_O (1;p_O)^T ]\|_{e,\infty}  \leq  \tilde{C} \sqrt{\log (|V_O|)/n}  \ \ \ \mbox{and} \ \ \  \| \En[ \varepsilon_O p_O^T H^{-T} ]\|_{e,\infty}  \leq \tilde{C} \sqrt{\log (|V_O|)/n},
 		\end{equation}
 		as well as
 		$\max_{i\in V_O }
 		\frac{1}{n}\sum_{t=1}^n|\varepsilon_i^{(t)} |^2  \leq \tilde{C}$
 		for some constant $\tilde{C}\geq 0$.
 	\end{lemma}
 	\proof{Proof of Lemma \ref{lemma:auxBoundsLinfty}.}
 	First note that
by Lemma \ref{lem:MInv0} $H^{-1}$ is a submatrix of $M^{-1}$.
Since the latter matrix has bounded absolute row/column sums, we conclude that   $ \| H^{-1}\|_\infty \leq C'_1$ for some constant $C_1'$.
 	Therefore, using Holder's inequality
 	and the fact that the infinity norm of a matrix yields the maximum absolute row sum,
 	we obtain
 	\begin{equation}\label{eq:auxBoundLinfStep1}
 	\begin{aligned}
 	\| \En[ \varepsilon_O p_O^T H^{-T} ]\|_
 	{e,\infty}
 	&\leq
 	\left(\| \En[ \varepsilon_O p_O^T  ]\|_{e,\infty}\right)
 	\left(
 	  \max_{i\in V_O} \| (H^{-T})_{\cdot,i}\|_1
 	  \right)
 	  \\
 	&\leq
 	\left(\| \En[ \varepsilon_O p_O^T  ]\|_{e,\infty}\right)
 	\left(
 	\max_{i\in V_O} \| (H^{-1})_{i,\cdot}\|_1
 	\right)
 	=
 \| \En[ \varepsilon_O p_O^T  ]\|_{e,\infty} \cdot \|H^{-1}\|_\infty
 	\\
 	&\leq C'_1 \| \En[ \varepsilon_O p_O^T  ]\|_{e,\infty}.
 	\end{aligned} 	
 	\end{equation}

 	Let $Z^{(t)}_{ij}= \varepsilon_i^{(t)} (1;p_O^{(t)})_j$
 	for $t\in[n]$, $i\in V_O, j\in \bar V_{O}$.
 	Observe that
 	by Assumption \ref{assumption:basicRandom},   we have $E[Z_{ij}^{(t)}]=0$.
 	Note that ${E}[(1;{p}_O^{(t)})
 	(1;{p}_O^{(t)})^T]$ is a positive semidefinite matrix.
 	Since Assumption \ref{assumption:basicRandom} implies that
 	eigenvalues of
 	$\bar \Psi:=E[(1;p_O^{(t)})(1;p_O^{(t)})^T]^{-1}$  are upper bounded, it follows that
 	eigenvalues of
 	${E}[(1;{p}_O^{(t)})
 	(1;{p}_O^{(t)})^T]$
 	are
 	lower bounded by a constant strictly greater than zero.
 	Hence
 	$E[(p_j^{(t)})^2]\geq c_1'>0$ for some constant $c_1'$.
 	In addition, Assumption \ref{assumption:basicRandom} implies that
 	$\min_{j\in V_O} E[(\varepsilon_j^{(t)})^2|p_O^{(t)}] \geq c$.
 	Using these observations
 	we  obtain
 	\begin{equation}\label{eq:1stEqBound}
 	\begin{aligned}
 	E[(Z^{(t)}_{ij})^2] = E \left[ E[(Z^{(t)}_{ij})^2 | p_O^{(t)}] \right]
 	&\geq
 	E\left[  (1;p_O^{(t)})_j^2
 	E[(\varepsilon_i^{(t)})^2| p_O^{(t)}]  ] \right]
 	\geq c E\left[ (1;p_O^{(t)})_j^2\right] \geq { c_2'}  \\
 	\end{aligned}
 	\end{equation}
 	for some constant $c_2'>0$.
 	Moreover, using the fact that the covariates are bounded by $\bar p$,
 	Assumption~\ref{assumption:basicRandom},
 	and Jensen's inequality,
 	we conclude
 	\begin{equation}\label{eq:2ndEqBound}
 	\begin{aligned}
 	E[|Z^{(t)}_{ij}|^3] = E \left[ E[|Z^{(t)}_{ij}|^3 | p_O^{(t)}] \right] &\leq
 	E\left[
 	\bar{p}^3
 	E[(\varepsilon_i^{(t)})^3| p_O^{(t)}]   \right]
 	\leq
 	\bar{p}^3
 	E \left[E[(\varepsilon_i^{(t)})^4| p_O^{(t)}]  ]^{3/4}  \right]
 	\leq
 	\bar{p}^3C^{3/4} \leq C_2',
 	\end{aligned}
 	\end{equation}
 	for some constant  $C_2' >0$.

 	For each $t\in[n]$ denote by
 	$X^{(t)}$   a  vector of length $|V_O| \times |\bar{V}_O|$ whose entries consist of
 	$\{Z^{(t)}_{ij}\}_{ i\in V_O,j \in \bar{V}_O }$.
 	By  Assumption \ref{assumption:basicRandom} (and Jensen's inequality) we have  that
 	$\{X^{(t)}\}$ are independent vectors and that
 	\begin{equation}
 	\begin{aligned}
 	E[\max_{t\in[n] } \|X^{(t)}\|_\infty^2 ] &= E[  \max_{t \in [n],i\in {V}_O,j\in \bar{V}_O}
 	(\varepsilon_i^{(t) }(1;p_O^{(t)})_j)^2 ]
 	\leq \bar{p}^2 E\left[
 	\max_{t \in [n],i\in {V}_O}
 	(\varepsilon_i^{(t) })^2
 	\right]\\
 	& \leq
 	\bar{p}^2 E\left[
 	\max_{t \in [n],i\in {V}_O}
 	(\varepsilon_i^{(t) })^4
 	\right]^{1/2}
 	\leq
 	\bar{p}^2 \sqrt{M_\varepsilon},
 	\end{aligned}
 	\end{equation}
 	where $M_\varepsilon$ is such that
 	$ {M}_\varepsilon \frac{\log |V_O|}{n} = o(1)$.
 	Since Assumption \ref{assumption:basicRandom} also implies that
 	$\frac{\log |V_O|}{n}=o(1)$, we also
 	obtain
 	$ \sqrt{M_\varepsilon} \frac{\log |V_O|}{n} = o(1)$.
 	Note that
 	Jensen's inequality and \eqref{eq:2ndEqBound}  also yield
 	\begin{equation}\label{eq:Lem12IntBound}
 	E[|Z^{(t)}_{ij}|^2] \leq
 	E[|Z^{(t)}_{ij}|^3]^{2/3} \leq (C_2')^{2/3},
 	\end{equation}
 	and hence
 	$\frac{1}{n}\sum_{t=1}^nE[|Z^{(t)}_{ij}|^2]\leq C'_3$ for some constant $C_3'>0$.	
 	Finally, $ C' \log |V_O| \geq  \log n$ by  Assumption \ref{assumption:basicRandom}.
 	These observations collectively imply  that Lemma \ref{lem:ozanVersion} applies (with $L=2$) and hence,
 	with probability
 	$1-o(1) $,
 	we have
 	\begin{equation}\label{eq:auxBoundLinfStep2_o}
 	\max_{i\in V_O,j\in{ \bar{V}}_O} |{\mathbb{E}_n[\varepsilon_i^{(t)}(1;p_O^{(t)})_j]
 	}|
 	\leq
 	\lambda  \max_{i\in V_O,j\in{ \bar{V}}_O}  \sqrt{  \En[(\varepsilon_i^{(t)} )^2 (1;p_O^{(t)})_j^2 ]}.
 	\end{equation}
 	
 	Note that	Lemma \ref{lem:ozanVersion} also implies  that with probability $1-o(1)$, we have
 	\begin{equation}\label{eq:lem12LastClaim}
 	C_4' \geq
 	\max_{i\in V_O,j\in{ \bar{V}}_O}
 	\frac{1}{n}\sum_{t=1}^n|Z^{(t)}_{ij}|^2 =
 	\max_{i\in V_O,j\in{ \bar{V}}_O}
 	\frac{1}{n}\sum_{t=1}^n|\varepsilon_i^{(t)} (1;p_O^{(t)})_j|^2
 	\geq
 	\max_{i\in V_O }
 	\frac{1}{n}\sum_{t=1}^n|\varepsilon_i^{(t)} |^2
 	\end{equation}
 	for some constant $C_4' \geq 0$. Hence, the last claim  follows.
 	Moreover,
\eqref{eq:auxBoundLinfStep2_o},
 	\eqref{eq:lem12LastClaim}, and
 	Lemma \ref{lem:lambdaScale} imply that with probability $1-o(1)$ we have
 	$$ \| \En[ \varepsilon_O (1;p_O)^T ]\|_{e,\infty} \leq \sqrt{C'_4} \lambda
 	\leq C'_5 \sqrt{\frac{\log |V_O|}{n}},$$
 	for some constant $C_5'>0$.
Finally,
 	using this inequality together with
 	\eqref{eq:auxBoundLinfStep1}, we also conclude that
 	$ \| \En[ \varepsilon_O p_O^T H^{-T} ]\|_{e,\infty}  \leq C'_{6} \sqrt{\log |V_O|/n}$
 	for some constant $C_{6}'> 0$, with probability $1-o(1)$.
 	\hfill\halmos
 	\endproof

\subsection{Assumption \ref{assumption:basicRandom} vs. Assumption \ref{Assumption:New}}
\label{subse:Assumptions}

In this section we establish that Assumption \ref{Assumption:New} is more general than (and hence implied by) Assumption \ref{assumption:basicRandom}.
The main result of this subsection  (Lemma \ref{lem:assBasicImpliesNew})
relies on the  following auxiliary lemma that relates to the restricted
eigenvalue condition (see \eqref{eq:kappaCond}), and is presented after the proof of this lemma.

\begin{lemma}\label{lemma:auxBoundsRE}
	Under Assumption \ref{assumption:basicRandom}
	with probability $1-o(1)$ we have that
	$\En[(1;p_O) (1;p_O)^T]$ has the restricted eigenvalue $\kappa_{\bar c}$
	satisfying
	$\kappa^2_{\bar c} \geq \tilde{c}$
	and that
	$\lambda M_n s_n /\kappa^2_{\bar{c}} \leq  1/8$,
	where $M_n = \sqrt{ 1\vee \max_{k\in V_O}\frac{1}{n}\sum_{t=1}^n(p_k^{(t)})^4 }$, and $\bar{c},\tilde{c}>0$ are constants.

\end{lemma}
\proof{Proof of Lemma \ref{lemma:auxBoundsRE}.}
Throughout the proof we assume that $s_n\geq 2$. This is so because
if  the result holds for $s_n\geq 2$ , then it readily follows from
the definition of restricted eigenvalues in
\eqref{eq:kappaCond} that it  holds for $s_n=1$ as well.

To prove the result, we first leverage
Lemma \ref{thm:RV34} to show that
$\En[\{(p_O- E[p_O])^T \beta\}^2]$ is lower bounded by a fraction of
$E[\{(p_O- E[p_O])^T \beta\}^2]$ (with probability $1-o(1)$),
for any vector $\beta$ with at most $s_n$ nonzero entries satisfying $\|\beta\|_2=1$. Then, we use this result with Lemma  \ref{lemma:Transfer} to obtain a characterization of
$\beta^T\En[p_Op_O^T]\beta$.
This characterization is in turn exploited to
derive the desired results on the restricted eigenvalues.

We start by providing a lower bound on
$\En[\{(p_O- E[p_O])^T \beta\}^2]$.
Using Lemma \ref{thm:RV34} with $X^{(t)}=p^{(t)}_O - E[p_O^{(t)}]$, and observing that
$(E[ \max_{t\in[n] }\|X^{(t)}\|_\infty^2])^{1/2} \leq \bar{p}$,  we obtain
\begin{equation} \label{eq:boundEvsAverageAux}
\begin{array}{rl}
&E\left[ \sup_{\|\beta \|_0\leq s_n, \|\beta \|_2 =1} \left| \En[ (\beta^T X)^2 - E[(\beta^T X)^2] ]\right|\right] \leq  C'_1 \delta_n^2 + C'_1 \delta_n \sup_{\|\beta\|_0\leq s_n, \|\beta\|_2 =1} \sqrt{\En E[(\beta^T X)^2]},
\end{array}
\end{equation}
where
$X=\{X^{(t)} \}_{t\in[n]}$,
$C'_1> 0$ is a constant,
$\| \beta \|_0$  stands for  the number of nonzero entries of vector $\beta$,
and
$$
\delta_n:= \frac{\bar{p} \sqrt{s_n}}{\sqrt n}\left(  \log^{1/2} |V_O| + (\log s_n) (\log^{1/2} |V_O|) (\log^{1/2} n) \right).
$$
Note that Holder's inequality implies that
\[
\sup_{\|\beta\|_0\leq s_n, \|\beta\|_2 =1} \sqrt{\En E[(\beta^TX)^2]} \leq
\sup_{\|\beta\|_0\leq s_n, \|\beta\|_2 =1} \sqrt{\En E[(\|\beta\|_1 \|X\|_\infty)^2]} \leq
\bar{p}
\sqrt{s_n}.
\]
Since
$(\log |V_O|) (\log n)^3  s_n^2  = o(n)$
we have $\log s_n =o(\log n)$, which, together with the previous inequality, implies that
\[
\delta_n \sup_{\|\beta\|_0\leq s_n, \|\beta\|_2 =1} \sqrt{\En E[(\beta^T X)^2]}\leq
\delta_n \bar{p}
\sqrt{s_n}
=\frac{\bar{p}^2 {s_n}}{\sqrt n}\left(  \log^{1/2} |V_O| + (\log s_n) (\log^{1/2} |V_O|) (\log^{1/2} n) \right) =o(1).
\]
Since $s_n \geq 1$, this  expression also implies that   $\delta_n^2 = o(1)$.
Combining these observations with \eqref{eq:boundEvsAverageAux}, we obtain
\begin{equation}\label{eq:boundEvsAverageAux2}
E\left[\sup_{ \|\beta\|_0\leq s_n,\|\beta\|_2=1} \left|\En[\{(p_O - E[p_O])^T\beta\}^2]- E[\{(p_O- E[p_O])^T\beta\}^2] \right| \right] =
O(\delta_n \sqrt{s_n})=
o(1).
\end{equation}

By Assumption \ref{assumption:basicRandom}
eigenvalues of
$\bar \Psi$ are upper bounded by a constant. Hence
eigenvalues of
$\bar \Psi^{-1}= E[(1;p_O)(1;p_O)^T]$
are bounded away from zero by a constant.
{We claim that this implies that the eigenvalues of $$\tilde{P}:=E[(p_O-E[p_O])(p_O-E[p_O])^T]
	=E[p_O p_O^T] - E[p_O]E[p_O^T]
	$$
	are also bounded away from zero.
	To see this, first note that this matrix is positive semidefinite. Suppose that for some vector $\bar{\beta}$, we have $\bar{\beta}^T \tilde{P}\bar{\beta}=0$
	or, equivalently,
	$\bar{\beta}^T E[p_O p_O^T] \bar{\beta} - \bar{\beta}^TE[p_O]E[p_O^T] \bar{\beta}=0$.
	On the other hand,  this implies that
	\begin{equation}
	\begin{aligned}
	(
	-\bar{\beta}^T E[p_O]
	;
	\bar{\beta})^T
	\bar \Psi^{-1}
	(-\bar{\beta}^T E[p_O];
	\bar{\beta})
	&=
	(
	-\bar{\beta}^T E[p_O]
	;
	\bar{\beta})^T
	E[(1;p_O)(1;p_O)^T]
	(-\bar{\beta}^T E[p_O];
	\bar{\beta})
	\\
	&=
	(\bar{\beta}^T E[p_O])^2 +
	\bar{\beta}^T E[p_O p_O^T] \bar{\beta}-
	2 (\bar{\beta}^T E[p_O])^2\\
	& =
	\bar{\beta}^T E[p_O p_O^T] \bar{\beta} - \bar{\beta}^TE[p_O]E[p_O^T] \bar{\beta}=0.
	\end{aligned}
	\end{equation}
	Hence, we obtain a contradiction to the fact that the eigenvalues of $\bar \Psi^{-1}$ are bounded away from zero.
	Therefore, the eigenvalues of $\tilde{P}$ are bounded away from zero.
}

This observation in turn implies that   for any $\beta$ such that $\|\beta\|_2=1$ we have
\begin{equation}\label{eq:auxResultEigenBound}
E[\{(p_O- E[p_O])^T \beta\}^2] \geq c'_1>0
\end{equation}
for some constant $c'_1$.
Thus,  \eqref{eq:boundEvsAverageAux2} together with Markov's inequality implies that
for any $\beta$ such that
$\|\beta\|_0\leq s_n $ and
$\|\beta\|_2=1$,
with probability $1-o(1)$ we have
\begin{equation}\label{eq:tempBoundEqRes}
\En[\{(p_O-E[p_O])^T\beta\}^2] \geq (1-\eta)E[\{(p_O-E[p_O])^T\beta\}^2],
\end{equation}
where
$\eta = \omega(\delta_n \sqrt{s_n}) =o(1)$.
Note that the fact that this inequality holds for  $\|\beta\|_2=1$ implies that it holds for all $\beta \in \mathbb{R}^{|{V}_O|}$ satisfying $\|\beta\|_0\leq s_n $.
Therefore, by Lemma \ref{lemma:Transfer} we conclude that
\begin{equation}\label{eq:intStepAux}
\En[\{(p_O-E[p_O])^T \beta\}^2] \geq (1-\eta)E[\{(p_O-E[p_O])^T \beta\}^2]- \frac{\|D^{1/2}\beta\|_1^2}{s_n-1}
\end{equation}
for any $\beta \in \mathbb{R}^{|{V}_O|}$. Here,
$D$ is a diagonal matrix with nonnegative diagonal entries that can be chosen so that
for all $j\in {V}_O$,
\begin{equation} \label{eq:DDefinition}
\begin{aligned}
D_{jj} &= \max_{i\in {V}_O} ~
\eta\En[ (p_O-E[p_O])_i^2 ]+
(1-\eta) \left(\En[ ( p_O-E[p_O])_i^2 ]-
E[ ( p_O-E[p_O])_i^2 ] \right).
\end{aligned}
\end{equation}
Note that by \eqref{eq:tempBoundEqRes} and \eqref{eq:DDefinition}, we have
$D_{jj}\geq 0$.
Recalling that $p_i^{(t)}\leq \bar{p}$, we have that
$[ ( p_O^{(t)}-E[p_O^{(t)}])_i^2 ] \leq C_2'$ for some constant $C_2'>0$ and for all $t\in[n], i\in V_O$.
Moreover, by using Hoeffding's inequality,
we obtain
\[
P
\left(\left| \En[ ( p_O-E[p_O])_i^2 ]-
E[ ( p_O-E[p_O])_i^2 ]  \right|
\geq  C_4' \frac{\sqrt{\log |V_O|}}{\sqrt{n}}
\right)
\leq C_3' \frac{1}{|V_O|^{k'}}
\]
for some constants $C_3',C_4'>0$ and $k'>2$.
Thus, using the union bound we conclude that with probability $1-o(1)$,
\begin{equation} \label{eq:boundPDiff}
\max_{i \in {V}_O}
\left| \En[ ( p_O-E[p_O])_i^2 ]-
E[ ( p_O-E[p_O])_i^2 ]  \right|
\leq C_4' \frac{\sqrt{\log |V_O|}}{\sqrt{n}}.
\end{equation}
Using these observations in \eqref{eq:DDefinition} we obtain
\begin{equation}\label{eq:dBound}
|D_{jj}|\leq C_5'n^{-1/2}\sqrt{\log V_O} + C_5'\eta
\end{equation}
for some constant $C_5'> 0$.
Similarly, by using  Hoeffding's inequality,
with probability $1-o(1)$ we have
\begin{equation} \label{eq:boundPDiff2}
\max_{i \in {V}_O}
\left| \En[  (p_O)_i ]-
E[ ( p_O)_i ]  \right|
\leq
C_6'
\frac{\sqrt{\log |V_O|}}{\sqrt{n}}
\end{equation}
for some $C_6'>0$.

Note that for any $\beta \in \mathbb{R}^{|V_O|}$
such that $\|\beta\|_0\leq s_n$,
with probability $1-o(1)$
we have
\begin{equation} \label{eq:firstBoundPo}
\begin{aligned}
\beta^T\En[p_Op_O^T]\beta & = \beta^T
\En[((p_O-E[p_O]) +E[p_O]) ((p_O-E[p_O]) +E[p_O])^T] \beta\\
& = \beta^T\En[(p_O-E[p_O])(p_O-E[p_O])^T]\beta+2\beta^T\En[p_O-E[p_O]]\beta^TE[p_O]+(\beta^TE[p_O])^2\\
& \geq (1-\eta)\beta^T\{{\rm var}(p_O)\}\beta - \frac{\|D^{1/2}\beta\|_1^2}{s_n-1} + 2\beta^T(\En[p_O]-E[p_O])\beta^TE[p_O]+(\beta^TE[p_O])^2\\
& \geq (1-\eta)\beta^T\{{\rm var}(p_O)\}\beta - \frac{\|D^{1/2}\beta\|_1^2}{s_n-1} -2 \bar{p} \|\beta\|_1^2\|\En[p_O]-E[p_O]\|_\infty +(\beta^TE[p_O])^2\\
& \geq (1-\eta)\beta^T\{{\rm var}(p_O)\}\beta - \frac{\|D^{1/2}\beta\|_1^2}{s_n-1}-C'_7\|\beta\|_1^2
\sqrt{\frac{\log |V_O|}{n}}
+(\beta^TE[p_O])^2,
\end{aligned}
\end{equation}
where
${\rm var}  (p_O):= E[(p_O-E[p_O])(p_O-E[p_O])^T]$
and $C_7'>0$ is a constant. Here,
the first inequality follows from \eqref{eq:intStepAux}, and
the second one follows from Holder's inequality and the fact that $p_i^{(t)} \leq\bar{p}$ for all $i\in V_O$ and $t\in [n]$.
Finally, the
last inequality follows from \eqref{eq:boundPDiff2}.

Moreover, we have
\begin{equation} \label{eq:secondBoundPo}
\begin{aligned}
(\tilde \beta; \beta)^T \En[(1;p_O)  (1;p_O)^T](\tilde \beta;\beta) &= \tilde \beta^2 + \beta^T\En[p_Op_O^T]\beta+2\tilde \beta \En[p_O^T]\beta\\
& \geq \tilde \beta^2 + (\beta^TE[p_O])^2 +2\tilde \beta \En[p_O^T]\beta \\
& \quad+  (1-\eta)\beta^T\{{\rm var}(p_O)\}\beta
- \frac{\|D^{1/2}\beta\|_1^2}{s_n-1}
-C'_7\|\beta\|_1^2\sqrt{\frac{\log |V_O|}{n}}\\
& \geq
(\tilde \beta +\beta^TE[p_O])^2 -2 |\tilde \beta| \|\beta\|_1\|\En[p_O]-E[p_O]\|_\infty \\
&\quad +
(1-\eta)\beta^T\{{\rm var}(p_O)\}\beta
- \frac{\|D^{1/2}\beta\|_1^2}{s_n-1}
-C'_7\|\beta\|_1^2\sqrt{\frac{\log |V_O|}{n}}.
\end{aligned}
\end{equation}
Here, the first inequality follows from \eqref{eq:firstBoundPo}, and the second one follows from Holder's inequality.
Observe that  by \eqref{eq:boundPDiff2}
and the fact that
$2|\tilde \beta|\|\beta\|_1 \leq \|(\tilde \beta; \beta)\|_1^2$
we have
$2 |\tilde \beta| \|\beta\|_1\|\En[p_O]-E[p_O]\|_\infty
\leq C_6'   \| (\tilde{\beta}; \beta) \|_1^2 \frac{\sqrt{\log |V_O|}}{\sqrt{n}}
$.
In addition, note that
\begin{equation}
\begin{aligned}
\beta^T\{{\rm var}(p_O)\}\beta +(\tilde \beta +\beta^TE[p_O])^2 &=
\beta^T\{E[p_O p_O^T] - E[p_O]E[p_O]^T\}\beta
+ \tilde{\beta}^2 + 2 \tilde{\beta} \beta^TE[p_O] +(\beta^TE[p_O])^2\\
&= \beta^T\{E[p_O p_O^T]\}\beta+
\tilde{\beta}^2 + 2 \tilde{\beta} \beta^TE[p_O]\\
&=(\tilde \beta;\beta)^TE[(1;p_O)(1;p_O)^T](\tilde \beta;\beta).
\end{aligned}
\end{equation}
Combining these observations and the fact that
$(\tilde \beta +\beta^TE[p_O])^2\geq 0$ with  \eqref{eq:secondBoundPo} yields

\begin{equation} \label{eq:thirdBoundPo}
\begin{aligned}
(\tilde \beta; \beta)^T \En[(1;p_O)  (1;p_O)^T](\tilde \beta;\beta)
&\geq
(1-\eta) (\tilde \beta;\beta)^TE[(1;p_O)(1;p_O)^T](\tilde \beta;\beta) \\
&\quad
- \frac{\|D^{1/2}\beta\|_1^2}{s_n-1}
-C'_8\|(\tilde \beta;\beta)\|_1^2\sqrt{\frac{\log |V_O|}{n}},
\end{aligned}
\end{equation}
for some constant $C_8'>0$.

Let $\theta= (\tilde \beta; \beta)$, and suppose that
$J\subset \bar{V}_O$ is such that
$|J|\leq s_n$ and  $\|\theta_{J^c}\|_1\leq \bar{c} \|\theta_{J}\|_1$.
Note that this implies that $\|\theta\|_1 =
\|\theta_J\|_1+\|\theta_{J^c}\|_1\leq (1+\bar{c})\|\theta_J\|_1$.
Using
this observation together with
\eqref{eq:dBound}
we obtain that
$$\begin{array}{rl}
\|D^{1/2}\beta\|_1^2 & \leq \{C_5'n^{-1/2}\sqrt{\log |V_O|} + C_5'\eta\}\|\beta\|_1^2 \\
& \leq \{C_5'n^{-1/2}\sqrt{\log |V_O|} + C_5'\eta\}(1+\bar{c})^2\|\theta_{J}\|_1^2 \\
& \leq \{C_5'n^{-1/2}\sqrt{\log |V_O|} + C_5'\eta\}(1+\bar{c})^2s_n\|\theta_J\|_2^2 \\
& \leq \{C_5'n^{-1/2}\sqrt{\log |V_O|} + C_5'\eta\}(1+\bar{c})^2s_n\|\theta\|_2^2.
\end{array}$$
Using this inequality together with
\eqref{eq:thirdBoundPo}
and  recalling that
eigenvalues of $E[(1;p_O)(1;p_O)^T]$ are lower bounded by some $c_2'>0$, $s_n\geq 2$,
$\eta=o(1)$,
and that $s_n \sqrt{\frac{\log |V_O|}{n}}  = o(1)$ by Assumption \ref{assumption:basicRandom},
it follows that
\begin{equation} \label{eq:fourthBoundPo}
\begin{aligned}
\theta^T \En[(1;p_O)  (1;p_O)^T]\theta
&\geq
(1-\eta) c_2' \|\theta\|_2^2
-
\{C_5'n^{-1/2}\sqrt{\log |V_O|} + C_5'\eta\}(1+\bar{c})^2\frac{s_n}{s_n-1}\|\theta\|_2^2 \\
&\qquad
-C'_8\|\theta\|_1^2\sqrt{\frac{\log |V_O|}{n}}\\
&\geq c_3' \|\theta \|_2^2  - C'_8\|\theta \|_1^2\sqrt{\frac{\log |V_O|}{n}},
\end{aligned}
\end{equation}
for some constant $c_3'>0$.
On the other hand,
\begin{equation} \label{eq:thetaBound}
\|\theta\|_2^2 \geq \|\theta_J\|_2^2 \geq \|\theta_J \|_1^2/s_n \geq \|\theta \|_1^2/({s_n(1+\bar{c})^2}).
\end{equation}
Combining this with \eqref{eq:fourthBoundPo}
and the fact that
$s_n \sqrt{\frac{\log |V_O|}{n}}  = o(1)$
yields
\begin{equation} \label{eq:fifthBoundPo}
\begin{aligned}
\theta^T \En[(1;p_O)  (1;p_O)^T]\theta
&\geq c_3' \|\theta \|_1^2/({s_n(1+\bar{c})^2})  - C'_8\|\theta\|_1^2\sqrt{\frac{\log |V_O|}{n}} \\
&\geq c_4' \|\theta\|_1^2/{s_n} ,
\end{aligned}
\end{equation}
for some $c_4'>0$.
Since this inequality holds for any $\theta \in \mathbb{R}^{|\bar{V}_O|}$
and
$J\subset \bar{V}_O$  such that
$|J|\leq s_n$ and  $\|\theta_{J^c}\|_1\leq \bar{c} \|\theta_{J}\|_1$,
it follows that
with probability $1-o(1)$ we have
$\kappa^2_{\bar{c}} \geq c_4'>0$, as claimed.

Observe that $M_n\leq \bar{p}^2$. Hence, by Lemma \ref{lem:lambdaScale} we have
$\lambda M_n s_n /\kappa^2_{\bar{c}} \leq C_9' \sqrt{\frac{\log|V_O|}{n}} s_n$ for some constant $C_9'>0$.
On the other hand, since
$s_n \sqrt{\frac{\log |V_O|}{n}}  = o(1)$, it follows that
$\lambda M_n s_n /\kappa^2_{\bar{c}}=o(1)$. Thus, we obtain that
$\lambda M_n s_n /\kappa^2_{\bar{c}} \leq  1/8$
with probability $1-o(1)$.
\hfill \halmos
\endproof

\begin{lemma}\label{lem:assBasicImpliesNew}
	Assumption \ref{assumption:basicRandom} implies Assumption \ref{Assumption:New}.
\end{lemma}
\proof{Proof.}
It can be readily seen that
Assumption \ref{Assumption:New}\ref{ass7.2}
follows from
Assumption \ref{assumption:basicRandom}\ref{ass3.3} and Assumption \ref{assumption:basicRandom}\ref{ass3.5},
and
Assumption \ref{Assumption:New}\ref{ass7.4}
follows from
Assumption \ref{assumption:basicRandom}\ref{ass3.5}.
Moreover, Lemma \ref{lemma:auxBoundsRE} implies that
Assumption \ref{Assumption:New}\ref{ass7.3}
follows from Assumption \ref{assumption:basicRandom}.
We proceed by establishing
that Assumption \ref{assumption:basicRandom} implies
Assumption~\ref{Assumption:New}\ref{ass7.1}

Observe that by
Assumption \ref{assumption:basicRandom}\ref{ass3.1} we have
\begin{equation}\label{Anew:p10}
{E}[({\varepsilon}^{(t)}_O) ({p}^{(t)}_O)^T ]
=
E\bigg[
{E}[({\varepsilon}^{(t)}_O) ({p}^{(t)}_O)^T
\mid p_O^{(t)}
]
\bigg]=
E\bigg[
{E}[({\varepsilon}^{(t)}_O)
\mid p_O^{(t)}
]
({p}^{(t)}_O)^T
\bigg]=0.
\end{equation}
Assumption \ref{assumption:basicRandom}\ref{ass3.2} and \ref{ass3.3} imply that
\begin{equation} \label{Anew:p11}
E\bigg[  \{
(\varepsilon_{k}^{(t)}) (1;p_O^{(t)})_j \}^2\bigg]
=
E\bigg[  E \bigg[ \{
(\varepsilon_{k}^{(t)}) (1;p_O^{(t)})_j \}^2 \mid p_O^{(t)} \bigg]  \bigg]
\geq c
E\bigg[  (1;p_O^{(t)})_j ^2\bigg] \geq  c/ c_\Psi>0,
\end{equation}
where $c_\Psi>0$ denotes an upper bound on the eigenvalues of $\bar\Psi=E[(1;p_O^{(t)})(1;p_O^{(t)})^T]^{-1}$.
The last inequality makes use of the fact that $1/c_\Psi$  is a lower bound on the eigenvalues (and hence diagonal entries) of
$E[(1;p_O^{(t)})(1;p_O^{(t)})^T]$.

Similarly, by Assumption \ref{assumption:basicRandom}\ref{ass3.2}  and the fact that $\bar{p}\geq p_j^{(t)}$ for all $j\in V_O$, we have
\begin{equation}\label{Anew:p12}
E\bigg[
\left|
(\varepsilon_{k}^{(t)}) (1;p_O^{(t)})_j \right|^4 \bigg]
=
\bar{p}^4
E\bigg[
\left|
(\varepsilon_{k}^{(t)})  \right|^4 \mid p_O^{(t)} \bigg]
\leq C \bar{p}^4.
\end{equation}
Finally, by Assumption \ref{assumption:basicRandom}\ref{ass3.2} we also have
\begin{equation}\label{Anew:p13}
E\big[ \max_{t\in [n]; k\in V_O} |
\varepsilon_{k}^{(t)} |^4\big]
=
E\big[
E\big[
\max_{t\in [n]; k\in V_O} |
\varepsilon_{k}^{(t)} |^4 \mid \{p_O^{(t)}\}\big]
\big] \leq M_\varepsilon.
\end{equation}
On the other hand, Assumption \ref{assumption:basicRandom} also implies that
$M_\varepsilon \frac{\log|V_O|}{n} =o(1)$.
This observation, together with
Assumption \ref{assumption:basicRandom}\ref{ass3.1},
\eqref{Anew:p10},
\eqref{Anew:p11}, and \eqref{Anew:p12}, implies
Assumption \ref{Assumption:New}\ref{ass7.1}.
\hfill\halmos
\endproof

 	\subsection{Concentration Bounds and Other Useful Results from the Literature}\label{subse:litThms}
 	
The following technical lemma is a concentration bound; see \cite{BRT2014} for a proof.
In this result, and in the remainder of this subsection, a universal constant refers to
a constant that is independent of the  primitives of the relevant setting.

\begin{lemma}\label{lem:m2bound}
	Let $\{X^{(t)}\}_{t\in[n]}$ be independent random vectors in $\mathbb{R}^p$, where $p\geq 3$. Define $\bar m_k := \max_{j\leq p }\frac{1}{n}\sum_{t\in[n]} {E}[|X^{(t)}_{j}|^k]$ and $M_{k} \geq {E}[ {\displaystyle \max_{t\in [n]}}\|X^{(t)}\|_\infty^k]$.
	Then
	{\small $${E}\left[\max_{j\leq p}\frac{1}{n}\left|\sum_{t\in [n]}|X_{j}^{(t)}|^k-{E}[|X_{j}^{(t)}|^k]\right|\right] \leq \tilde{C}^2 \frac{\log p}{n}M_{k}+ \tilde{C}\sqrt{\frac{\log p}{n}}M_{k}^{1/2}\bar m_k^{1/2}, $$}
	$${E}\left[\max_{j\leq p}\frac{1}{n}\sum_{t\in[n]}|X_{j}^{(t)}|^k\right] \leq \tilde CM_{k} n^{-1}\log p+ \tilde C\bar m_k $$
	for some universal constant $\tilde C$.%
\end{lemma}

For the next two lemmas, we use the notation $\|\beta \|_0$ to denote the number of nonzero entries of a given vector $\beta$.
The first lemma is a variant of the main result in \cite{RudelsonVershynin2008}; see \cite{BCCW-ManyProcesses} for a proof.

\begin{lemma}\label{thm:RV34}
	Let  $X^{(t)}$, $t\in[n]$, be independent (across $t$) random vectors such that $X^{(t)}  \in \mathbb{R}^p$ with $p\geq 2$ and $(E[ \max_{t\in[n]}\|X^{(t)}\|_\infty^2])^{1/2} \leq K$. Furthermore, for $k\geq 1$, define
	$$
	\delta_n:= \frac{K \sqrt{k}}{\sqrt n}\left( \log^{1/2} p + (\log k) (\log^{1/2}p) (\log^{1/2} n) \right).
	$$
	Then
	$$
	E\left[ \sup_{\|\theta\|_0\leq k, \|\theta\|_2 =1} \left| \En[ (\theta^T X)^2 - E[(\theta^T X)^2] ]\right|\right]
	 \leq  \tilde C\delta_n^2 +  \tilde  C\delta_n \sup_{\|\theta\|_0\leq k, \|\theta\|_2 =1 } \sqrt{\En E[(\theta^TX)^2]}
	$$	
	for some universal constant $\tilde C>0$.
\end{lemma}

The next lemma,
whose  proof is based on Maurey's empirical method,
is due to \cite{oliveira2016}.

\begin{lemma}[Transfer principle]\label{lemma:Transfer} Suppose that $\hat \Sigma$ and $\Sigma$ are matrices with nonnegative diagonal entries, and assume that $\eta \in (0,1)$, $d\in[p]$ are such that
	$$ \forall v \in \mathbb{R}^p \ \ \mbox{with} \ \|v\|_0 \leq d, \ \ v^T\hat\Sigma v\geq (1-\eta)v^T\Sigma v.$$
	Assume that $D$ is a diagonal matrix whose elements $D_{jj}\geq 0$ and satisfy $D_{jj}\geq \hat \Sigma_{jj}-(1-\eta) \Sigma_{jj}$. Then
	$$ \forall x \in \mathbb{R}^p,  \ x^T\hat\Sigma x\geq (1-\eta)x^T\Sigma x - \frac{\|D^{1/2}x\|_1^2}{d-1}.$$
\end{lemma}

The last result, due to
\cite{jing2003self},
 is leveraged in
analyzing the performance of our algorithm.
\begin{lemma}[Moderate deviations for self-normalized sums]\label{Lemma: MDSN} Let $Z_{1}$,$\ldots$, $Z_{n}$ be independent, zero-mean random variables. Let
	$S_{n} = \sum_{i=1}^n Z_i,  \ \ V^2_{n} = \sum_{i=1}^nZ^2_i,$ $$M_n= \left \{\frac{1}{n} \sum_{i=1}^n E [ Z_i^2 ] \right \}^{1/2} \Big / \left \{\frac{1}{n} \sum_{i=1}^n E[|Z_i|^{3}] \right\}^{1/3}>0$$
	and $0< \ell_n \leq n^{\frac{1}{6}}M_n$. Then for some universal
	constant $A$,
	$$
	\left |\frac{P(|S_{n}/V_{n}|  \geq x) }{ 2 (1-\Phi(x))} - 1 \right |  \leq
	\frac{A}{ \ell_n^{3}},  \ \ 0 \leq  x \leq  n^{\frac{1}{6}}\frac{M_n}{\ell_n}-1. \\
	$$
\end{lemma}

    	\end{APPENDIX}

\end{document}